\documentclass[longauth]{aa}
\def \eg {e.g.}

\def \ie {i.e.}

\def \omegam {{\hbox{$\Omega_{\rm m}$}}}
\def \omegal {{\hbox{$\Omega_\Lambda$}}}
\def \hzero {{\hbox{$H_0$}}}
\def \arcmin {\hbox{$^\prime$}}
\def \arcsec {\hbox{$^{\prime\prime}$}}
\def \deg {\hbox{$^\circ$}}

\def \msun {\hbox{${\rm M_\odot}$}}

\def \mfive {\hbox{$M_{500}$}}

\def \rfive {\hbox{$r_{500}$}}

\newcommand{\kmsmpc }{\mbox{km s$^{-1}$ Mpc$^{-1}$}}

\newcommand{\mjyb }{\mbox{mJy beam$^{-1}$}}
\newcommand{\mujyb }{\mbox{$\mu$Jy beam$^{-1}$}}
\newcommand{\mujyarcsecsq }{\mbox{$\mu$Jy arcsec$^{-2}$}}
\newcommand{\muG }{\mbox{$\mu$G}}
\newcommand{\whz }{\mbox{W Hz$^{-1}$}}

\newcommand{\epic }{EPIC}
\newcommand{\epicE }{European Photon Imaging Camera}
\newcommand{\obsid }{ObsID}

\newcommand{\uv }{\textit{uv}}

\newcommand{\prefactor }{\textsc{prefactor}}

\newcommand{\wsclean }{\textsc{WSClean}}

\newcommand{\killms }{\textsc{killMS}}
\newcommand{\ddfacet }{\textsc{DDFacet}}

\newcommand{\mcmc }{MCMC}
\newcommand{\mcmcE }{Markov chain Monte Carlo}

\newcommand{\ciao }{\textsc{ciao}}
\newcommand{\ciaoE }{\textit{Chandra} Interactive Analysis of Observations}

\newcommand{\sas }{\textsc{sas}}

\newcommand{\halofdca }{\textsc{Halo-FDCA}}
\newcommand{\halofdcaE }{Halo-Flux Density CAlculator}

\newcommand{\xmm }{{\em XMM-Newton}}
\newcommand{\chandra }{{\em Chandra}}

\newcommand{\planck }{{\em Planck}}

\newcommand{\erosita }{{\em eROSITA}}

\newcommand{\gmrt }{GMRT}
\newcommand{\gmrtE }{Giant Metrewave Radio Telescope}
\newcommand{\ugmrt }{uGMRT}
\newcommand{\ugmrtE }{upgraded Giant Metrewave Radio Telescope}

\newcommand{\vla }{VLA}
\newcommand{\vlaE }{Very Large Array}
\newcommand{\jvlaE }{Karl G. Jansky Very Large Array}
\newcommand{\lofar }{LOFAR}
\newcommand{\lofarE }{LOw Frequency ARray}

\newcommand{\mwa }{MWA}
\newcommand{\mwaE }{Murchison Widefield Array}
\newcommand{\askap }{ASKAP}
\newcommand{\askapE }{Australian Square Kilometre Array Pathfinder}
\newcommand{\meerkat }{MeerKAT}
\newcommand{\wsrt }{WSRT}
\newcommand{\wsrtE }{Westerbork Synthesis Radio Telescope}

\newcommand{\lotss }{LoTSS}
\newcommand{\lotssE }{LOFAR Two-meter Sky Survey}
\newcommand{\lolss }{LoLSS}
\newcommand{\lolssE }{LOFAR LBA Sky Survey}
\newcommand{\lodess }{LoDeSS}

\newcommand{\apertif }{Apertif}
\newcommand{\apertifE }{APERture Tile In Focus}

\newcommand{\nvss }{NVSS}
\newcommand{\nvssE }{NRAO VLA Sky Survey}
\newcommand{\wenss }{WENSS}
\newcommand{\wenssE }{WEsterbork Northern Sky Survey}

\newcommand{\vlass }{VLASS}
\newcommand{\vlassE }{VLA Sky Survey}

\newcommand{\panstarrs }{Pan-STARRS}
\newcommand{\panstarrsE }{Panoramic Survey Telescope and Rapid Response System}

\usepackage{graphicx}
\usepackage{placeins}
\usepackage{amssymb}
\usepackage{multirow,bigdelim}
\usepackage{txfonts}
\usepackage[export]{adjustbox}
\usepackage{longtable}
\usepackage{lscape}
\usepackage{comment}
\usepackage{color}

\usepackage{hyperref}
\hypersetup{colorlinks,citecolor=blue,filecolor=black,linkcolor=red,urlcolor=black}
\begin{document} 

\title{The \planck\ clusters in the \lofar\ sky}
\subtitle{I. \lotss-DR2: New detections and sample overview}

\authorrunning{A. Botteon et al.} 
\titlerunning{The \planck\ clusters in the \lofar\ sky. I.}

\author{A. Botteon\inst{\ref{leiden}}, T. W. Shimwell\inst{\ref{astron},\ref{leiden}}, R. Cassano\inst{\ref{ira}}, V. Cuciti\inst{\ref{hamburg}}, X. Zhang\inst{\ref{leiden},\ref{sron}}, L. Bruno\inst{\ref{ira},\ref{unibo}}, L. Camillini\inst{\ref{iasf},\ref{unimi}}, R. Natale\inst{\ref{iasf},\ref{unimi}}, A. Jones\inst{\ref{hamburg}}, F. Gastaldello\inst{\ref{iasf}}, A. Simionescu\inst{\ref{sron},\ref{leiden},\ref{kavli}}, M. Rossetti\inst{\ref{iasf}}, H. Akamatsu\inst{\ref{sron}}, R. J. van Weeren\inst{\ref{leiden}}, G. Brunetti\inst{\ref{ira}}, M. Br\"uggen\inst{\ref{hamburg}}, C. Groeneveld\inst{\ref{leiden}}, D. N. Hoang\inst{\ref{hamburg}}, M. J. Hardcastle\inst{\ref{herts}}, A. Ignesti\inst{\ref{oapd}}, G. Di Gennaro\inst{\ref{hamburg}}, A. Bonafede\inst{\ref{unibo},\ref{ira}}, A. Drabent\inst{\ref{tls}}, H. J. A. R\"ottgering\inst{\ref{leiden}}, M. Hoeft\inst{\ref{tls}} and F. de Gasperin\inst{\ref{hamburg},\ref{ira}}}

\institute{
Leiden Observatory, Leiden University, PO Box 9513, NL-2300 RA Leiden, The Netherlands \label{leiden} \\
\email{botteon@strw.leidenuniv.nl} 
\and
ASTRON, the Netherlands Institute for Radio Astronomy, Postbus 2, NL-7990 AA Dwingeloo, The Netherlands \label{astron}
\and
INAF - IRA, via P.~Gobetti 101, I-40129 Bologna, Italy \label{ira}
\and
Hamburger Sternwarte, Universit\"{a}t Hamburg, Gojenbergsweg 112, D-21029 Hamburg, Germany \label{hamburg}
\and
SRON Netherlands Institute for Space Research, Niels Bohrweg 4, NL-2333 CA Leiden, The Netherlands  \label{sron}\and
Dipartimento di Fisica e Astronomia, Universit\`{a} di Bologna, via P.~Gobetti 93/2, I-40129 Bologna, Italy \label{unibo}
\and
INAF - IASF Milano, via A.~Corti 12, I-20133 Milano, Italy \label{iasf}
\and
Dipartimento di Fisica, Universit\`{a}  degli Studi di Milano, via Celoria 16 , I-20133 Milano, Italy \label{unimi}
\and
Kavli Institute for the Physics and Mathematics of the Universe (WPI), The University of Tokyo, Kashiwa, Chiba 277-8583, Japan\label{kavli}
\and
Centre for Astrophysics Research, University of Hertfordshire, College Lane, Hatfield AL10 9AB, UK \label{herts}
\and
INAF - Astronomical Observatory of Padova, vicolo dell'Osservatorio 5, I-35122 Padova, Italy \label{oapd}
\and
Th\"{u}ringer Landessternwarte, Sternwarte 5, D-07778 Tautenburg, Germany \label{tls}
}

\date{Received XXX; accepted YYY}

\abstract
{Relativistic electrons and magnetic fields permeate the intra-cluster medium (ICM) and manifest themselves as diffuse sources of synchrotron emission observable at radio wavelengths, namely radio halos and radio relics. Although there is broad consensus that the formation of these sources is connected to turbulence and shocks in the ICM, the details of the required particle acceleration, the strength and morphology of the magnetic field in the cluster volume, and the influence of other sources of high-energy particles are poorly known.}
{Sufficiently large samples of radio halos and relics, which would allow us to examine the variation among the source population  and pinpoint their commonalities and differences, are still missing. At present, due to the physical properties of the sources and the capabilities of existing facilities, large numbers of these sources are easiest to detect at low radio frequencies, where they shine brightly.}
{We examined the low-frequency radio emission from all 309 clusters in the second catalog of \planck\ Sunyaev Zel'dovich detected sources that lie within the 5634 deg$^2$ covered by the Second Data Release of the \lotssE\ (\lotss-DR2). We produced \lofar\ images at different resolutions, with and without discrete sources subtracted, and created overlays with optical and X-ray images before classifying the diffuse sources in the ICM, guided by a decision tree.}
{Overall, we found 83 clusters that host a radio halo and 26 that host one or more radio relics (including candidates). About half of them are new discoveries. The detection rate of clusters that host a radio halo and one or more relics in our sample is $30\pm11$\% and $10\pm6$\%, respectively. Extrapolating these numbers, we anticipate that once \lotss\ covers the entire northern sky it will provide the detection of $251\pm92$ clusters that host a halo and $83\pm50$ clusters that host at least one relic from \planck\ clusters alone. All images and results produced in this work are publicly available via the project website.}
{}


\maketitle
%

\section{Introduction}

Radio emission associated with galaxy clusters and their member galaxies is mainly related either to radio galaxies that are powered by a central active galactic nucleus (AGN) or to nonthermal components residing in the intra-cluster medium (ICM). While AGN are often bright sources that contribute the majority of the radio flux from a cluster at gigahertz frequencies, the diffuse sources generated by relativistic electrons (Lorentz factors of $\gamma_{\rm L} > 1000$) that propagate in the ICM magnetic field ($\sim$\muG\ level) have remained somewhat elusive, despite many extensive searches, due to their lower surface brightness and rapidly declining flux density with increasing frequency \citep[\eg,][for reviews]{feretti12rev, vanweeren19rev}. The observed levels of diffuse radio emission from clusters suggest that a few percent of the energy of a cluster merger is dissipated by shocks and turbulence in the ICM and transferred to nonthermal components (the largest amount goes into ICM heating; see \citealt{markevitch07rev}). However, the details of the particle acceleration and magnetic field amplification mechanisms on cluster scales are still poorly understood \citep[\eg,][for a review]{brunetti14rev}. Thus, by studying the emission associated with the ICM we can probe the fundamental physics of particle acceleration in highly rarefied plasmas that are beyond the reach of those that can be studied in laboratories, and more generally we can provide insights into large-scale structure formation and evolution. \\
\indent
Diffuse cluster sources are typically classified as radio halos, mini-halos, relics, and revived fossil plasma sources (or phoenixes) according to their location in the cluster, morphology, size, and radio spectral properties. Observations with many facilities, such as the \vlaE\ \citep[\vla;][]{thompson80}, the \wsrtE\ \citep[\wsrt;][]{hogbom74}, and the \gmrtE\ \citep[\gmrt;][]{swarup91}, have played a crucial role in the discovery of new cluster radio sources and in constraining their main properties \citep[\eg,][]{giovannini99, giovannini06, giovannini00, kempner01, venturi07, venturi08, rudnick09, vanweeren09gmrt, vanweeren11nvss}. These instruments, in combination with X-ray observations, have provided conclusive evidence that diffuse (up to megaparsec-scale) radio sources in the ICM are connected to the dynamical motions of the ICM. Proton-proton collisions in the ICM represent an alternative process for producing (secondary) electrons in clusters \citep[\eg,][]{dennison80, blasi99}; however, their contribution is likely not dominant enough to explain extended emission on megaparsec scales \citep[\eg,][]{jeltema11, zandanel14coma, brunetti17, adam21}. The search for correlations between diffuse radio sources and host cluster properties \citep[\eg,][]{liang00bullet, cassano07, cassano08revised, brunetti09evolution, degasperin14, yuan15} as well as their connection with the cluster dynamical state \citep[\eg,][]{buote01, cassano10connection, wen13, cuciti15, giacintucci17} is fundamental to unveiling the origin of these objects. However, until recently, many of these studies were hampered by the sensitivity of the observations, which has limited the number of detections of diffuse emission to about a hundred and the statistical analysis to very massive systems. This has challenged the overall interpretation of the population of these sources through theoretical models \citep[\eg,][]{cassano05, cassano06, nuza12, nuza17, bruggen20}. \\
\indent
Thanks to the increased sensitivity to these diffuse sources that has been made possible due to upgrades to facilities such as the \jvlaE\ \citep{perley11} and the \ugmrtE\ \citep[\ugmrt;][]{gupta17}, as well as the advent of new-generation interferometers, such as the \lofarE\ \citep[\lofar;][]{vanhaarlem13}, the \mwaE\ \citep[\mwa;][]{tingay13}, the \askapE\ \citep[\askap;][]{hotan21}, and \meerkat\ \citep{jonas09}, it is now possible to search for diffuse radio sources in clusters with a number of complementary and sensitive instruments. In particular, the leap forward in the capabilities of low-frequency interferometers allows us to study diffuse cluster sources in a regime where they are brighter due to their steep synchrotron spectra ($\alpha > 1$, with $S_\nu \propto \nu^{-\alpha}$, where $S_\nu$ is the flux density at frequency $\nu$ and $\alpha$ is the spectral index). In this respect, \lofar\ has recently enabled the first detailed observations of galaxy clusters at frequencies of $<$200 MHz thanks to the unprecedented high sensitivity and high resolution in its operational frequency range. This potential has already been demonstrated as \lofar\ has proved to be very fruitful in investigating different aspects of nonthermal phenomena in the ICM, allowing us: to discover new instances of diffuse sources in clusters \citep[\eg,][]{shimwell16, savini18planck139, savini19, wilber19}, including ultra-steep spectrum emission \citep[\eg,][]{bruggen18, wilber18a1132, mandal20, biava21rxj1720} and very large-scale emission outside the central cluster region \citep[\eg,][]{govoni19, botteon19lyra, botteon20a1758, bonafede21, hoeft21, hoang21clg0217}, as well as new faint halos and relics \citep[\eg,][]{botteon19lyra, botteon21ant, locatelli20, hoang21a990} and high-$z$ systems \citep[\eg,][]{cassano19, digennaro21fast}; to pinpoint the complex interplay between tailed cluster AGN and ICM motions \citep[\eg,][]{degasperin17gentle, clarke19, hardcastle19, botteon20a2255, botteon21a1775, ignesti21arx}; to study the central cluster AGN duty cycle, structure, and interaction with the hot ICM \citep[\eg,][]{brienza20, birzan20, biava21agn, timmerman22}; and to detect extended, extraplanar emission from star-forming galaxies infalling into clusters \citep[\eg,][]{ignesti20a2626, ignesti22jw100, roberts21a, roberts21b, roberts22perseus}. \\
\indent
Sensitive wide-area searches for diffuse cluster sources require significant observational time and are arguably most efficient at low frequencies due to the higher survey speed resulting from the larger field-of-view (FoV) of the instruments. In addition, much of the undiscovered population is thought to have very steep spectra \citep[$\alpha > 1.5$; \eg,][]{cassano05, cassano06}. This implies that, of the planned wide-area surveys, those at low frequencies are anticipated to make the largest number of discoveries \citep[\eg,][]{cassano10lofar, cassano12, nuza12}. In this respect, \lofar\ is carrying out wide and deep surveys, and it is currently observing the entire northern sky at 120$-$168 MHz and 42$-$66 MHz in the context of the \lotssE\ \citep[\lotss;][]{shimwell17} and the \lolssE\ \citep[\lolss;][]{degasperin21}, respectively. These surveys have enormous discovery potential in many fields of astrophysics, including galaxy cluster science, offering the opportunity to study large samples of objects in synergy with surveys performed at other wavelengths. As it is currently believed that the cluster mass is a key parameter for the formation of the most extended radio sources in the ICM (namely, halos and relics), catalogs of clusters detected via the Sunyaev-Zel'dovich (SZ) effect \citep{sunyaev72} are particularly interesting as they provide unbiased samples that are almost mass-selected and that are ideal to be cross-matched with the \lofar\ surveys. \\
\indent
This is Paper I of a series dedicated to the study of diffuse radio emission in the ICM of galaxy clusters selected from the second \planck\ catalog of SZ sources \citep[PSZ2;][]{planck16xxvii} that have also been covered by the Second \lotss\ Data Release \citep[\lotss-DR2;][]{shimwell22}. It represents an extension of our previous work \citep{vanweeren21}, which was based on the galaxy clusters covered by the First \lotss\ Data Release \citep[\lotss-DR1;][]{shimwell19}. Here, we present the new sample (Sect.~\ref{sec:sample}), describe the methods and data used (Sect.~\ref{sec:methods}), classify the cluster radio sources (Sect.~\ref{sec:classification}), provide the quantities used for the analysis that will be performed in subsequent papers, and present the new detections and the results of our study (Sects.~\ref{sec:fluxdensity} and ~\ref{sec:results}). In Cassano et al. (in preparation) and Cuciti et al. (in preparation), we discuss the occurrence and the scaling relations of radio halos in the sample, while in Jones et al. (in preparation) we focus on radio relics. Other papers dedicated to the study of the X-ray properties of the sample (Zhang et al., in preparation) and to the methods developed to derive upper limits to the diffuse cluster radio emission (Bruno et al., in preparation) are also forthcoming. \\
\indent
Hereafter, we adopt a $\Lambda$ cold dark matter cosmology, with $\omegal = 0.7$, $\omegam = 0.3$, and $\hzero = 70$ \kmsmpc.

\section{Cluster sample}\label{sec:sample}

\begin{figure}
 \centering
 \includegraphics[width=\hsize,trim={0cm 0cm 0cm 1cm},clip]{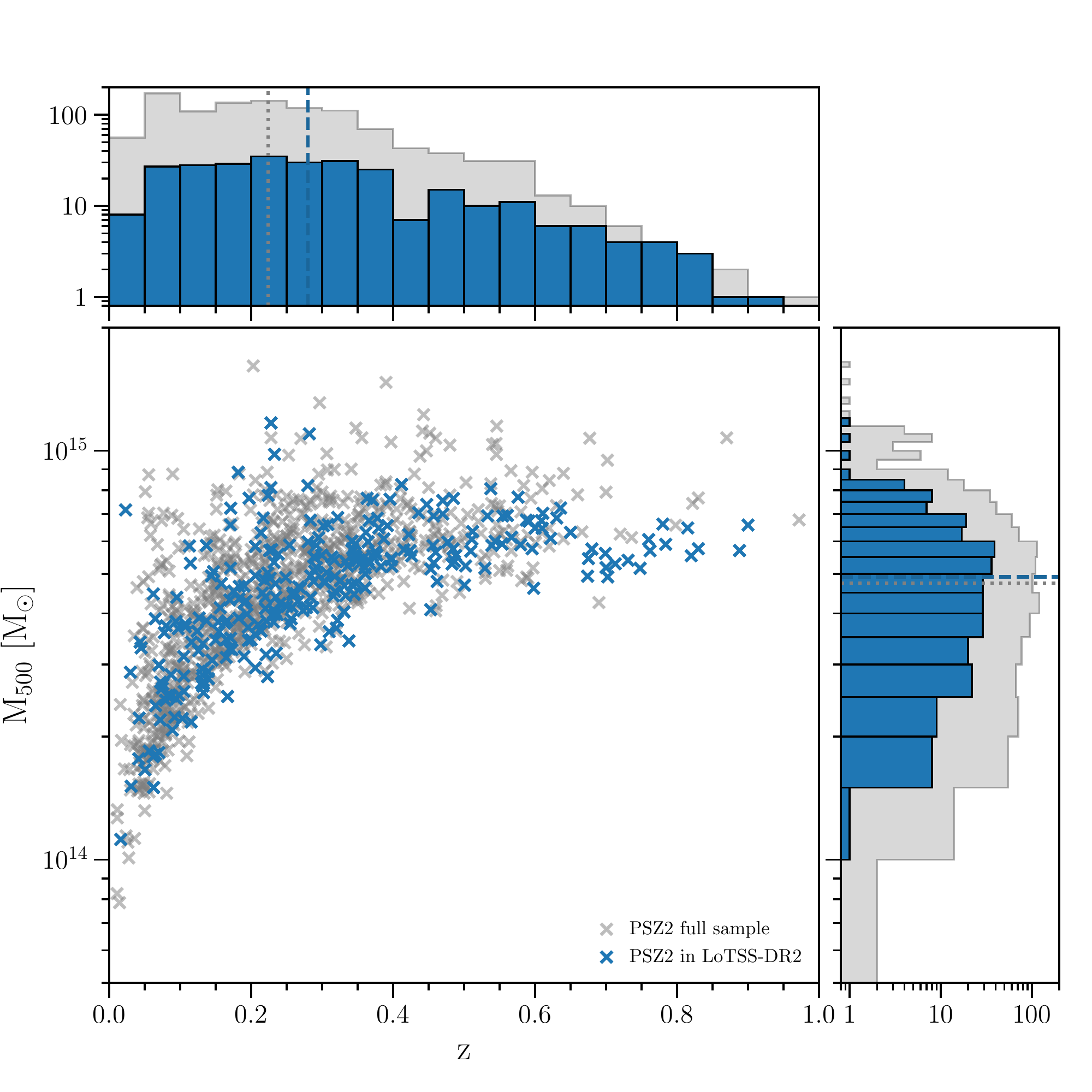}
  \caption{Redshift-mass distribution of PSZ2 sources. Clusters that are located in the \lotss-DR2 area are indicated in blue. The histograms show the number of clusters at various redshifts and masses; the dashed and dotted lines mark the median values of the \lotss-DR2 sample and the full PSZ2 sample, respectively. Similarly to the full PSZ2 sample, our sample spans a wide range of redshifts and masses.}
 \label{fig:dr_sample}
\end{figure}

\begin{figure*}
 \centering
 \includegraphics[width=\hsize,trim={0cm 0cm 0cm 0cm},clip]{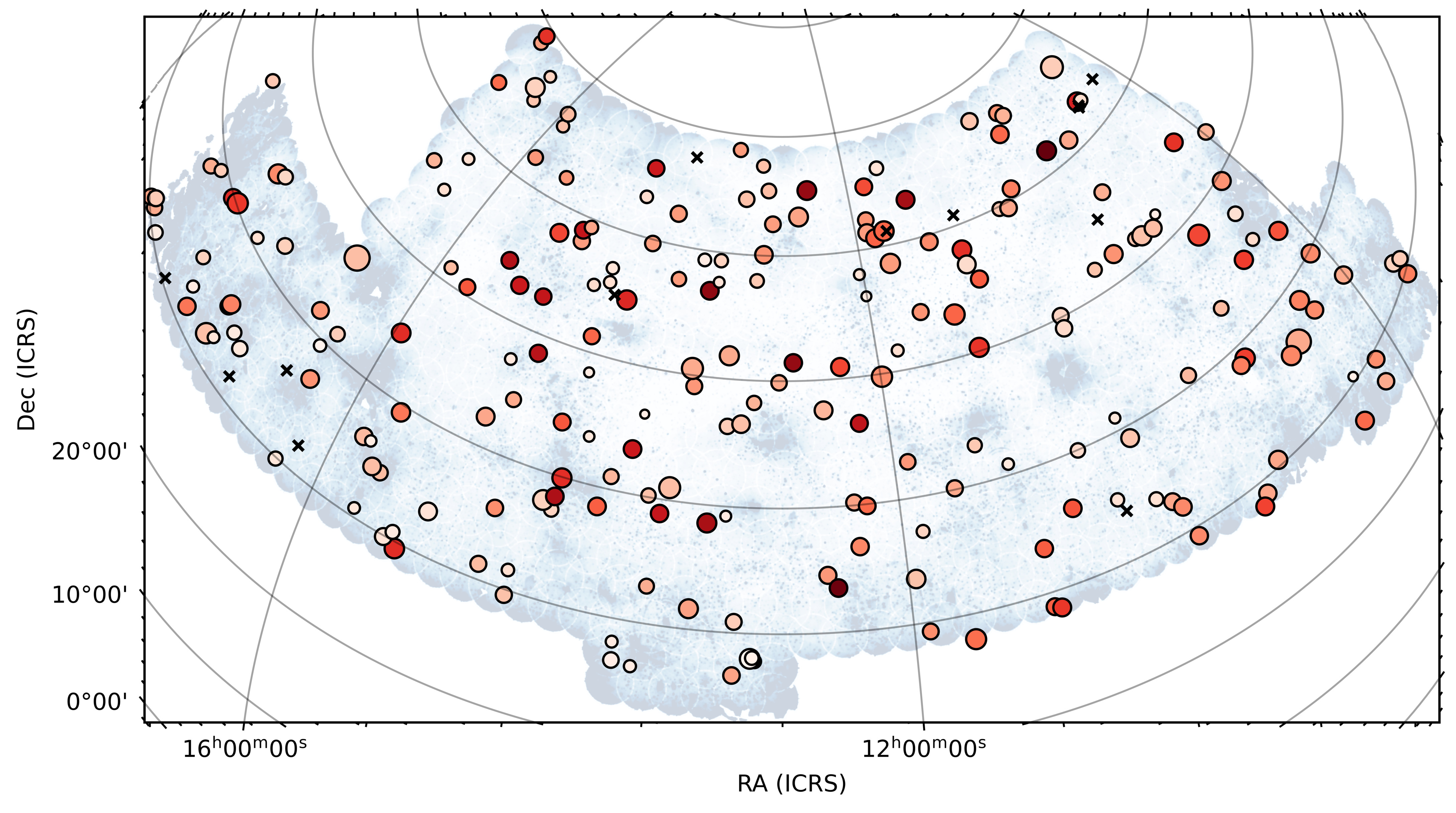}
 \includegraphics[width=\hsize,trim={0cm 0cm 0cm 0cm},clip]{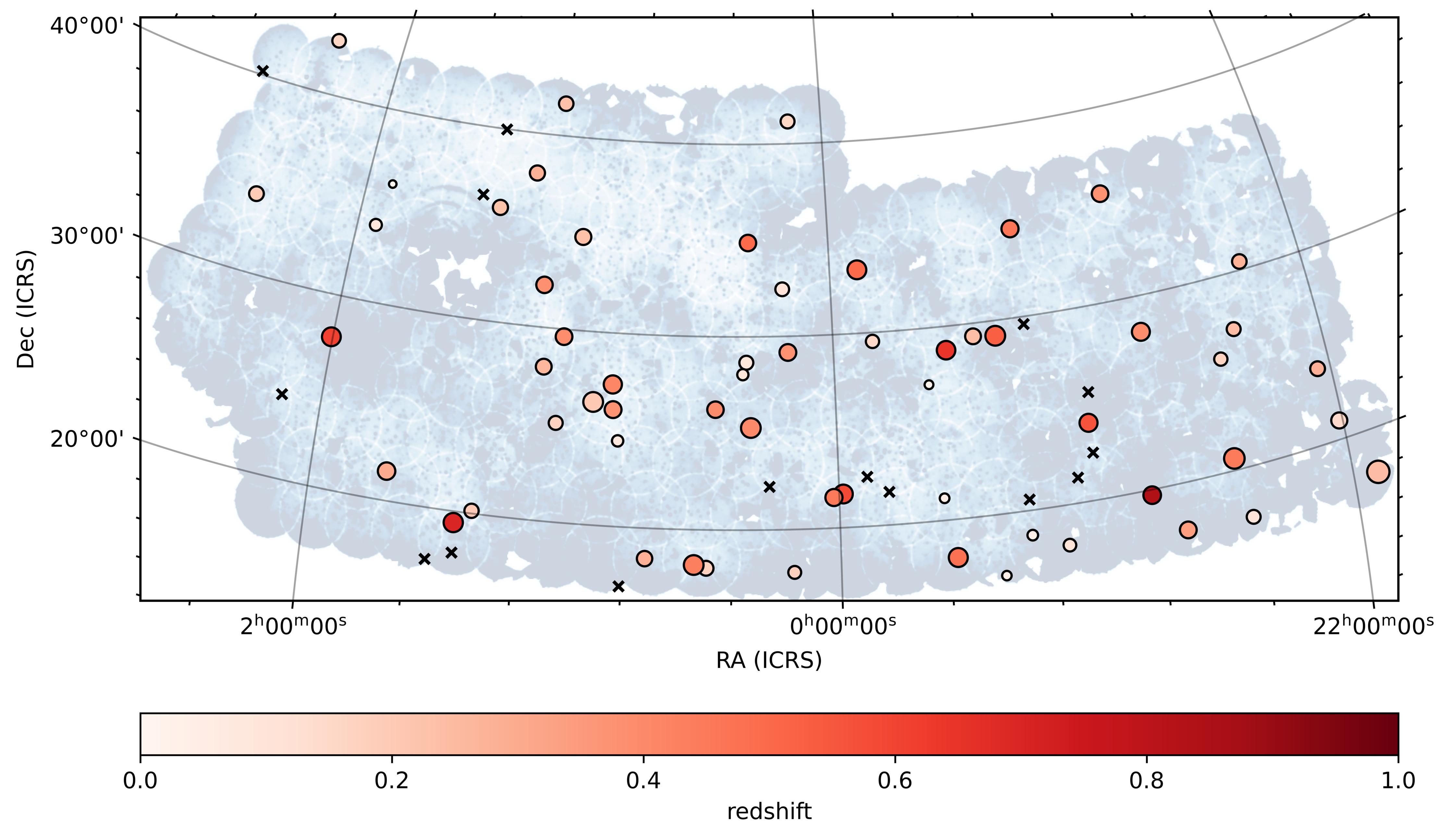}
 \caption{Position of the PSZ2 clusters in the RA-13 (\textit{top}) and RA-1 (\textit{bottom}) regions covered by \lotss-DR2. The color code indicates the redshift of the cluster. The radius of the circle is proportional to \mfive. Clusters without redshift and mass are reported as black crosses. The background image represents the noise variations in \lotss-DR2 (darker colors denote higher noise values) and is reproduced from \citet{shimwell22}.}
 \label{fig:dr2_area}
\end{figure*}

The PSZ2 catalog \citep{planck16xxvii} contains 1653 SZ sources detected over the entire sky. Here, we focus on the 309 entries listed in Table~\ref{tab:sample} that lie in the \lotss-DR2 footprint. This comprises two regions covering 5634 deg$^2$ that are centered at approximately 12h45m00s +44\deg30\arcmin00\arcsec\ and 01h00m00s +28\deg00\arcmin00\arcsec. In the PSZ2 catalog, 63 entries out of 309 are without redshift and mass estimates, meaning that they were not confirmed detections at the time of their publication. However, in follow-up optical studies by \citet{buddendiek15}, \citet{burenin17}, \citet{burenin18}, \citet{barrena18}, \citet{streblyanska18}, \citet{streblyanska19}, \citet{aguadobarahona19}, \citet{boada19}, and \citet{zohren19}, redshifts have been obtained for 35 of these 63 \planck\ detections. From these redshifts we computed \mfive\ by interpolating the \mfive\ versus $z$ curves provided in the PSZ2 individual algorithm catalogs for each detection \citep[see Appendix D in][]{planck16xxvii}. There are 28 remaining PSZ2 detections without redshift confirmation in the \lotss-DR2 area. For simplicity, in the paper we refer to all 309 entries in Table~\ref{tab:sample} as ``galaxy clusters'', even if 28 of them should formally be referred to as ``SZ detections''. In the end, our sample consists of galaxy clusters that are known to span at least the redshift and mass ranges of $0.016 < z < 0.9$ (median of 0.280) and $1.1 \times 10^{14}$ \msun\ $ < \mfive < 11.7 \times 10^{14}$ \msun\ (median of $4.9 \times 10^{14}$ \msun). As shown in Fig.~\ref{fig:dr_sample}, the distribution of redshift and mass in the sample of clusters included in our study provides a qualitatively good representation of the full PSZ2 population. Nonetheless, we note that our sample was selected only based on right ascension and declination cuts. For a better assessment of the similarity between the two samples, we performed a two-sample Kolmogorov-Smirnov test on \mfive\ and $z$, and found $p$-values of the null hypothesis (that the two samples are drawn from the same distribution) of 0.25974 and 0.00148, respectively. The absolute differences between the median values of the two samples are $0.17 \times 10^{14}$ \msun\ (for the mass) and 0.056 (for the redshift). These numbers indicate that the mass distributions of the two samples are in agreement. Concerning the redshift distributions, the low $p$-value and the slightly higher median value of the \lotss-DR2 sample are related to the fact that our sample includes clusters that were confirmed with optical follow-ups after the publication of the \planck\ catalog. These clusters did not have a redshift in the original PSZ2 catalog and they are mostly clusters at high $z$ (30 out of the 35 clusters confirmed by optical follow-ups have redshift higher than the median value of the full PSZ2 sample). The distribution of the clusters within the \lotss-DR2 area is shown in Fig.~\ref{fig:dr2_area}.

\section{Methods and data analysis}\label{sec:methods}

\subsection{Data reduction}\label{sec:datareduction}

\lotss\ is an ongoing radio survey that employs \lofar\ High Band Antennas (HBA) to observe the entire northern sky in the frequency range 120$-$168 MHz. \lotss\ observations are generally 8~hr long, the nominal central frequency of the survey is 144 MHz, and the typical root-mean-square (rms) noise $\sigma$ is $\sim$0.1 \mjyb. More details on \lotss, such as its design and scientific goals, can be found in \citet{shimwell17, shimwell19}. Here, we use the data from the \lotss-DR2 \citep{shimwell22}, which covers an area that is a factor of $\sim$13 larger than \lotss-DR1 and has additional improvements to image fidelity particularly for faint diffuse structures. \lotss-DR2 pointings are processed with fully automated pipelines developed by the \lofar\ Surveys Key Science Project team that aim to correct for direction-independent and direction-dependent effects that are present in the data. These pipelines are \prefactor\footnote{\url{https://github.com/lofar-astron/prefactor}} \citep{vanweeren16calibration, williams16, degasperin19} and \texttt{ddf-pipeline}\footnote{\url{https://github.com/mhardcastle/ddf-pipeline}} \citep{tasse21}. The latter employs \killms\ \citep{tasse14, tasse14arx, smirnov15} and \ddfacet\ \citep{tasse18} to perform direction-dependent self-calibration of the entire \lofar\ FoV, and has been significantly improved compared to the version used to process \lotss-DR1 \citep{shimwell17}. We refer the reader to \citet{tasse21} and \citet{shimwell22} for more details. \\
\indent
In order to further improve the image quality toward the targets in our sample while also allowing for more flexible imaging, we adopted the ``extraction + recalibration'' scheme described by \citet{vanweeren21}, which was also used for the analysis of the galaxy clusters in the \lotss-DR1 region. This method consists of the subtraction of the sources outside a small square region of the sky (typically, $\sim$0.3$-$0.7 deg$^2$) containing the target from the \uv\  data and using the direction-dependent calibration solutions and sky model  derived from \texttt{ddf-pipeline}. The extracted data sets are then phase-shifted to the center of the region, averaged, and corrected for the \lofar\ station beam in this direction. Finally, the calibration of the data is refined by performing a series of typically 4 phase and 6 phase and amplitude calibration loops. \lotss\ pointings have a full width at half maximum of $3.96\deg$ at 144 MHz and are separated by $\sim$2.6\deg, so usually a specific target is covered by multiple pointings, which are combined and analyzed together. We typically extract the visibility data from pointings that are $<$2.2\deg\ from the center of the extracting region. \\
\indent
Among the 309 PSZ2 sources in the \lotss-DR2 area, we were not able to apply this method to 5 targets. This included the Coma cluster (PSZ2 G057.80+88.00) whose radio emission is too large for us to approximate the ionospheric and beam errors with single solutions as is done in the extraction + recalibration scheme, requiring a special treatment \citep[see][]{bonafede21}. Embedded within the Coma cluster radio halo, there are a further two clusters (PSZ2 G056.62+88.42 and PSZ2 G061.75+88.11) where we are unable to differentiate their emission from that of Coma. Finally, PSZ2 G060.10+15.59 and PSZ2 G075.08+19.83 are located in regions where the direction-dependent calibration with \texttt{ddf-pipeline} failed likely due to very poor ionospheric conditions. These 5 targets were excluded from the analysis. A collection of the \lofar\ images of our PSZ2 sample is shown in Fig.~\ref{fig:dr2_collection}.

\subsection{Alignment of the flux scale}\label{sec:fluxscale}

Because of inaccuracies in the \lofar\ beam model, transferring amplitude solutions from the calibrator field data to the target field data may introduce offsets in the flux density scale of the target field \citep[\eg,][]{hardcastle16}. For this reason, as described by \citet{hardcastle21} and \citet{shimwell22}, when constructing final \lotss-DR2 catalogs and mosaics the images are scaled to align the flux density scale with the \citet{roger73} scale. This procedure involves cross-matching catalogs derived from each \lotss-DR2 observation with the \nvssE\ \citep[\nvss;][]{condon98} catalog and assuming a global scaling relationship between \nvss\ and the 6C catalog \citep{hales88, hales90}, which is thought to be consistent with \citet{roger73} to 5\%. In \lotss-DR2 the derived scaling factors are applied to the images during the mosaicing and not to the visibilities directly \citep{shimwell22}. Hence, for our processing, which uses the archived \uv\  data (Sect.~\ref{sec:datareduction}), we instead adopt a procedure where we align catalogs created from the images obtained from the extracted data sets with the final \lotss-DR2 catalog in which the scaling factors have been applied. For this we perform a simple cross-match between the two catalogs (5 arcsec) and use the criteria given in \citet{shimwell22} to select only compact sources and remove those that are not (nearest neighbor within 30 arcsec) and those at low signal-to-noise (less than 7). As outliers can still exist in this cross-matched catalog, we used three different fitting methods (\citealt{sen68}, \citealt{huber81}, and regular linear regression). All three are available within the \texttt{scikit-learn} package\footnote{\url{https://scikit-learn.org/stable/}} \citep{pedregosa12arx}, and all have different outlier rejection criteria, ranging from a more robust median calculation to a less robust simple linear regression. About 90\% of the time the derived values from the different methods give results that are consistent within 10\%. The remaining cases are those where outliers are more prominent and are rejected differently by the adopted fitting methods. Thus, for each fit we calculated the mean absolute error and selected the method with the lowest value, which we then used to scale our images for that particular object and align it with the \lotss-DR2 scale.

\begin{figure*}
  \centering
  \includegraphics[width=.24\hsize,trim={0cm 0cm 0cm 0cm},clip,valign=c]{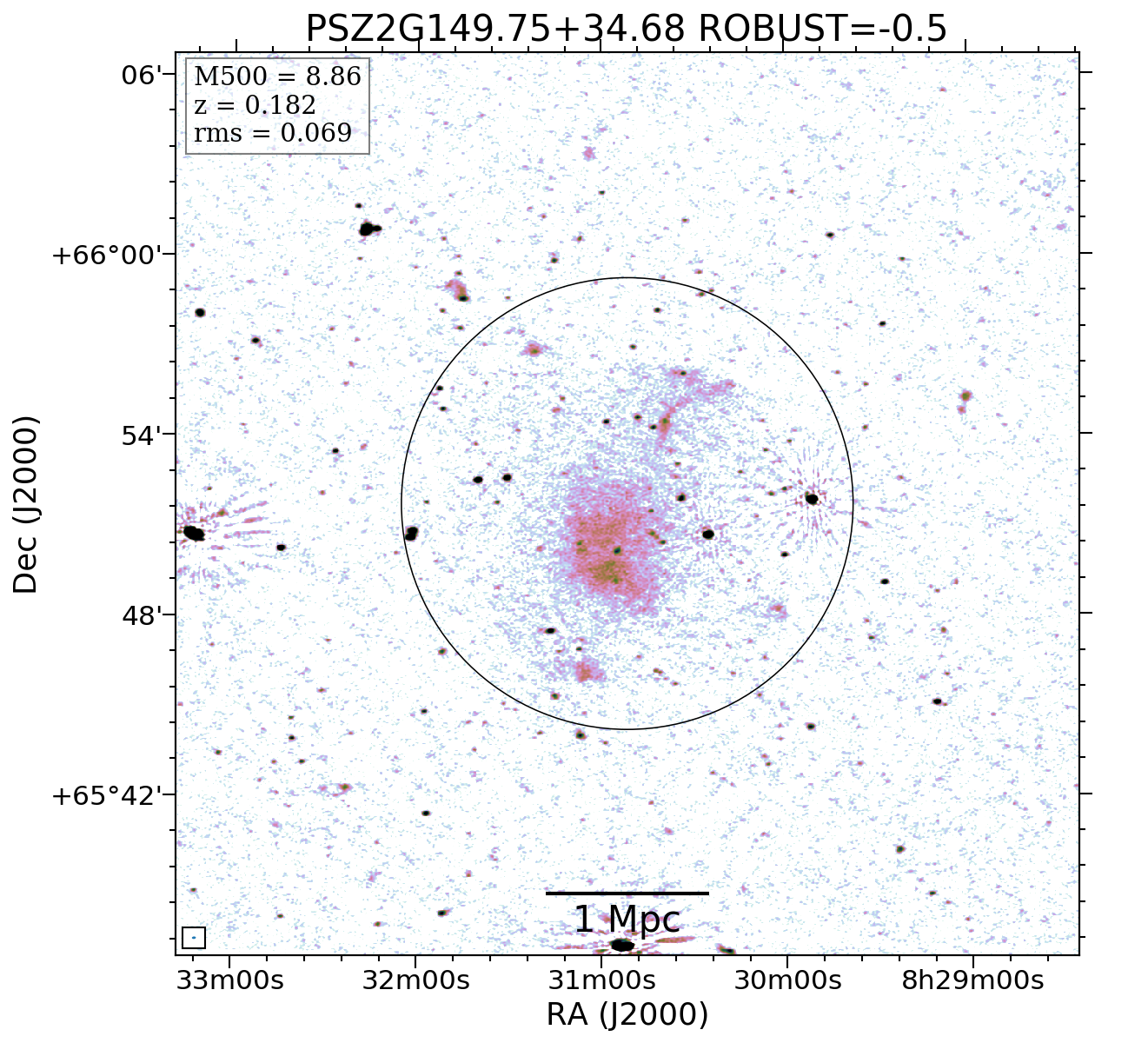}
  \includegraphics[width=.24\hsize,trim={0cm 0cm 0cm 0cm},clip,valign=c]{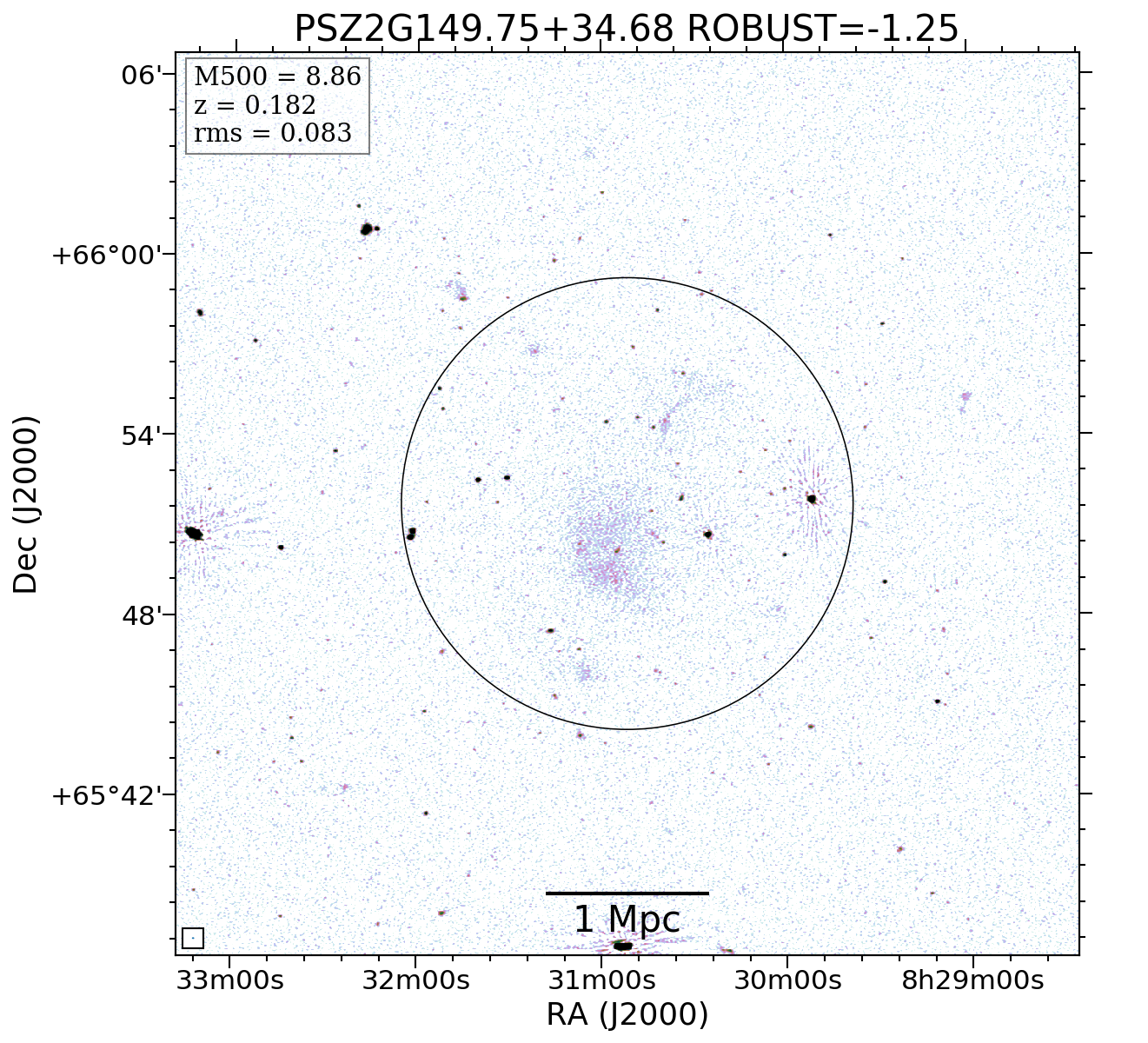}
  \includegraphics[width=.24\hsize,trim={0cm 0cm 0cm 0cm},clip,valign=c]{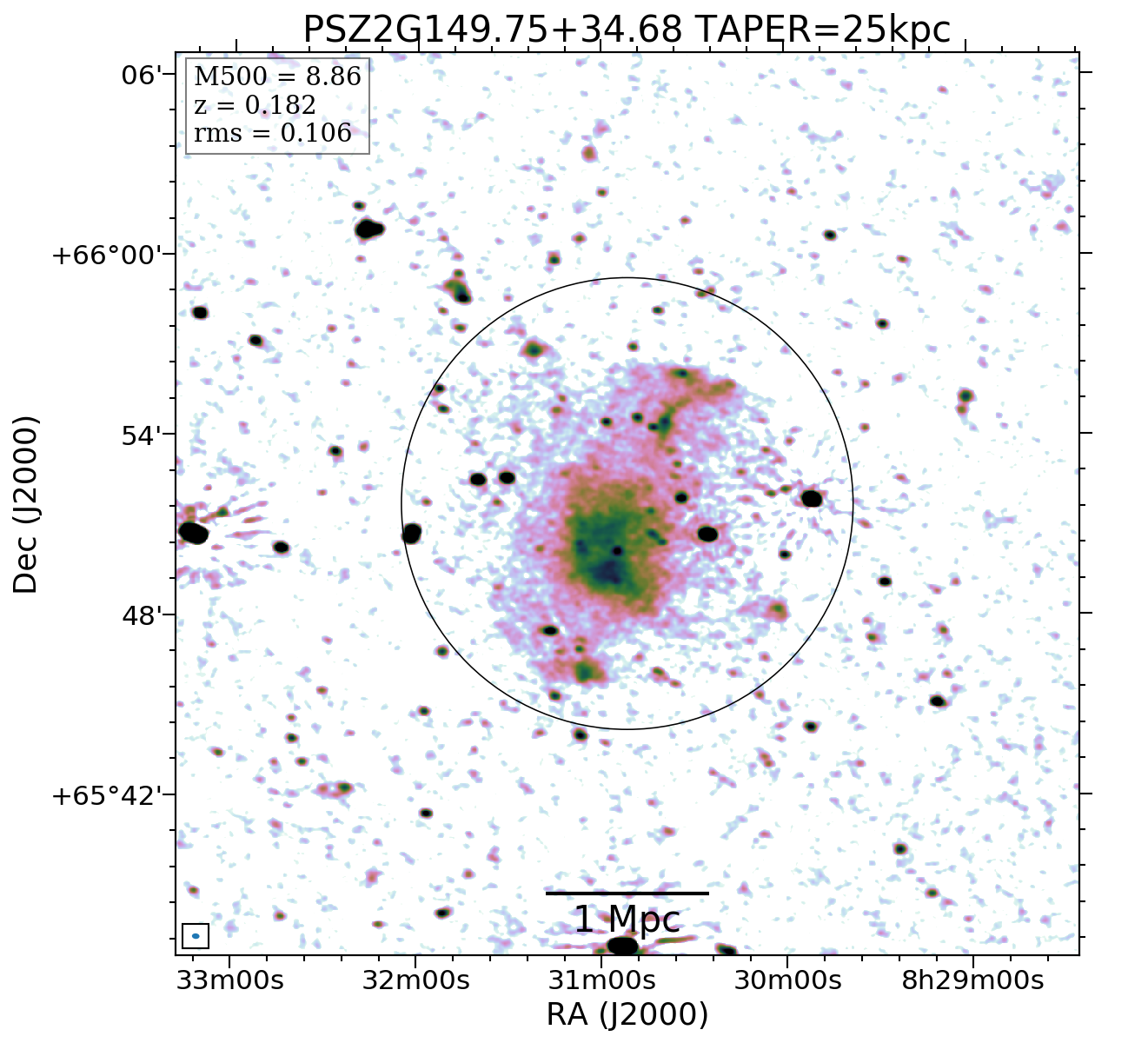}
  \includegraphics[width=.24\hsize,trim={0cm 0cm 0cm 0cm},clip,valign=c]{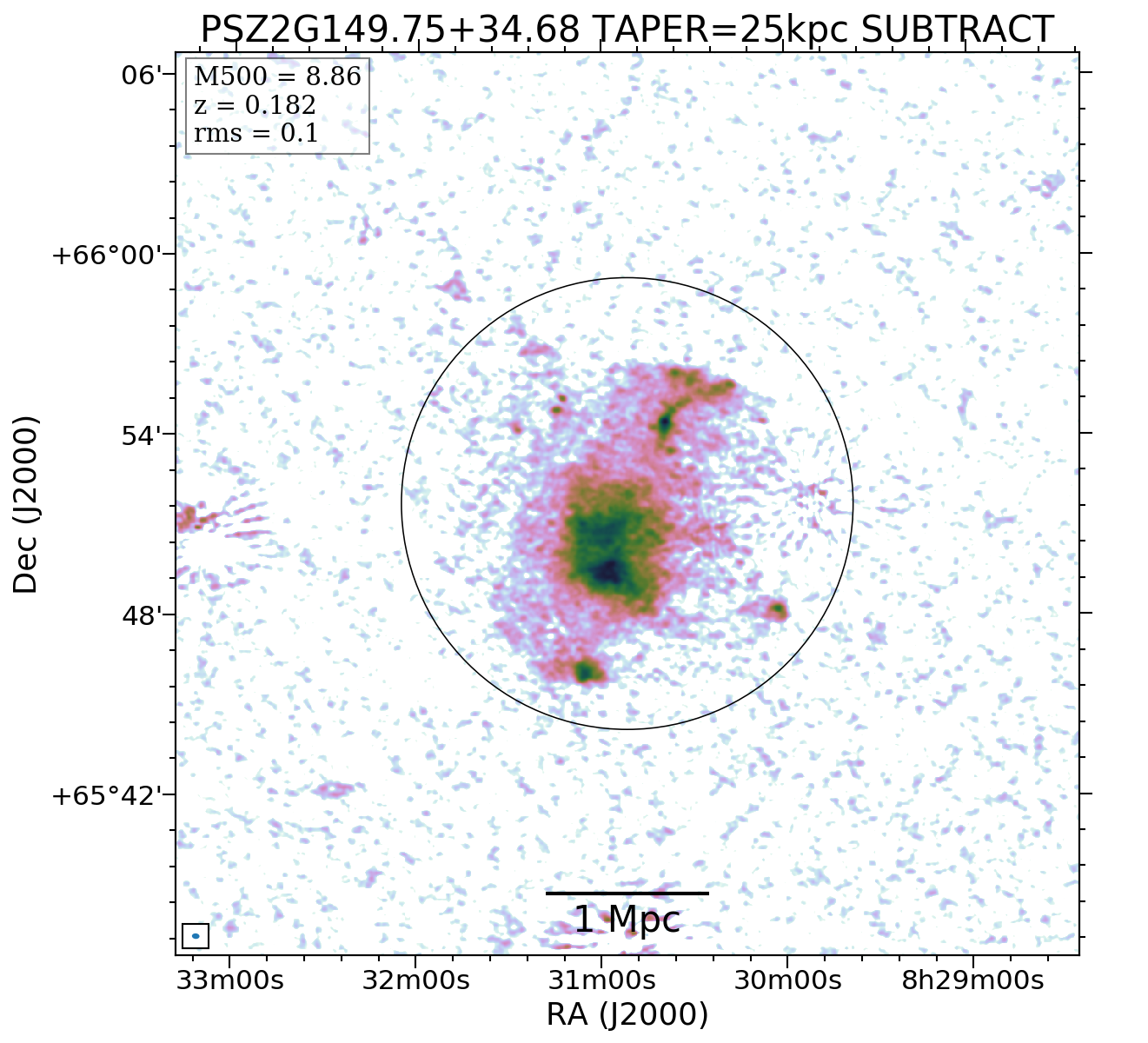}
  \includegraphics[width=.24\hsize,trim={0cm 0cm 0cm 0cm},clip,valign=c]{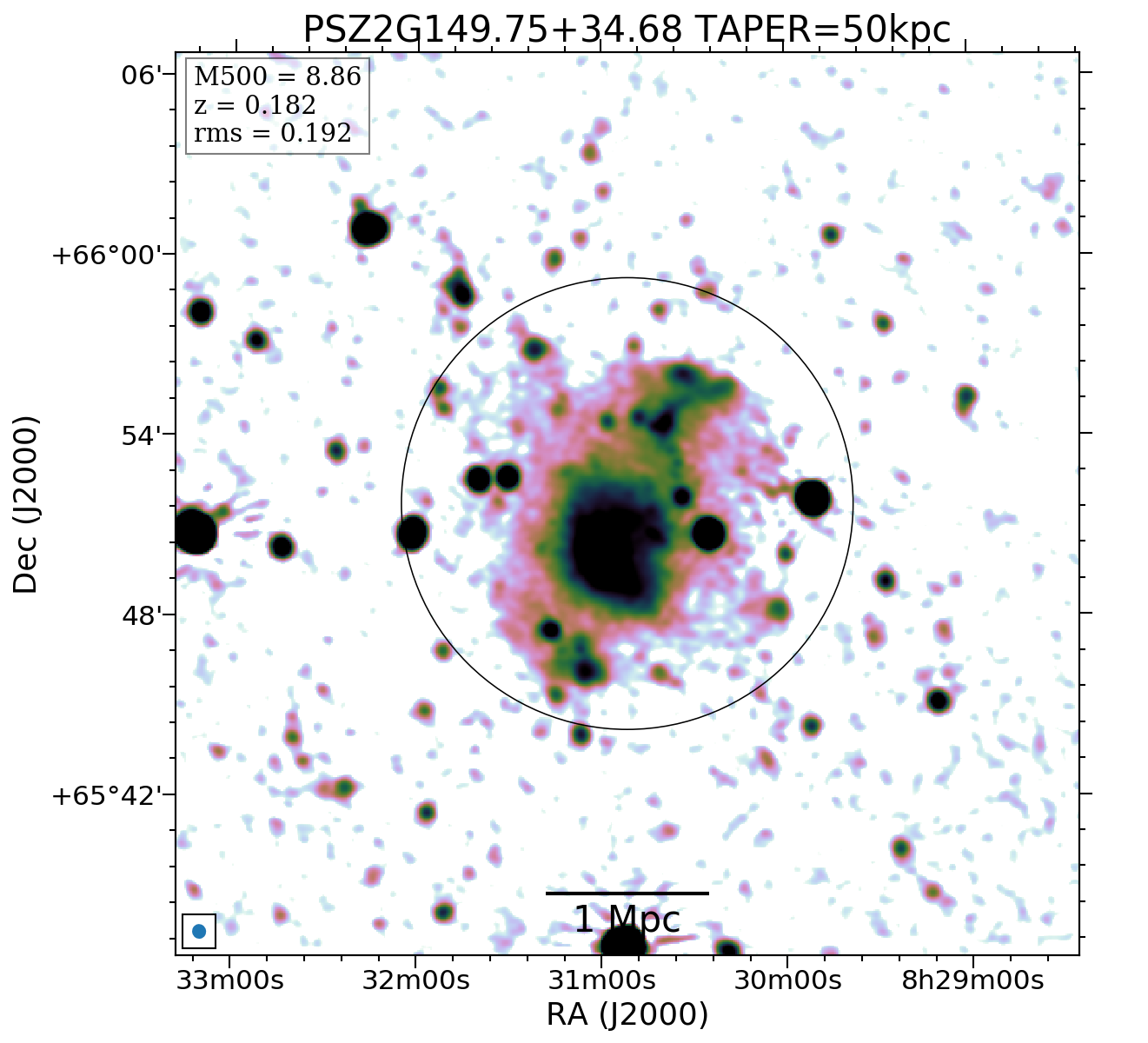}
  \includegraphics[width=.24\hsize,trim={0cm 0cm 0cm 0cm},clip,valign=c]{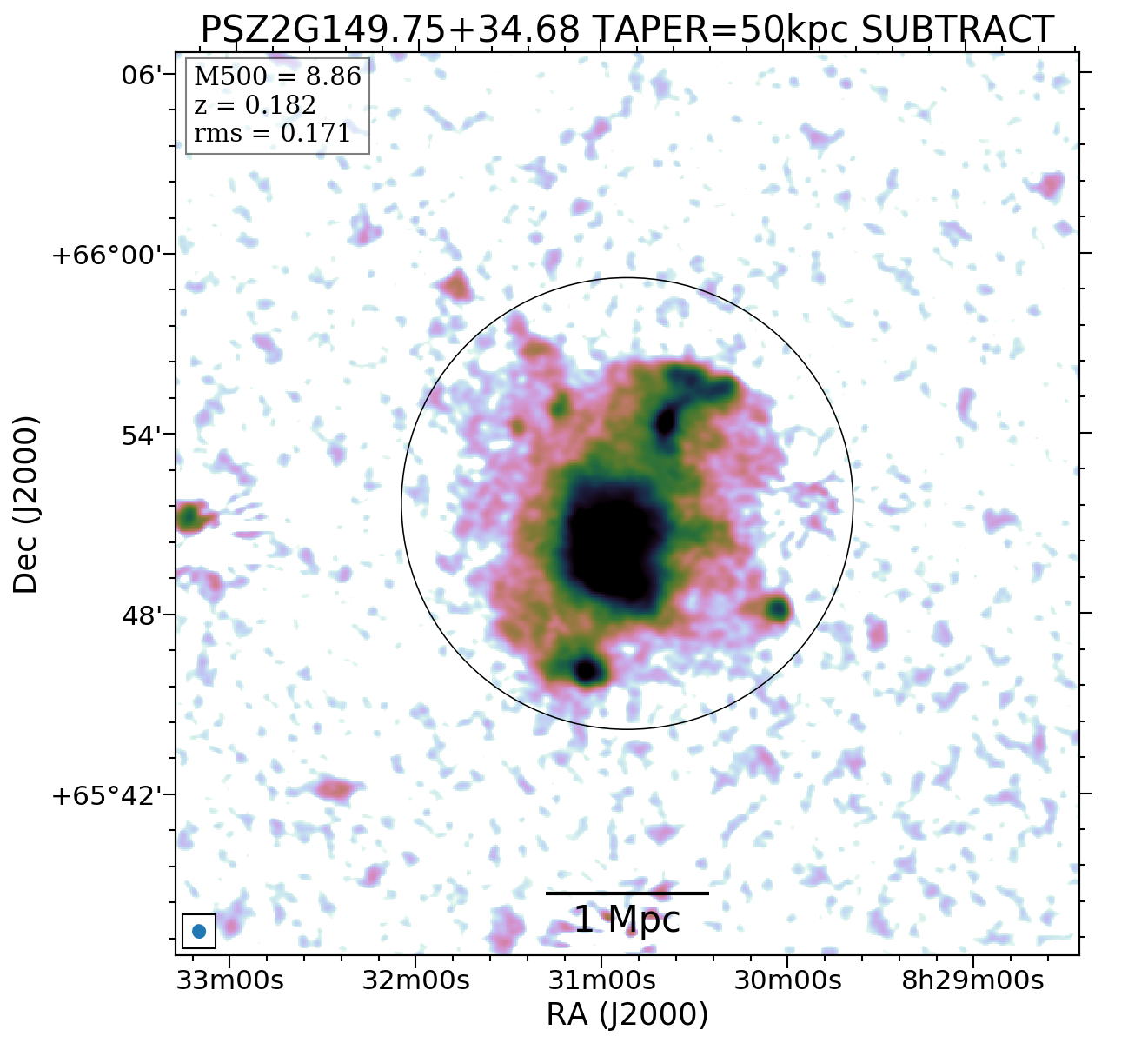}
  \includegraphics[width=.24\hsize,trim={0cm 0cm 0cm 0cm},clip,valign=c]{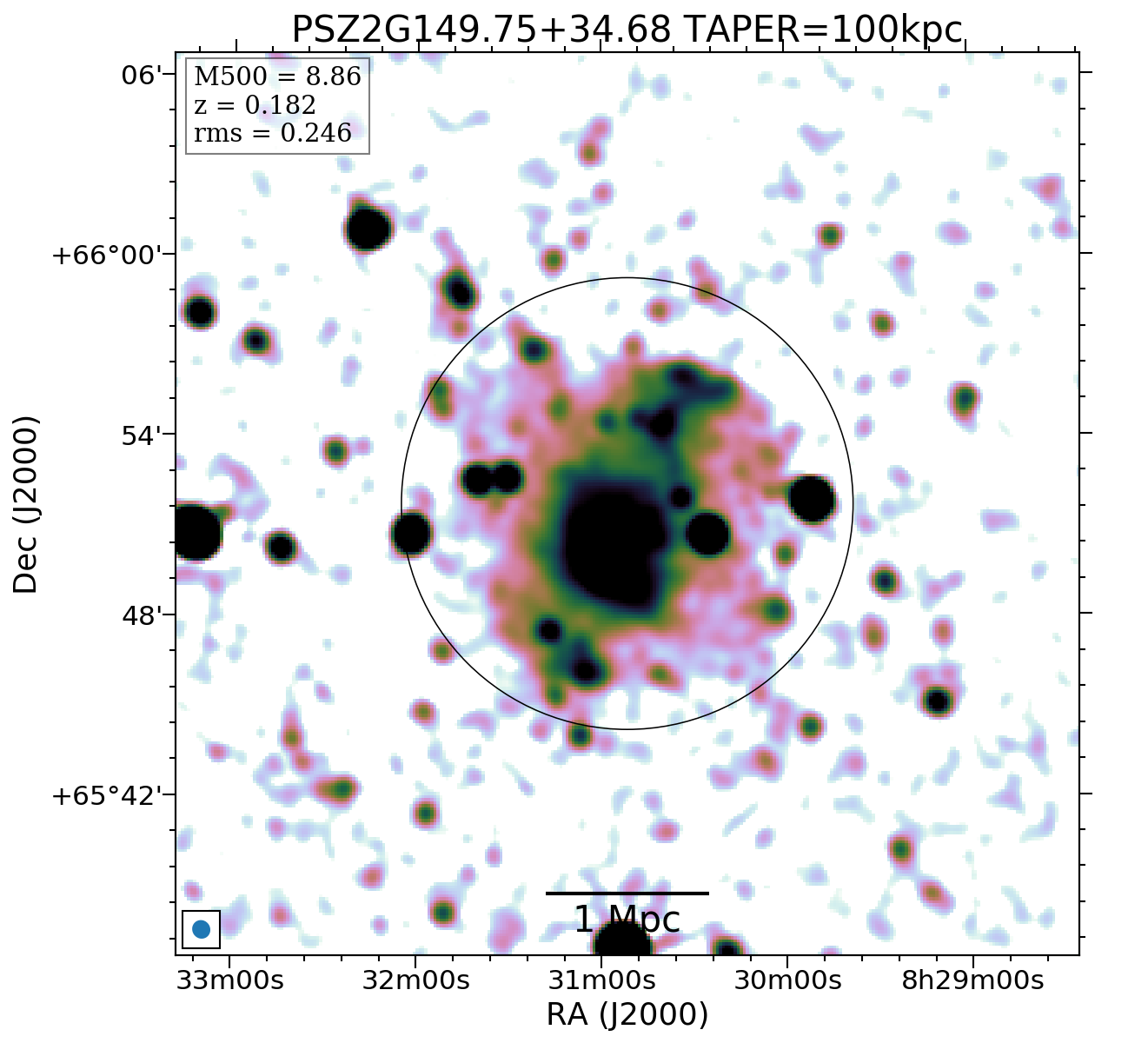}
  \includegraphics[width=.24\hsize,trim={0cm 0cm 0cm 0cm},clip,valign=c]{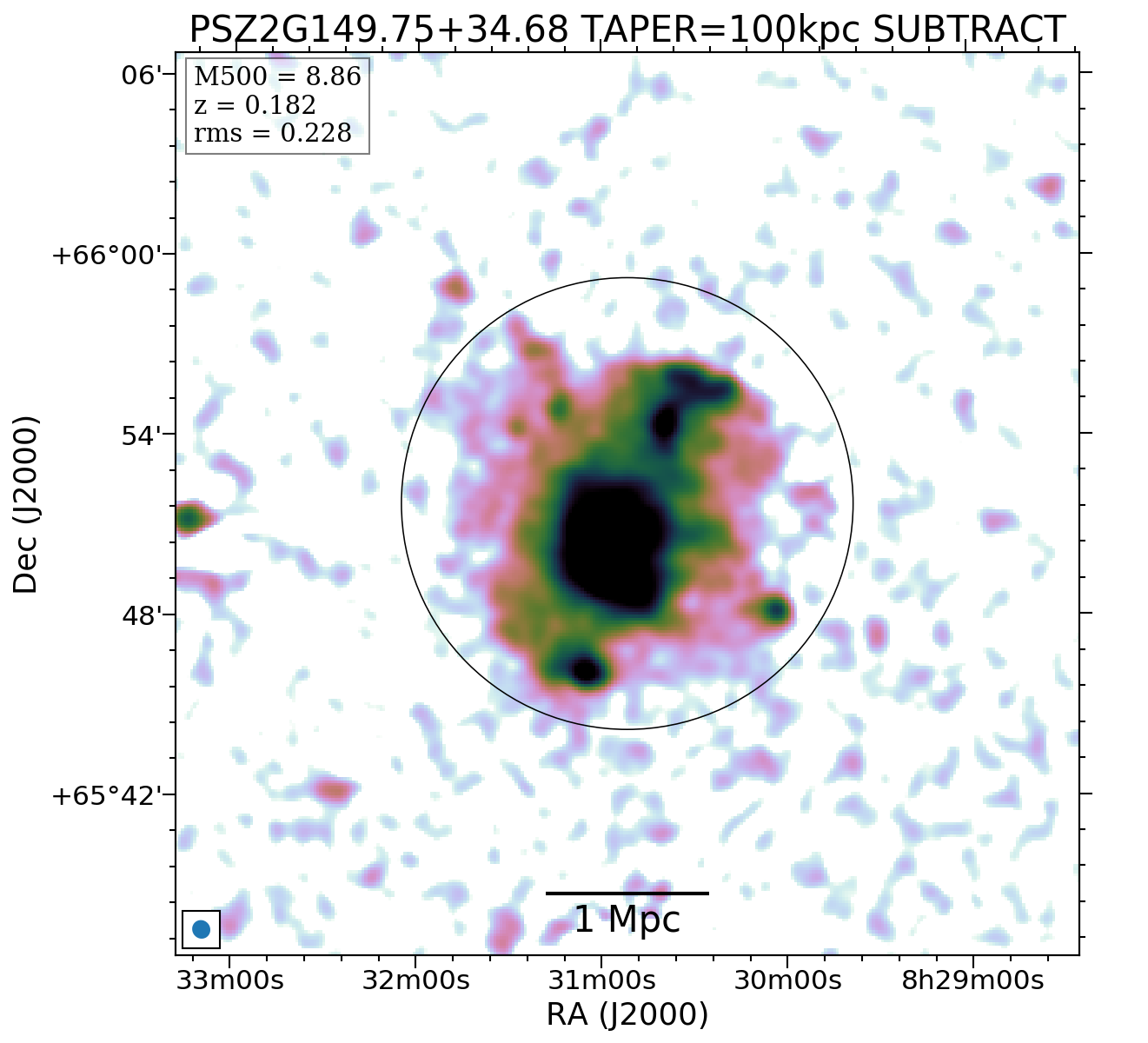}
  \includegraphics[width=.24\hsize,trim={0cm 0cm 0cm 0cm},clip,valign=c]{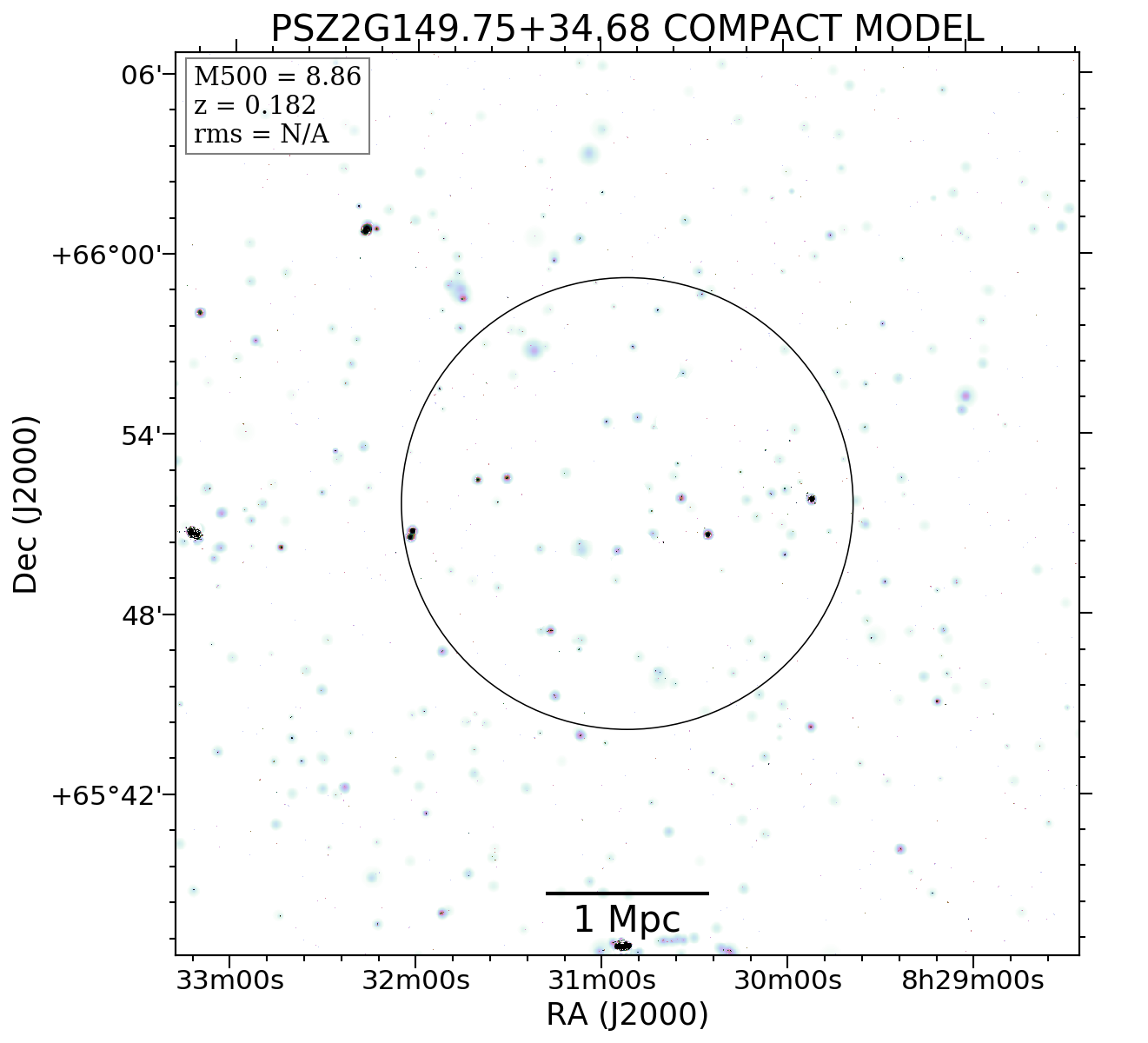}
  \includegraphics[width=.24\hsize,trim={0cm 0cm 0cm 0cm},clip,valign=c]{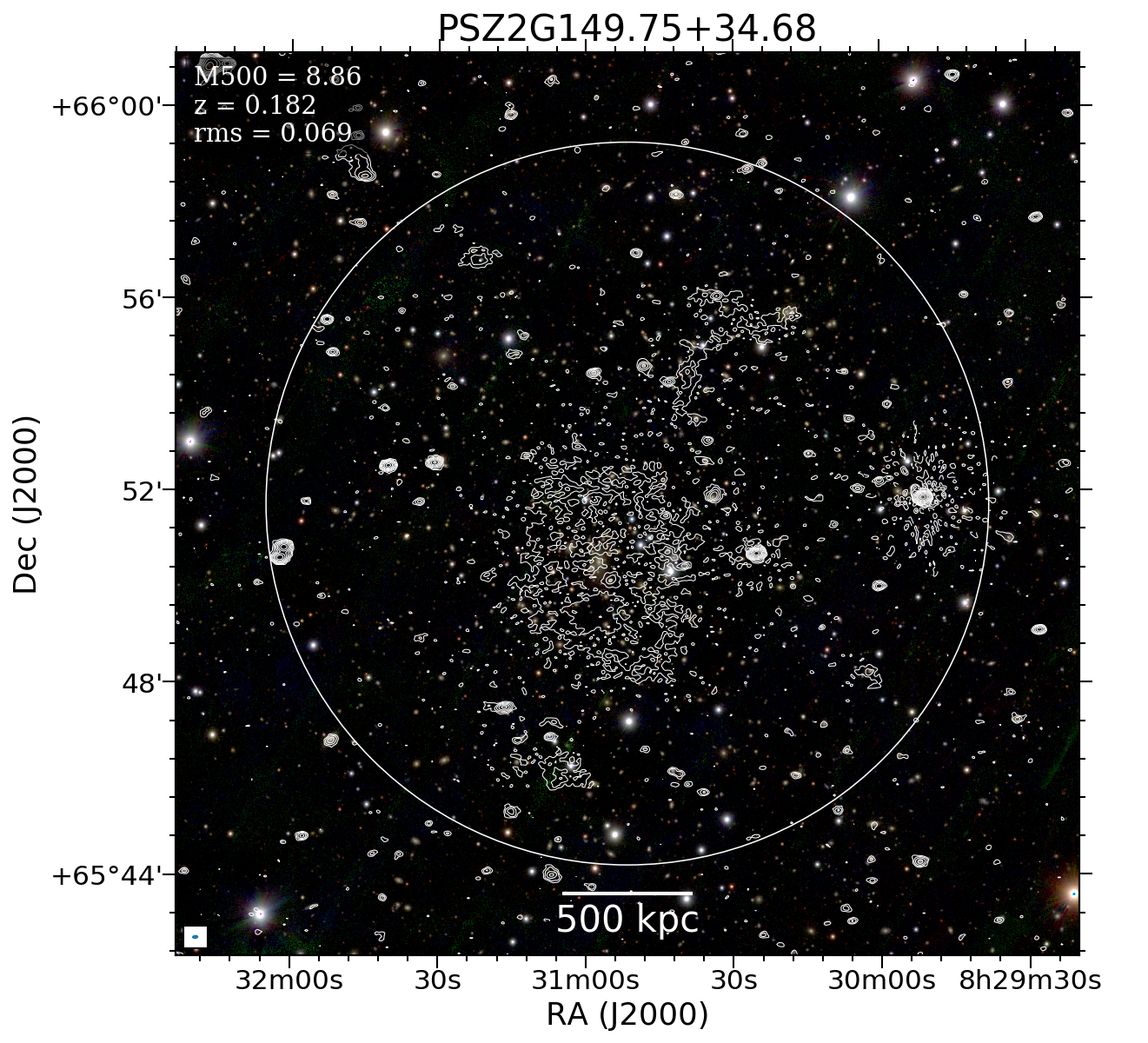}
  \includegraphics[width=.24\hsize,trim={0cm 0cm 0cm 0cm},clip,valign=c]{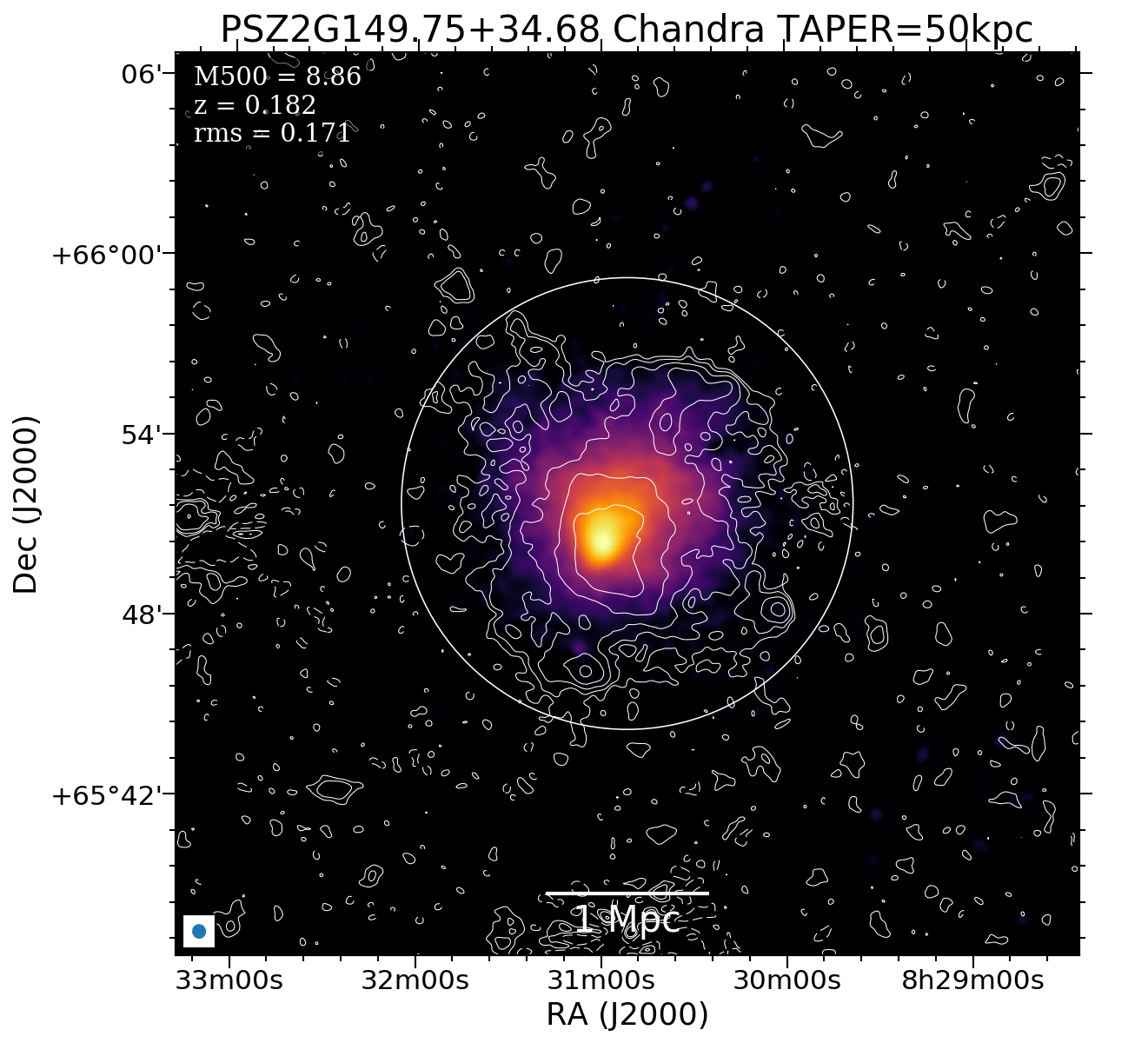}
  \includegraphics[width=.24\hsize,trim={0cm 0cm 0cm 0cm},clip,valign=c]{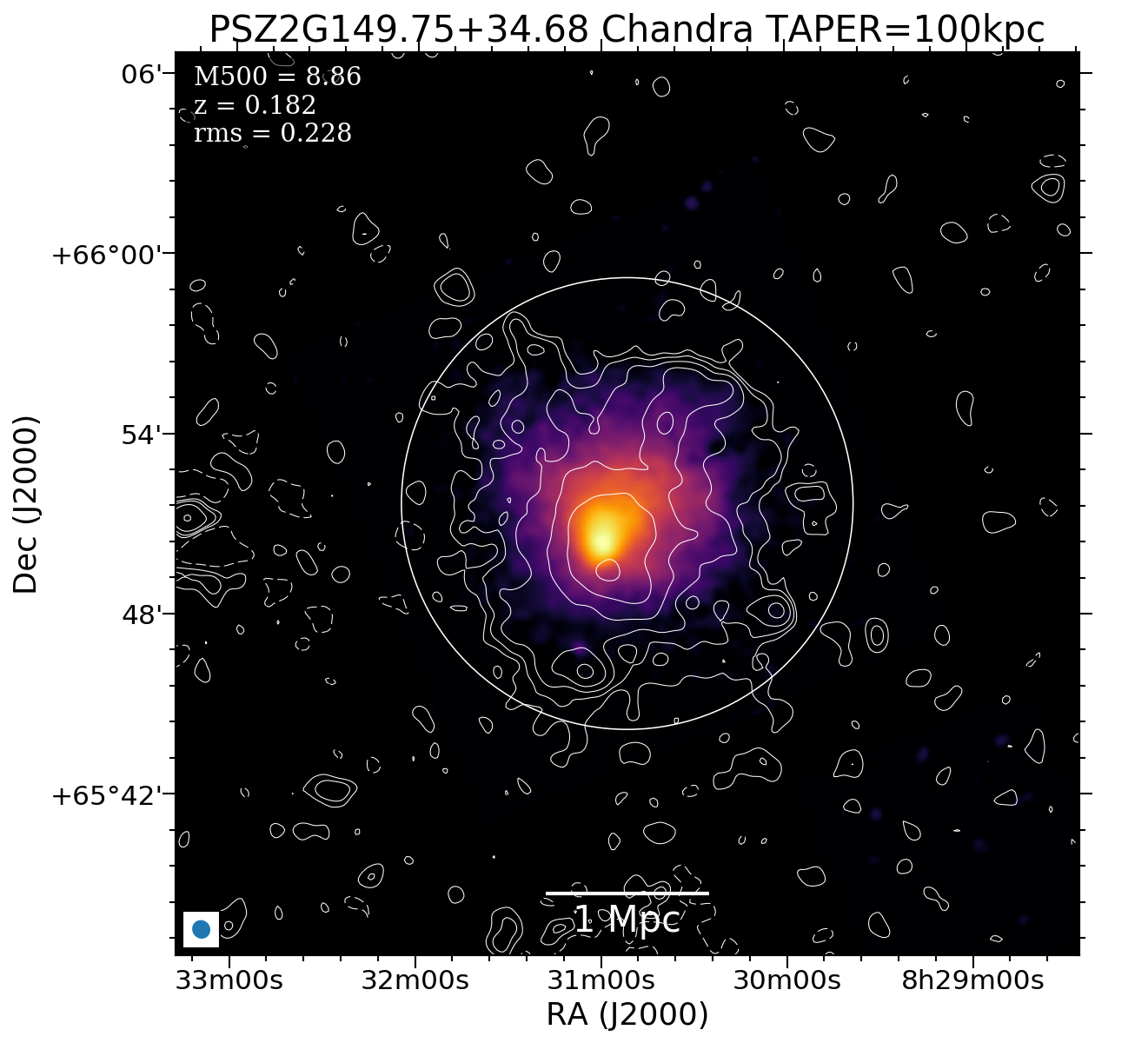}
  \caption{Example of the set of images that we produced for each object in our sample. The displayed cluster is PSZ2 G149.75+34.68, and the reported panels show (from left to right and from top to bottom) the reference image (\texttt{robust=-0.5}), the high-resolution image (\texttt{robust=-1.25}), taper 25, 50, and 100 kpc images with and without discrete sources, the clean model components used for the source subtraction, contours from the reference radio image starting from $3\sigma$ overlaid on an optical image (\panstarrs\ \textit{g,r,i}), and taper 50 and 100 kpc discrete-source-subtracted contours starting from $2\sigma$ overlaid on an X-ray image (\chandra\ or \xmm). Contours are always spaced by a factor of 2. The beam is shown in the bottom-left corner, and the mass (\mfive, in $\times 10^{14}$ \msun\ units), redshift ($z$), and image noise (rms, in \mjyb\ units) are reported in the top-left corner. The circle denotes \rfive\ and is centered at the coordinates reported in the PSZ2 catalog. The images are available at full resolution as well as in FITS format on the project website, \url{https://lofar-surveys.org/planck_dr2.html}.}
  \label{fig:example_reimaging}
\end{figure*}

\subsection{Radio images}\label{sec:imaging}

For each cluster, we produced images at different resolutions to search for diffuse radio emission in the ICM and perform the subsequent analysis. The imaging was done with \wsclean\ v2.8 \citep{offringa14} adopting the \citet{briggs95} weighting scheme with \texttt{robust=-0.5}, and applying Gaussian \uv\  tapers in arcsec equivalent approximately to 25, 50, and 100 kpc at the cluster redshift. For the PSZ2 entries without redshift, the Gaussian \uv\  taper was set to 15, 30, and 60 arcsec to span a wide range of resolutions. The 60 arcsec tapered images were produced only with discrete sources subtracted (see below). For each cluster we also produced a higher-resolution image by using \texttt{robust=-1.25}, which leads to a resolution typically of 5.0 arcsec $\times$ 3.5 arcsec. The multi-scale multifrequency deconvolution option \citep{offringa17} was enabled in \wsclean\ adopting fixed scales (\texttt{-multiscale-scales 0,4,8,16,32,64}) and subdividing the bandwidth into 6 channels for all imaging runs. \\
\indent
We used the images obtained with \texttt{robust=-0.5} and no \uv\  taper as reference images to assess the quality of the data sets, which were visually inspected and graded according to: 1 for high quality images; 2 for images that are partially affected by calibration artifacts or higher rms levels; and 3 for low quality images where the scientific analysis is not possible due to strong calibration artifacts or very high rms noise levels (\ie,\ $>$0.3 \mjyb). This image quality is reported in Table~\ref{tab:sample}. Radio contours from the reference images were overlaid on optical \panstarrsE\ \citep[\panstarrs;][]{chambers16arx} mosaics using the \textit{g,r,i} filters to verify the presence of optical counterparts. \\
\indent
To better study the diffuse emission, we removed the contribution of discrete sources by imaging the data sets with a \uv\  cut corresponding to a physical scale of 250 kpc at the cluster redshift (for PSZ2 entries without redshift we arbitrarily adopted a \uv\  cut of $2722\lambda$, corresponding to 2.82 arcmin or 250 kpc at $z=0.2$) and subtracting their clean components from the visibility data. The new visibility data were then imaged with the same values of \uv\  taper adopted for the images obtained before discrete source subtraction. The low-resolution radio contours with discrete sources subtracted were overlaid onto the \chandra\ and/or \xmm\ X-ray images (when available) smoothed to 30 kpc by a Gaussian function (for more details on the X-ray images, see Sect.~\ref{sec:xray}). \\
\indent
An example of the set of images produced for each cluster is shown in Fig.~\ref{fig:example_reimaging}. All images are available for download in PNG and FITS format on the project website\footnote{\url{https://lofar-surveys.org/planck_dr2.html}}.

\subsection{X-ray images and morphological parameters}\label{sec:xray}

As we will describe in Sect.~\ref{sec:classification}, for the purpose of classifying the detected cluster diffuse radio sources, it is crucial to compare the position of the extended radio emission with the other cluster components and especially the ICM, which can be traced by its X-ray emission and by the SZ effect. Since all clusters in our sample have been detected by \planck, two-dimensional maps of the SZ signal are available for all of them \citep{planck16xxii} but their use is hampered by their 10 arcmin resolution, which does not allow us to spatially resolve most of the targets of the sample. We thus decided to map the ICM distribution with the X-ray images obtained by the current generation X-ray telescopes (\chandra\ and \xmm), whose spatial resolution is higher or comparable to that of our radio images (Sect.~\ref{sec:imaging}). We searched the \chandra\ and \xmm\ archives for observations of the targets in our sample and we retrieved the data for 115 and 100 clusters, respectively (72 targets have been observed both by \chandra\ and \xmm). The procedures used to prepare the X-ray images for each instrument are described in Sects.~\ref{sec:chandra} and \ref{sec:xmm}. \\
\indent
The ICM distribution is very sensitive to the dynamical history of the clusters, and therefore quantitative measurements of the morphology of the X-ray emission of galaxy clusters have proved to be an effective way to characterize the dynamical state of large samples of galaxy clusters \citep[e.g.,][and references therein]{buote01, santos08, cassano10connection, rasia13rev, parekh15, rossetti17, lovisari17}. The use of a set of morphological parameters of the X-ray emission of clusters is only an approximation to the daunting task of assessing the dynamical state of the cluster. However, the combination of two morphological parameters is effective to provide a relatively robust classification. In particular the optimal choice is to combine a parameter sensitive to the presence of substructure, the centroid shift, with a parameter more sensitive to the core properties, the concentration parameter \citep{lovisari17}. In fact this has been the usual choice made in previous studies of the classification of radio sources \citep[for recent examples, see][]{cuciti15, cuciti21a}. The physical scale over which the morphological parameters are measured is also an important factor: here, following previous studies \citep{cassano10connection, cuciti21a}, we analyze an aperture of $R_{\rm ap} = 500$ kpc centered on the peak of the X-ray emission. \\
\indent
The concentration parameter has been introduced by \citet{santos08} as the ratio of the flux within two circular apertures to effectively identify cool cores even at high redshift. Here we adopt the choice of apertures made by \citet{cassano10connection}

\begin{equation}
 c=\frac{F(r <100\, \rm{kpc})}{F(r <R_{\rm ap})}\:,
\end{equation} 

\noindent
where $F(r <100\, \rm{kpc})$ is the flux within 100 kpc and $F(r <R_{\rm ap})$
is the flux within the aperture of 500 kpc. The error on this parameter is obtained by taking into account the Poisson noise in both the source and background images. \\
\indent
The centroid shift \citep{mohr93, poole06} is defined as the variance of the separation between the X-ray peak and the centroid of the emission obtained within a number $N$ of apertures of increasing radius out to $R_{\rm ap}$,

\begin{equation}
 w=\left[\frac{1}{N-1}\sum_i(\Delta_i-\overline{\Delta})^2 \right]^{\frac{1}{2}}\frac{1}{R_{\rm ap}}\:,
\end{equation}

\noindent
where $\Delta_i$ is the distance between the X-ray peak and the centroid of the $i$-th aperture. It traces the variation in the position of the centroid introduced by the presence of substructures in the X-ray emission. The number $N$ of apertures is fixed at 20 in the \xmm\ analysis and it is given by the number of annuli of fixed 5 arcsec width within 500 kpc for the \chandra\ analysis. The error on this parameter is obtained by using a Monte Carlo approach: for the \chandra\ analysis we simulated 100 realizations of the X-ray images obtained by resampling the counts per pixel according to their Poisson error, performed the measurement on the simulated image and estimated the standard deviation of the distribution of $w$ thus obtained; for the \xmm\ analysis we simulated 10000 realizations of the centroids of the 20 apertures, sampled within their statistical errors. \\
\indent
We measured the concentration parameter and the centroid shift for 105 PSZ2 objects with \chandra\ and for 98 PSZ2 objects with \xmm\ as a result of the following four selections:
(i) a low redshift cut to accommodate the aperture of 500 kpc within the FoV of each respective detector ($z > 0.065$ for ACIS-I, $z > 0.072$ for ACIS-S, and $z > 0.035$ for \xmm);
(ii) PSZ2 G165.95+41.01 does not have a measurement because of a possible incorrect redshift estimate: the X-ray and optical images suggest that this object is at a higher $z$ compared to the value of $z=0.062$ reported in the PSZ2 catalog (see the note in the catalog that discusses a superposition with a $z=0.21$ object); (iii) PSZ2 G067.52+34.75 does not have a \chandra\ morphological measurement because the observation is performed in a sub-array mode that does not allow a 500 kpc aperture to be accommodated; and (iv) PSZ2 G126.27+51.61 does not have a measurement because the available \chandra\ observation is too shallow for the faint emission of this high redshift ($z=0.815$) cluster. \\
\indent
There are 63 objects that have both \chandra\ and \xmm\ measurements and the total number of PSZ2 clusters in our sample for which we have X-ray morphological parameters is 140. For objects for which different clumps of X-ray emission could be clearly distinguished we measured morphological parameters for each component, labeling them according to their position in the sky. This explains why Table~\ref{tab:cw}, where we list the $c$ and $w$, has 150 entries. 
For the 63 objects (65 measurements including the objects with multiple clumps) with both \chandra\ and \xmm\ measurements we provide combined morphological parameters according to the following equations:

\begin{equation}
 \mathcal{P}_\mathrm{combined} = \frac{1}{2}\times\left(\mathcal{P}_\mathrm{xmm} + \mathcal{P}_\mathrm{chandra}\right)\:,
\end{equation}

\noindent
where $\mathcal{P}$ is either $c$ or $w$ and the error $\sigma_{\mathcal{P}}$ on this combined parameter is given by the sum of the statistical $\sigma_{\mathcal{P},\mathrm{stat}}$ and systematic $\sigma_{\mathcal{P},\mathrm{sys}}$ error,

\begin{align}
 \sigma_{\mathcal{P}}^{2}=&\sigma_{\mathcal{P},\mathrm{stat}}^2+\sigma_{\mathcal{P},\mathrm{sys}}^2 \nonumber\\
 =&\frac{1}{4}\times\left(\sigma_{\mathcal{P},\mathrm{xmm}}^2+\sigma_{\mathcal{P},\mathrm{chandra}}^2+2\sigma_{\mathcal{P},\mathrm{xmm}}\sigma_{\mathcal{P},\mathrm{chandra}}\right)\nonumber\\
 &+\frac{1}{4}\times\left(\mathcal{P}_\mathrm{xmm}-\mathcal{P}_\mathrm{chandra}\right)^2\:,
\end{align}

\noindent
where we also take into account the covariance with the term $2\sigma_{\mathcal{P},\mathrm{xmm}}\sigma_{\mathcal{P},\mathrm{chandra}}$. These values are used to discuss the occurrence of diffuse radio emission with the cluster dynamical state in Cassano et al. (in preparation). A thorough comparison between parameters derived from \xmm\ and \chandra\ as well as the detailed analysis on the X-ray data will be presented in Zhang et al. (in preparation). \\
\indent
We describe the reduction and analysis steps used for the \chandra\ and \xmm\ data in the following subsections.

\subsubsection{\chandra\ data reduction and analysis}\label{sec:chandra}

We analyzed \chandra\ data with the \ciaoE\ (\ciao) software v4.13 using CALDB v4.9.4 \citep{fruscione06}, reprocessing data from the level 1 event files and following the standard data reduction threads\footnote{\url{http://cxc.harvard.edu/ciao/threads/index.html}}. 
We reprocessed event files using the {\verb chandra_repro } tool and soft proton flares were excluded with the {\verb deflare } task with the {\verb lc_clean } routine analyzing the light curves extracted from the S2 chip when in ACIS-I configuration and from the S3 chip when in ACIS-S configuration.
We used the  {\verb fluximage } tool to produce images in the 0.5$-$7.0 keV bands and the appropriate exposure and point spread function maps. For the purpose of background subtraction we used the {\verb blanksky } and {\verb blankskyimage } tools to provide a corresponding background image to be subtracted.
We detected the point sources using the {\verb wavdetect } tool, and by means of {\verb dmfilth } we replaced their emission with the mean count rate in a surrounding annulus. 
Images were smoothed to a resolution of 30 kpc at the cluster redshift to minimize the effect of having different physical sizes for the same pixel scale given the broad redshift range of our sample \citep[see the discussion in][]{yuan20}. 
The peak of the X-ray image used as the center of the cluster has been selected as the brightest pixels in the smoothed image after masking the point sources.
When multiple observations were available for the same object, we used the observation with the longest available exposure time. For all cases, our single \obsid\ images have at least 500 counts, which is a safe limit to have a robust measurement of the X-ray morphological parameters \citep[\eg,][]{nurgaliev13}.

\subsubsection{\xmm\ data reduction and analysis}\label{sec:xmm}

We used \xmm\ Science Analysis System (\sas) v18.0.0 for \xmm\ \epicE\ (\epic) data reduction \citep{gabriel04}. MOS and pn event files are obtained from the observation data files with the tasks \texttt{emproc} and \texttt{epproc}. The out-of-time (OoT) event files of pn are produced by \texttt{epproc} as well. We extracted count images in the 0.5$-$2.0 keV band. The OoT count maps are directly subtracted from each pn count image following the user guide\footnote{\url{http://xmm-tools.cosmos.esa.int/external/xmm_user_support/documentation/sas_usg/USG/removingOoTimg.html}}. The corresponding exposure maps were generated using task \texttt{eexpmap} with parameter \texttt{withvignetting=yes}. Each exposure map was then multiplied by the on-axis effective area calculated by \texttt{arfgen}. \\
\indent
We used stacked filter wheel closed (FWC) event files as non-X-ray background (NXB). For each ObsID, the FWC event files were re-projected using the task \texttt{evproject} to match the observation. For the two MOS detectors, the NXB count maps were scaled using the blank regions out-of-FoV. For the pn detector, because of the contamination in out-of-FoV regions, we estimated the scaling factor based on the long-term NXB variation due to solar activities, which will be detailed described in Zhang et al. (in preparation). \\
\indent
Point source detection and removal procedures are the same as the \chandra\ data analysis. For each object, we stacked the point-source-removed NXB-subtracted count images and exposure maps, respectively. The stacked net count image was then divided by the stacked exposure map to obtain the final point source free flux map for morphological analysis. The X-ray peak of each object is determined from the 30 kpc Gaussian smoothed flux map. 

\begin{figure*}
 \centering
 \includegraphics[width=.7\hsize,trim={0cm 0cm 0cm 0cm},clip]{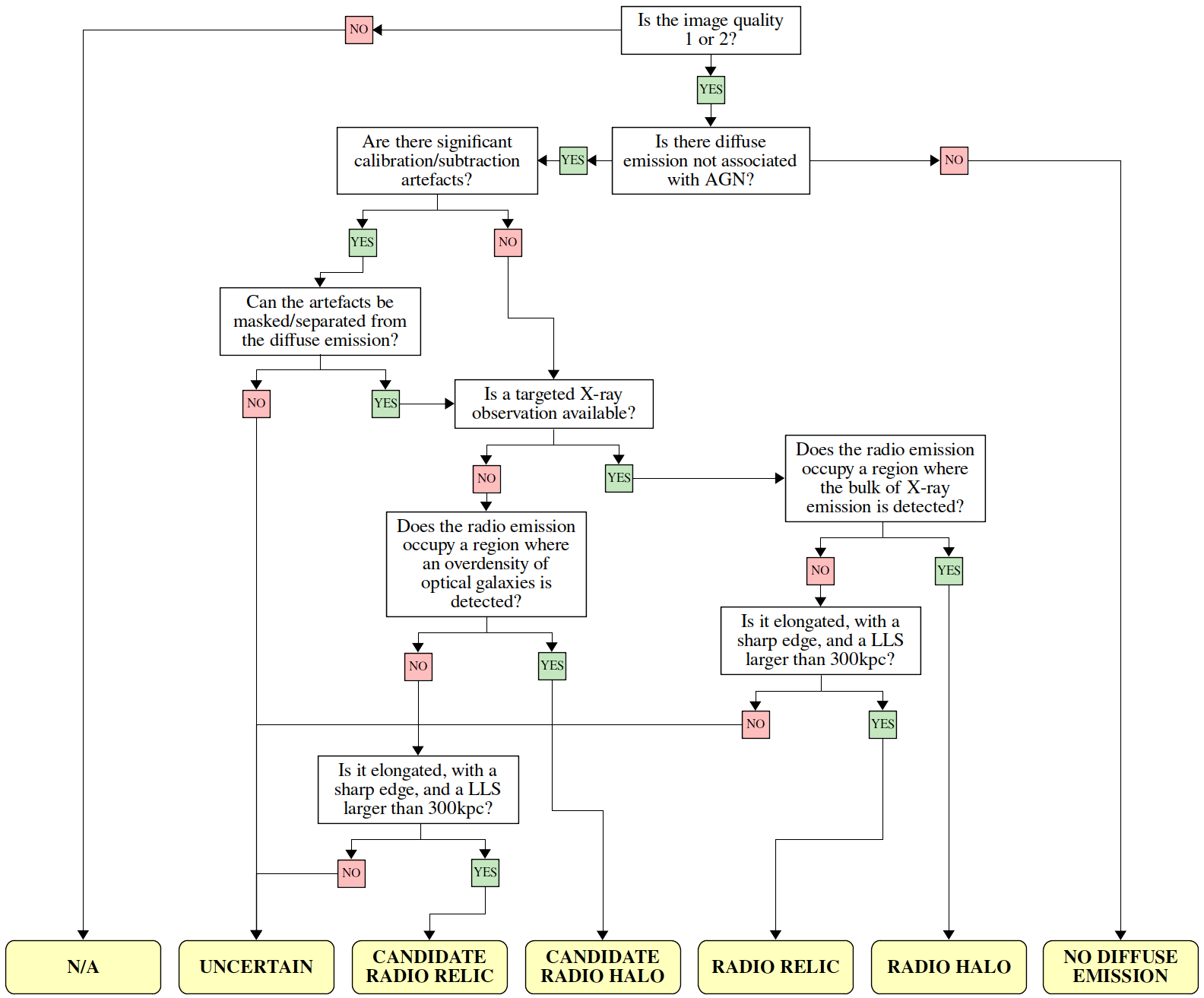}
 \caption{Decision tree used to classify the diffuse radio sources in the ICM. The classes of objects that form the end points of the decision tree are described in Sect.~\ref{sec:classification}.}
 \label{fig:decision_tree}
\end{figure*}

\section{Classification of radio sources}\label{sec:classification}

\begin{figure*}
  \centering
  \includegraphics[width=.33\hsize,trim={0cm 0cm 0cm 0cm},clip,valign=c]{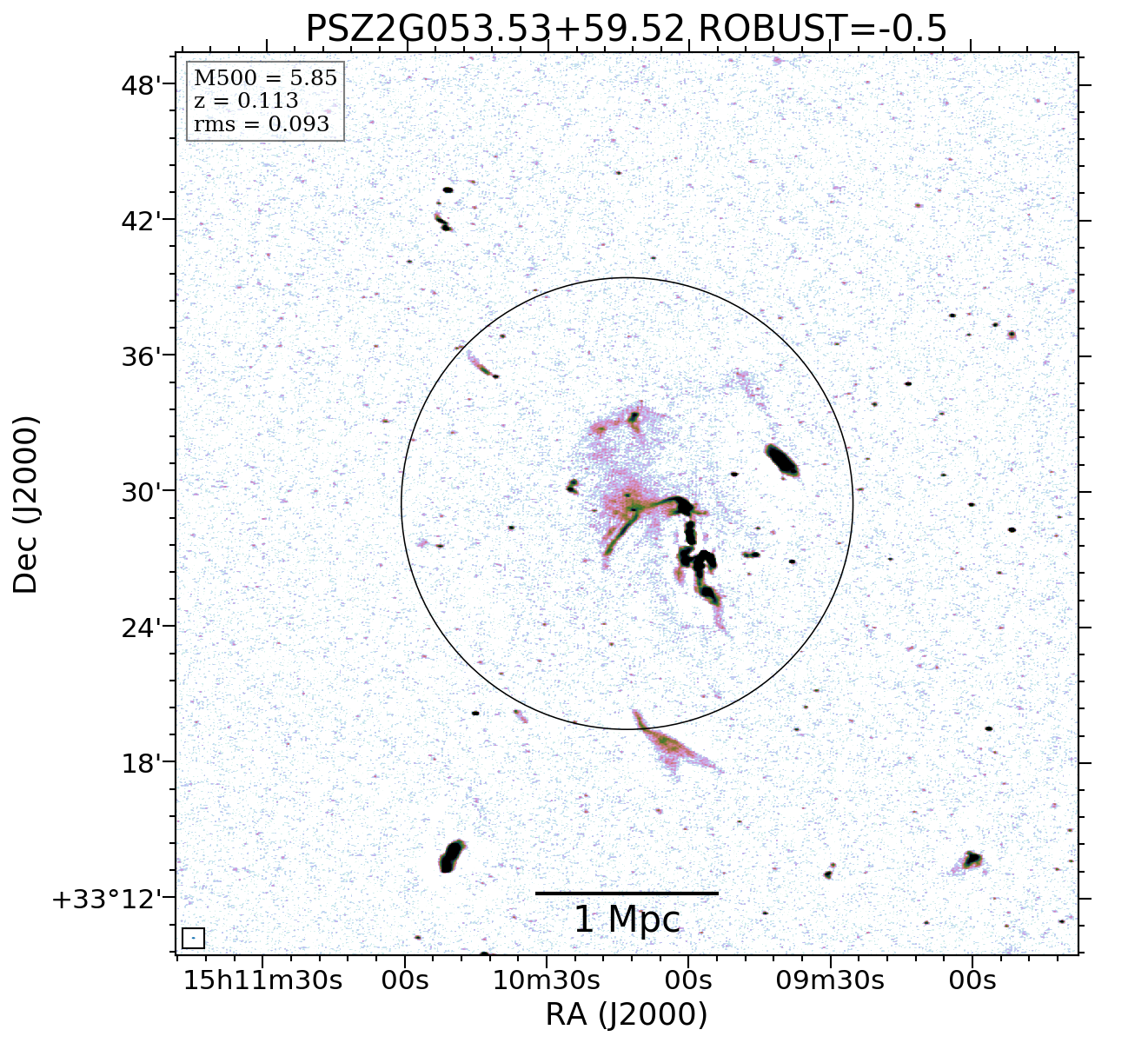}
  \includegraphics[width=.33\hsize,trim={0cm 0cm 0cm 0cm},clip,valign=c]{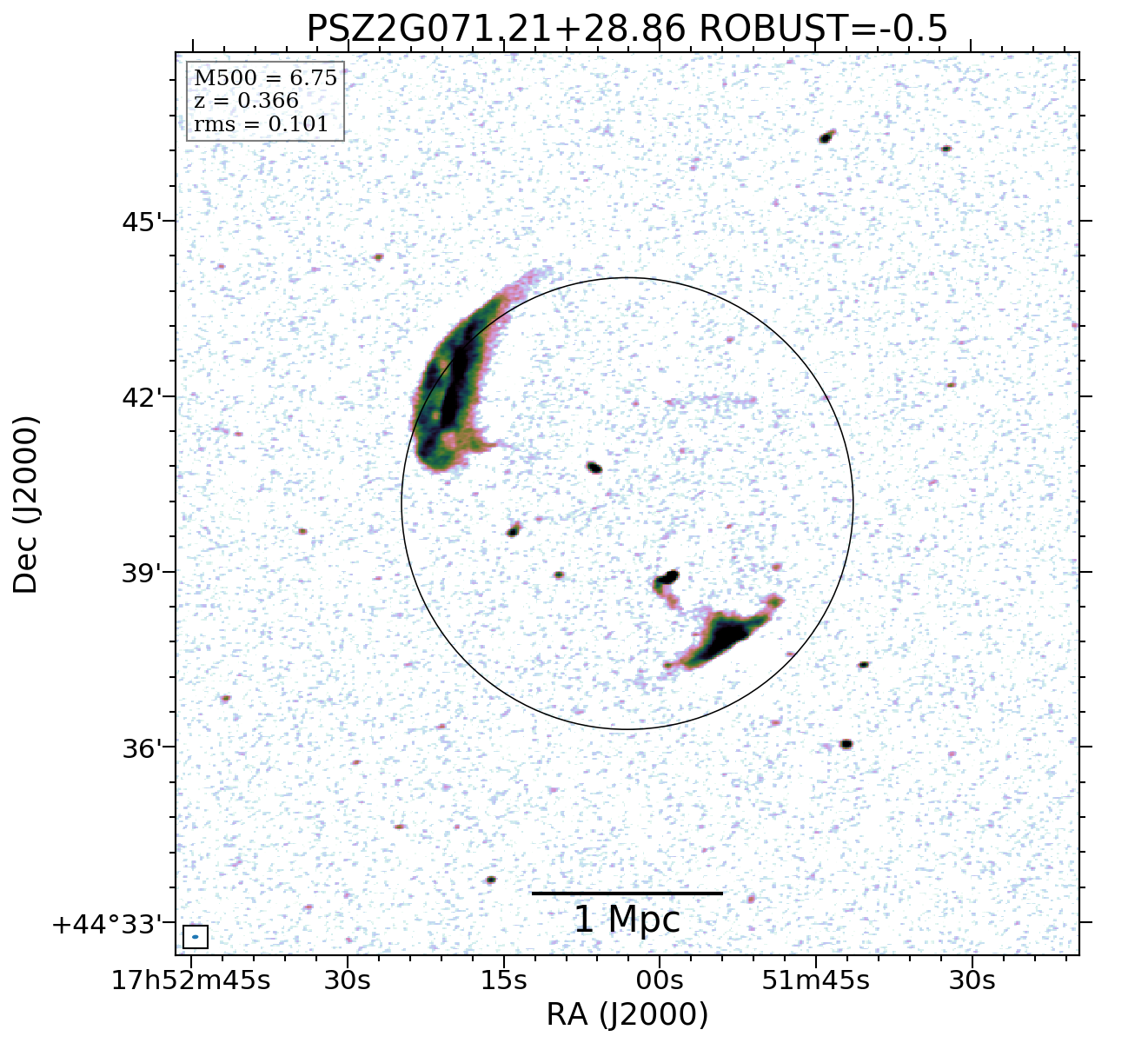}
  \includegraphics[width=.33\hsize,trim={0cm 0cm 0cm 0cm},clip,valign=c]{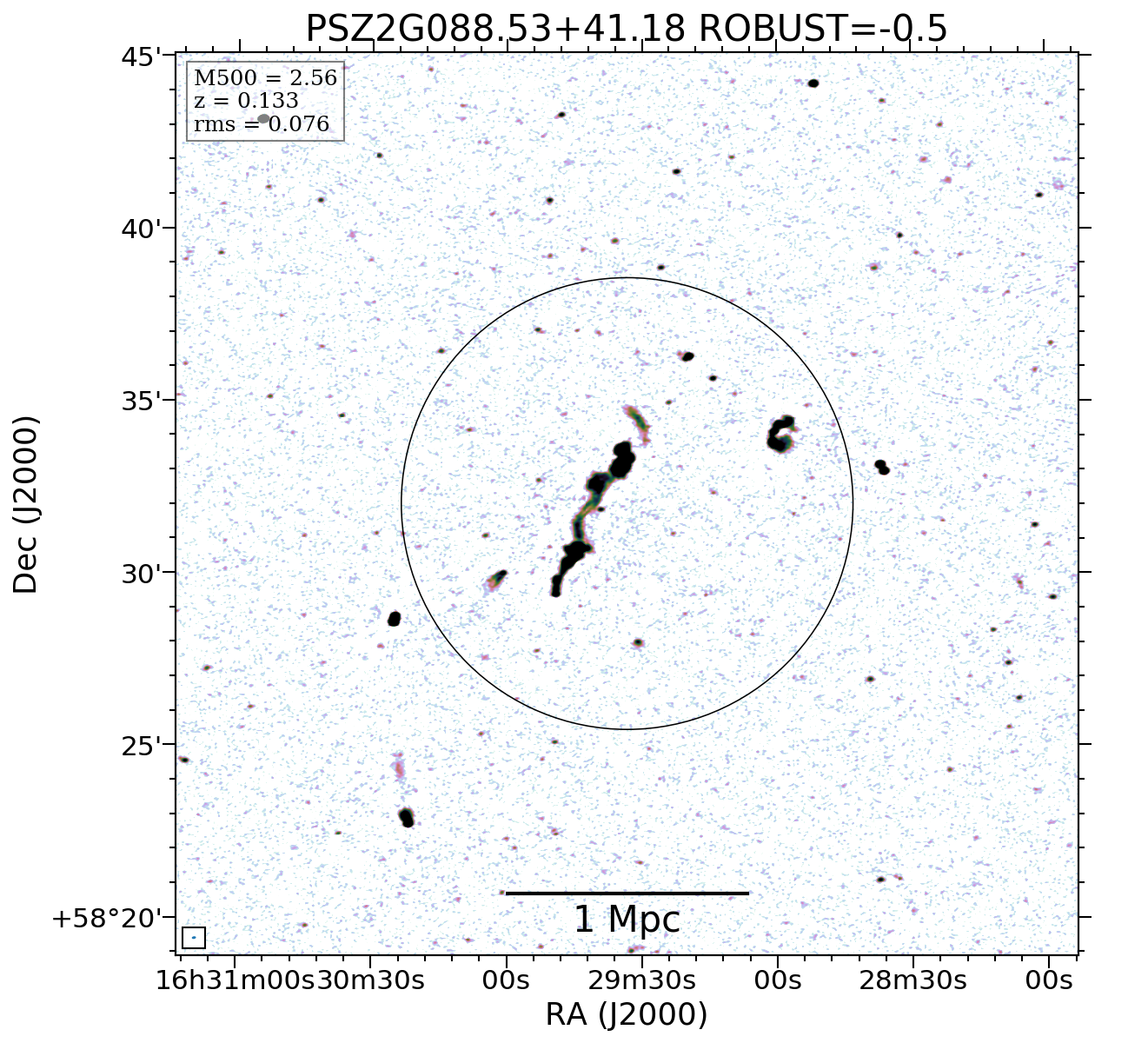}
  \includegraphics[width=.33\hsize,trim={0cm 0cm 0cm 0cm},clip,valign=c]{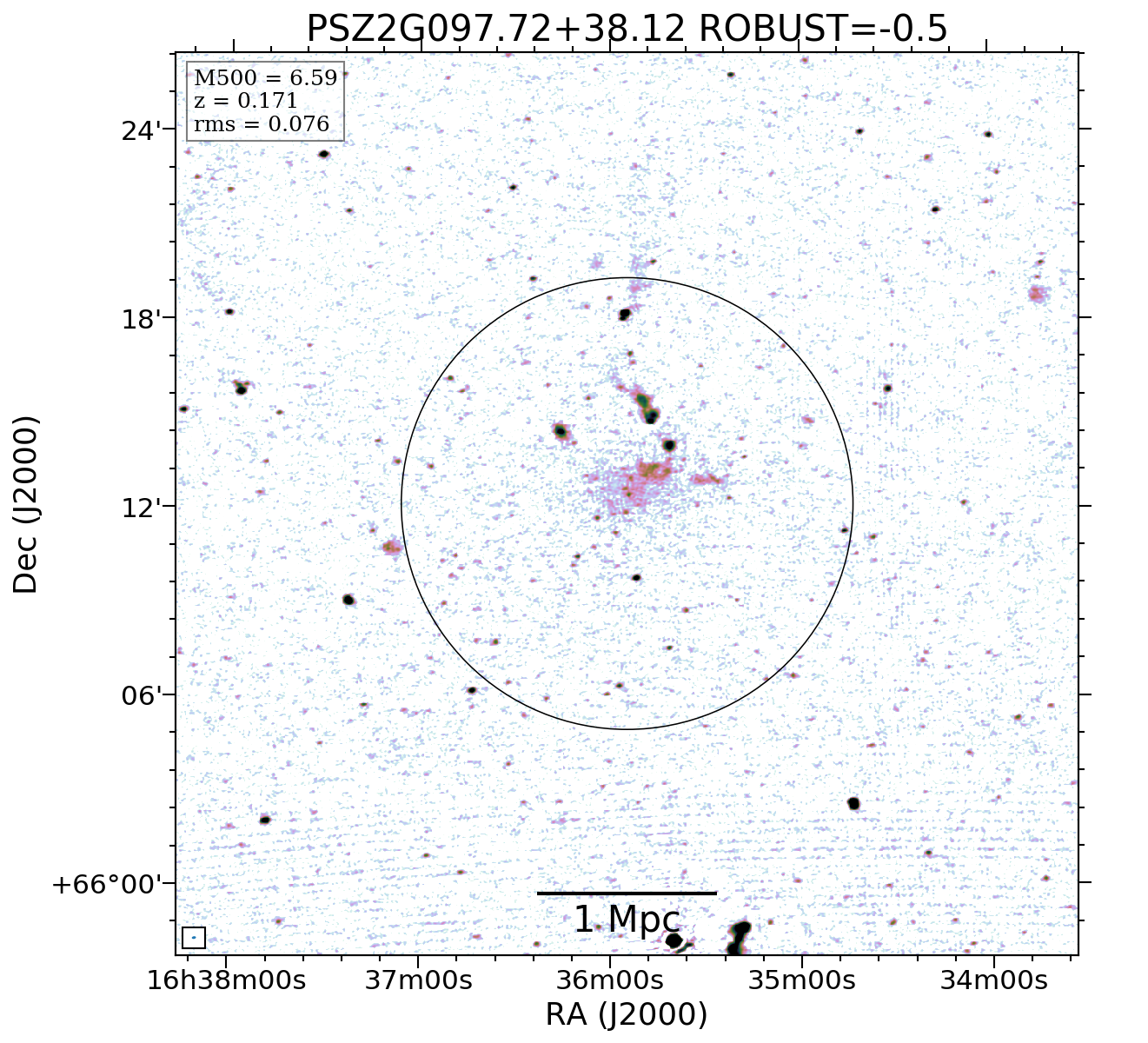}
  \includegraphics[width=.33\hsize,trim={0cm 0cm 0cm 0cm},clip,valign=c]{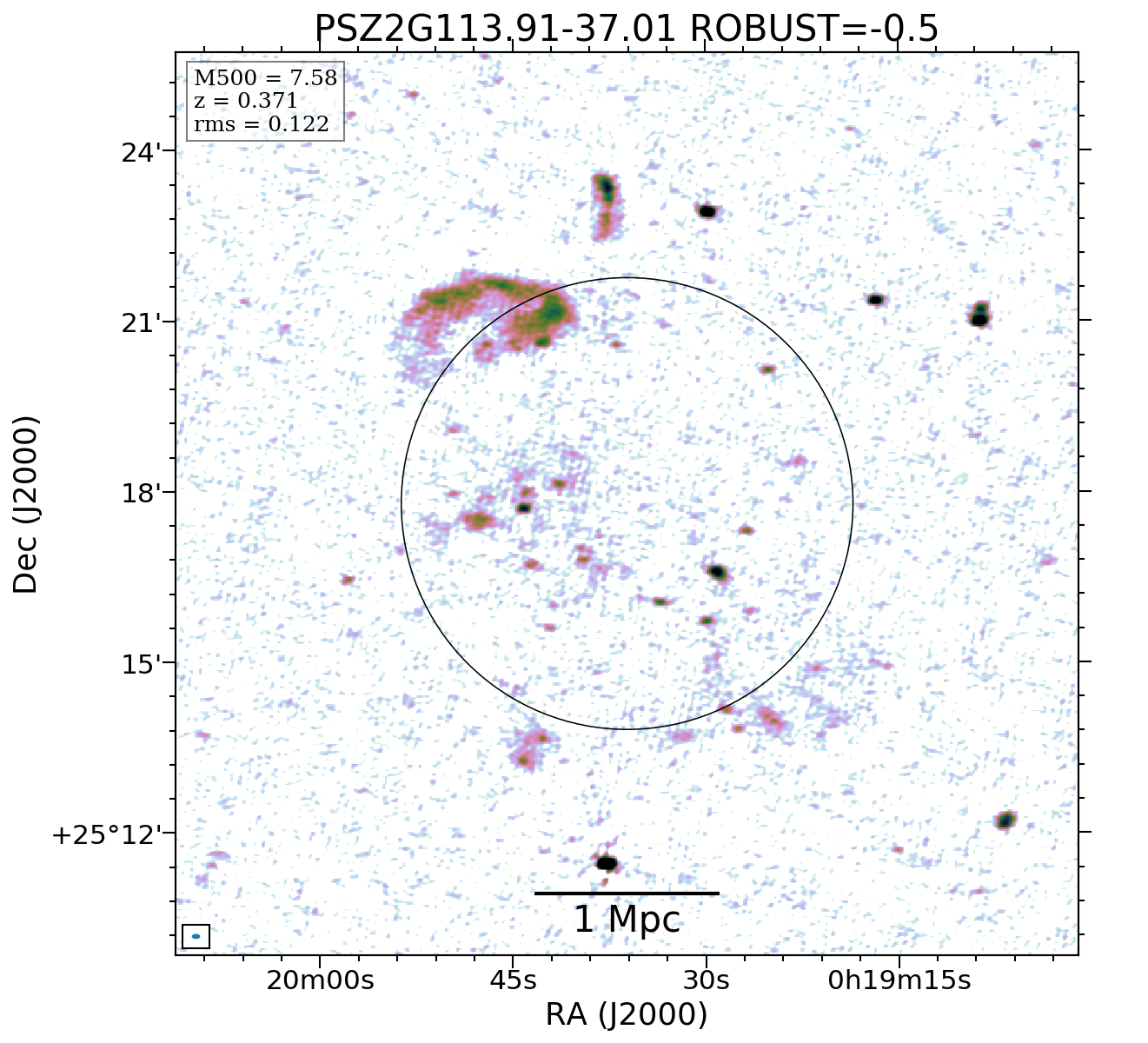}
  \includegraphics[width=.33\hsize,trim={0cm 0cm 0cm 0cm},clip,valign=c]{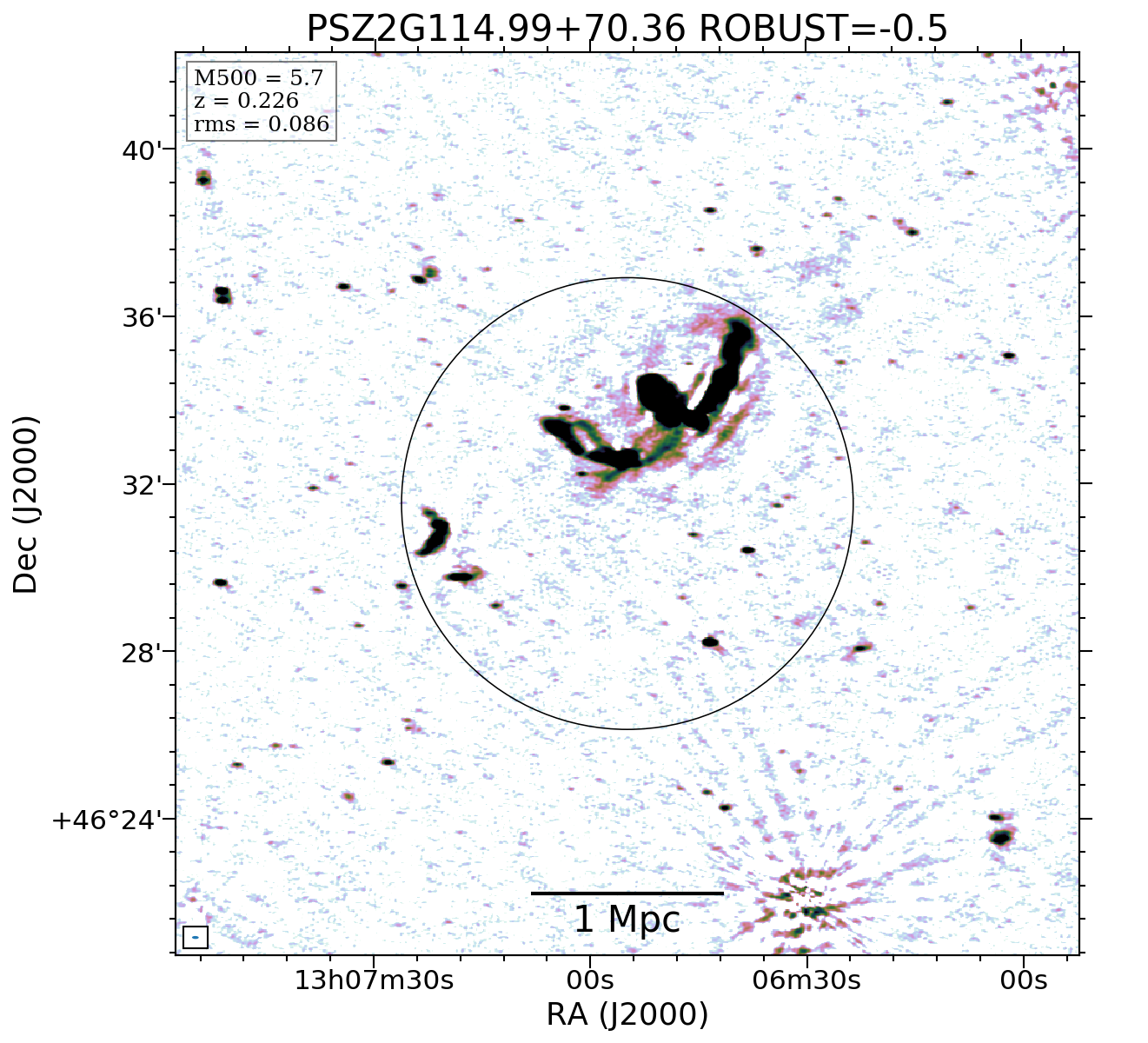}
  \includegraphics[width=.33\hsize,trim={0cm 0cm 0cm 0cm},clip,valign=c]{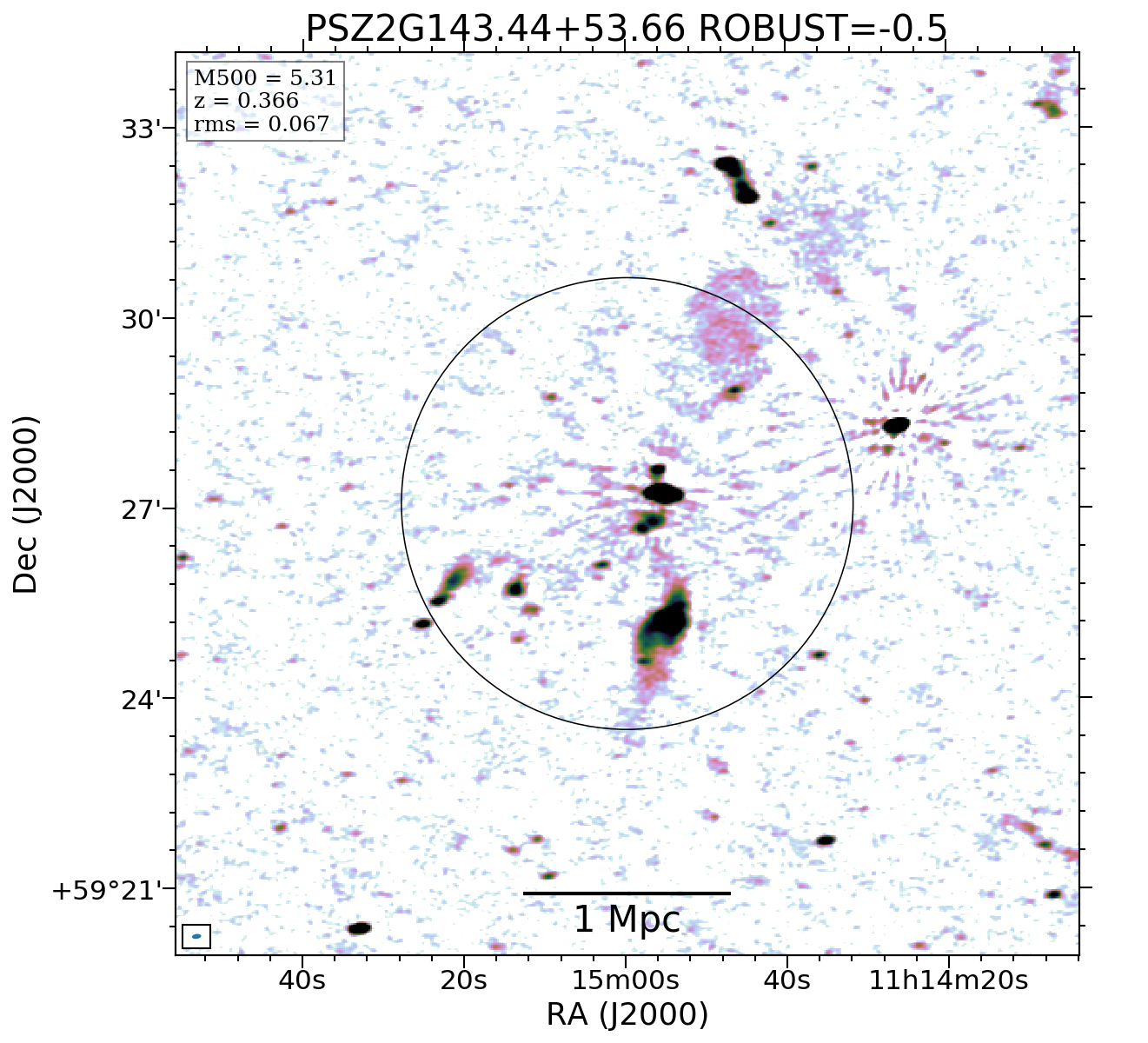}
  \includegraphics[width=.33\hsize,trim={0cm 0cm 0cm 0cm},clip,valign=c]{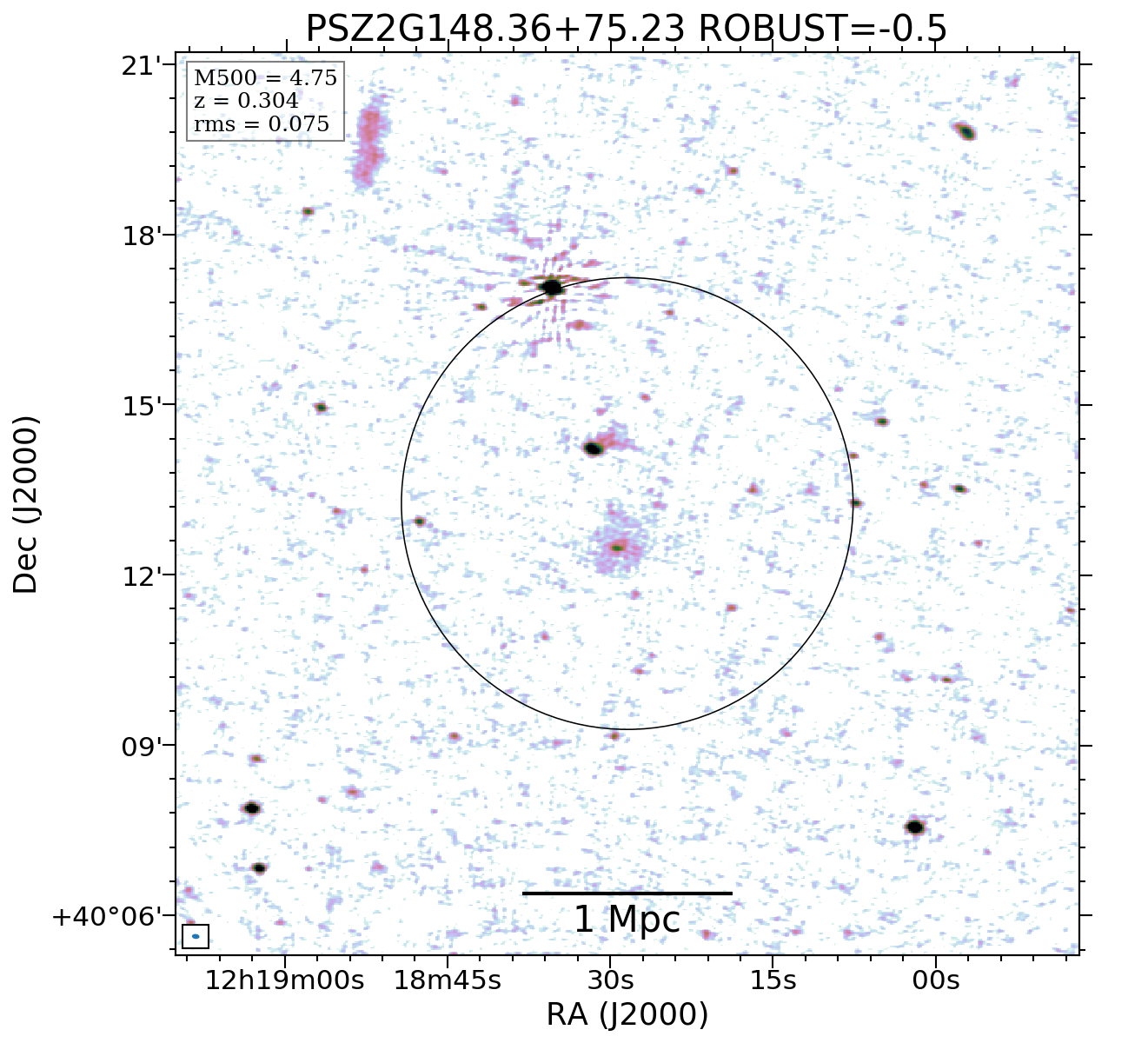}
  \includegraphics[width=.33\hsize,trim={0cm 0cm 0cm 0cm},clip,valign=c]{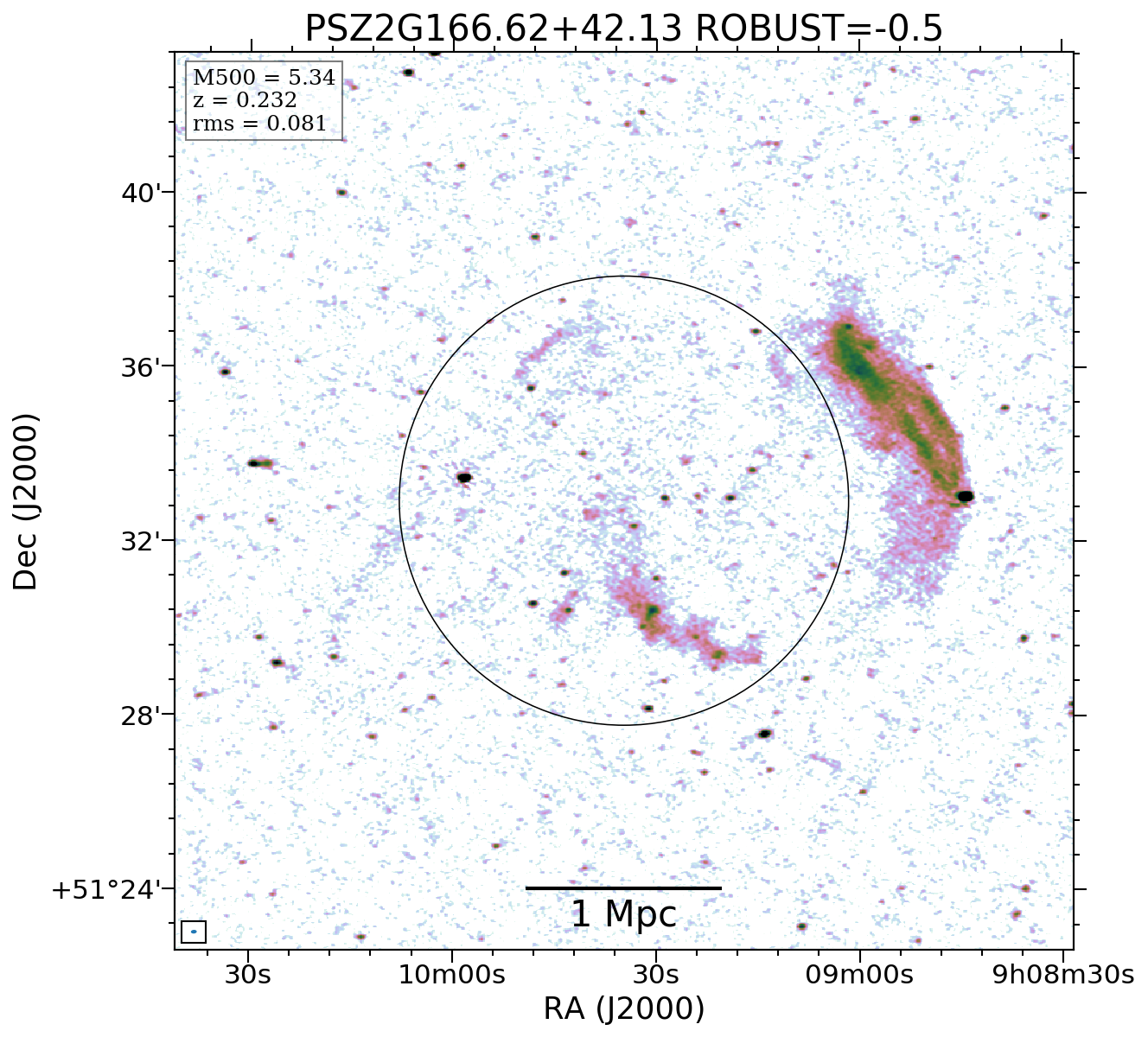}  
  \caption{Collection of clusters that show several types of radio emission. PSZ2 G053.53+59.52 has a central radio halo and a number of sources of uncertain origin. PSZ2G071.21+28.86 has a double radio relic. PSZ2 G088.53+41.18 is a system without diffuse cluster emission as the only emission detected is associated with an optical galaxy. PSZ2 G097.72+38.12 has a radio halo. PSZ2 G113.91-37.01 has a radio halo and two relics. PSZ2 G114.99+70.36 has emission of uncertain origin. PSZ2 G143.44+53.66 has emission of uncertain origin. PSZ2 G148.36+75.23 has a radio halo. PSZ2 G166.62+42.13 has a radio halo and multiple relics. We remark that the classification was done by inspecting all the sets of images available, while the images displayed in this gallery are only the reference ones (\ie,\ those with \texttt{robust=-0.5} and no \uv\  taper). For a more complete picture of the large variety of radio structures observed in our sample, we refer the reader to Fig.~\ref{fig:dr2_collection} and to the project website.}
  \label{fig:showcases}
\end{figure*}

For each object listed in Table~\ref{tab:sample} we searched for diffuse radio sources in the ICM that are not clearly associated with any AGN by visually inspecting the set of \lofar\ images at different resolutions (with and without source subtraction) together with the optical and X-ray overlay images. To make the classification of the radio emission as objective as practical at present and easily reproducible, we created a decision tree (Fig.~\ref{fig:decision_tree}) that we followed during the inspection of the images and classified each cluster. Below we define the six classes of objects that form the end points of our decision tree. \\
\indent
``Radio halos'' (RH) are extended sources that occupy the region where the bulk of the X-ray emission from the ICM is detected. Historically, they were divided according to their sizes: giant halos extended on cluster-scale and mini-halos covering the cluster central region. Since in this work we are dealing with a large sample of clusters with masses spanning over one order of magnitude of difference, we prefer to not separate mini-halos from giant-radio halos based on the size of the radio emission; instead, we refer generically to radio halos. \\
\indent
``Radio relics'' (RR) are elongated sources whose position is offset from the bulk of the X-ray emission from the ICM and consistent with lying in cluster outskirts. We also checked for a sharp edge in the radio emission and a largest-linear size (LLS) $\gtrsim$ 300 kpc.  \\
\indent
The classification ``candidate radio halo'' or ``candidate radio relic'' (cRH or cRR) is used when \chandra\ or \xmm\ X-ray observations are not available and as such the presence of a halo or relic cannot be firmly claimed because we are uncertain if the emission is in the central or outer region of the ICM. However, it is possible to make an assessment based on the position of the radio emission with respect to the apparent overdensity of galaxies in the optical image. We thus classify the emission as being a candidate radio halo or radio relic if it is consistent with the other properties of these types of sources but colocated with or offset from the overdensity of galaxies rather than the X-ray emission from the ICM. \\
\indent
The ``uncertain'' (U) classification is for objects whose emission was significantly affected by calibration and/or subtraction artifacts or which did not show a morphology, size, and/or position consistent with the categories of radio halos and relics. \\
\indent
``No diffuse emission'' (NDE) indicates that these objects do not show the presence of diffuse emission that could be attributed to the ICM (although they may show the presence of lobes or tails from AGN in the field). \\
\indent
Finally, ``not applicable'' (N/A) is used if the image quality is 3 (see Sect.~\ref{sec:imaging}) and as such the object cannot be adequately classified because of the poor data quality. \\
\indent
The classification for each target in the sample performed following the decision tree of Fig.~\ref{fig:decision_tree}, where the endpoints reflect the definitions above, is reported in Table~\ref{tab:sample}. If a target showed more than one diffuse source not related with any AGN, we repeated the decision tree for each source separately. All the answers to the questions of the decision tree are available on the project website. A gallery highlighting the wide variety of different types of emission is shown in Fig.~\ref{fig:showcases}. \\
\indent
We note that during our classification we did not attempt to identify radio phoenixes. These complex sources trace AGN radio plasma that has an ultra-steep ($\alpha \gtrsim 1.8$) spectrum and that is thought to have been reenergized through processes in the ICM, unrelated to the radio galaxy itself. New instances of this class of objects are rapidly emerging thanks to sensitive observations at low frequencies \citep[\eg,][for recent works]{degasperin17gentle, kale18, mandal19, mandal20, duchesne20a1127, duchesne21palaeontology, duchesne22arx, botteon21a1775, hodgson21jellyfish}. Our choice of not yet attempting to classify radio phoenixes is twofold. First, their definition is tightly connected to the spectral index, requiring the analysis of multifrequency observations. Second, their amorphous morphology and small sizes (at most a few hundred kiloparsecs) make a robust classification even more challenging. That said, in Table~\ref{tab:sample} we do highlight several instances where a suspected radio phoenix was noted to form a very prominent part of the cluster emission (for example, when it is the dominant emission in the cluster volume).

\section{Flux density measurements}\label{sec:fluxdensity}

\begin{figure*}
    \centering
    \includegraphics[width=\hsize,trim={0cm 0cm 0cm 0cm},clip]{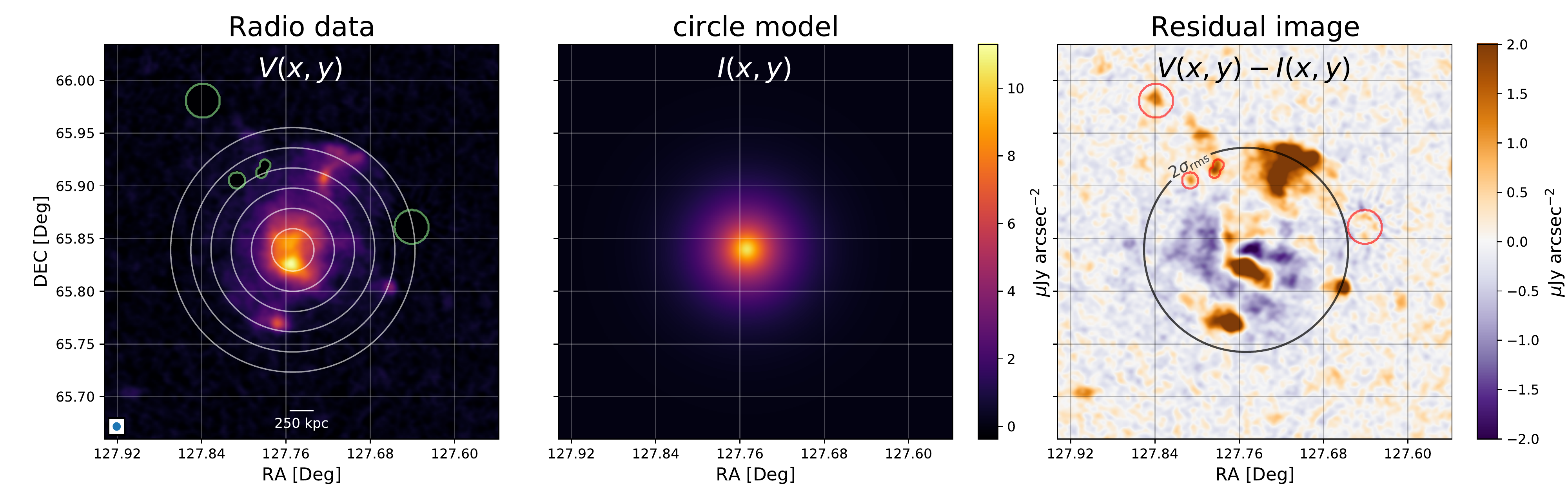}
    \begin{minipage}[t]{0.65\linewidth}
     \includegraphics[width=\hsize,trim={0cm 0cm 0cm 0cm},clip]{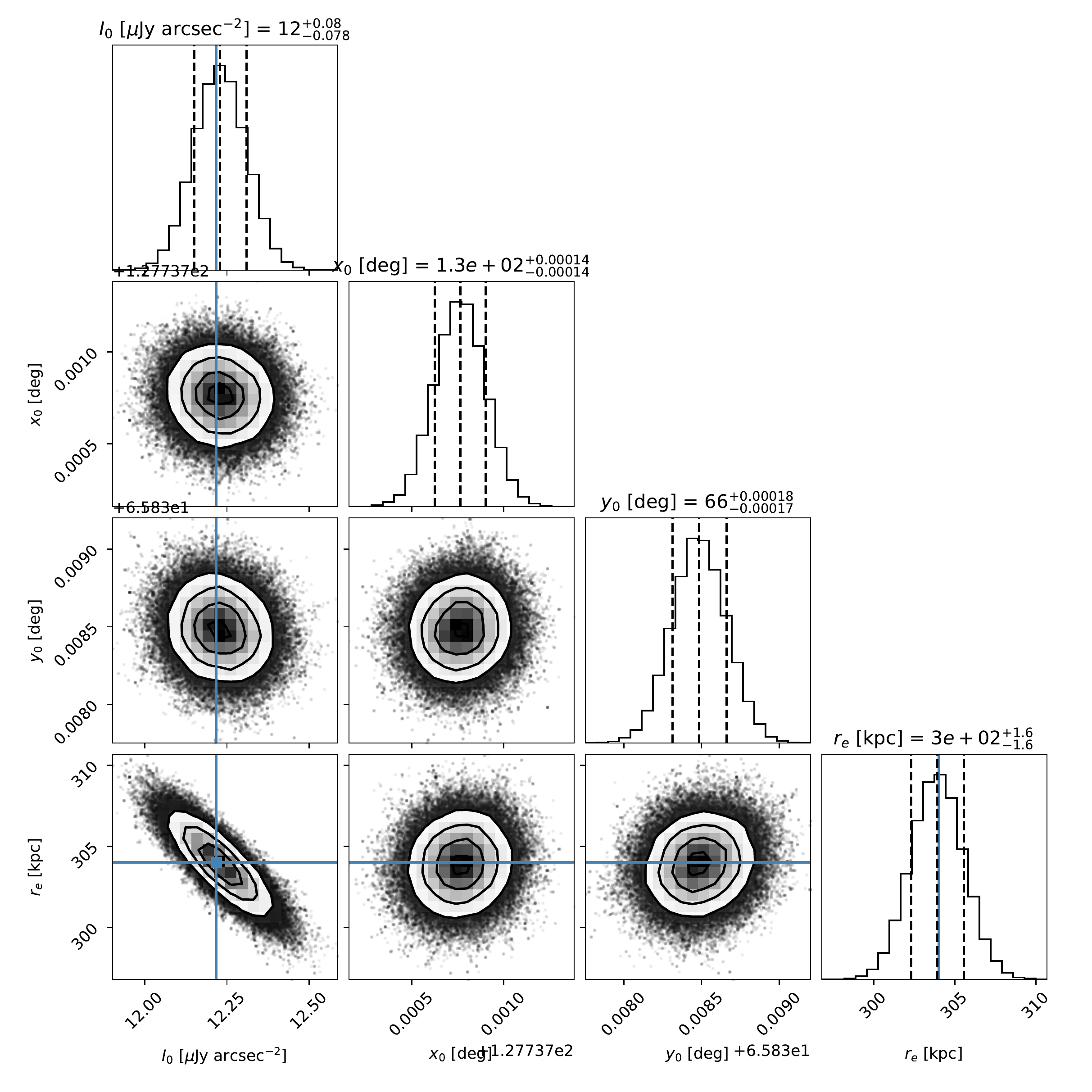}
    \end{minipage}
    \hfill
    \begin{minipage}[b]{0.3\linewidth}
      \caption{Results obtained by fitting the radio halo shown in Fig.~\ref{fig:example_reimaging} with \halofdca\ \citep{boxelaar21}.  \textit{Top figures}: Image used for the fit with overlaid: the contours (white circles) of the best-fit circular model drawn at $[1, 2, 4, 8, \dots] \times \sigma$ (\textit{left} panel), the image of the best-fit model (\textit{central} panel), and the residual image of the fit, with the $2\sigma$ model contour denoted by the black circle (\textit{right} panel). Contaminating sources are masked out and are highlighted by the green and red regions (\textit{left} and \textit{right} panels, respectively). \textit{Bottom figure}: MCMC corner plot presenting the distribution of the posteriors of each fitted parameter \citep{foremanmackey13}.}
      \label{fig:a665_fdca}
    \end{minipage}
\end{figure*}

\begin{figure}
 \centering
 \includegraphics[width=\hsize,trim={0cm 0cm 0cm 0cm},clip]{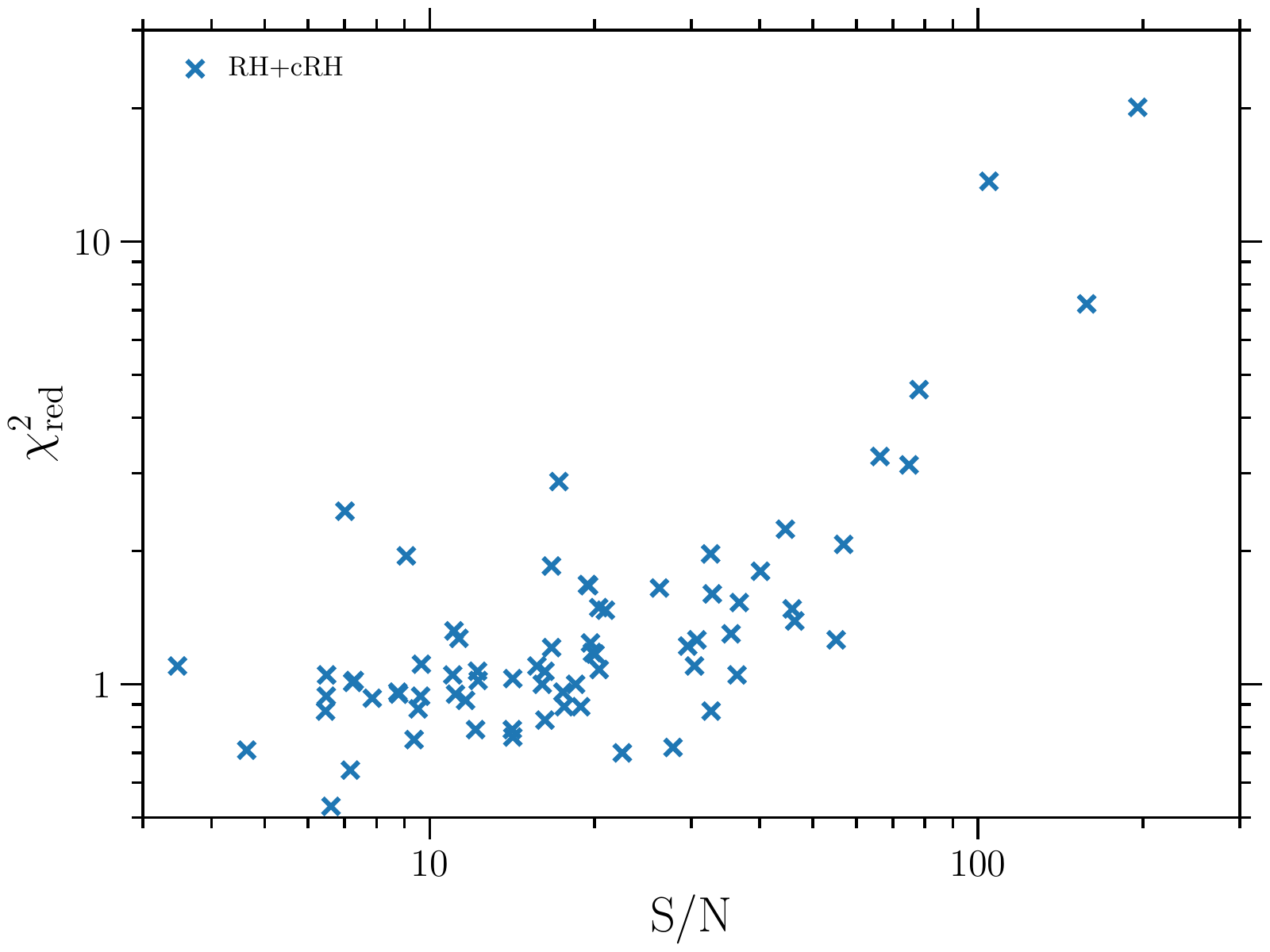}
 \caption{$\chi^2_{\rm red}$ vs. S/N (\ie,\ the ratio between the integrated flux density and its fitting uncertainty) for the radio halos and candidate radio halos in our sample. The positive trend indicates that deviations from the exponential model become evident only for the halos that are detected with sufficient significance.}
 \label{fig:SNR_vs_chi2red}
\end{figure}

In the following subsections, we describe how we computed the total flux density $S_\nu$ of the radio halos and relics in our sample. These values were used to calculate the $k$-corrected radio power of the sources at 150 MHz as

\begin{equation}\label{eq:power}
 P_{150} = 4\pi S_{150} D_{\rm L}^2 (1+z)^{\alpha-1}\:,
\end{equation}

\noindent
where $D_{\rm L}$ is the luminosity distance and $S_{150} = S_{144} \left(\frac{144\:\rm{MHz}}{150\:\rm{MHz}} \right)^\alpha$. Because our integrated flux density measurements are obtained at 144 MHz, $S_{150}$ is only marginally affected by the adopted (unknown) radio spectral index. In this respect, we assumed $\alpha=1.3$ for radio halos and $\alpha=1.0$ for radio relics, which are typical values found in the literature for these kinds of sources \citep[see, \eg,][]{feretti12rev, vanweeren19rev}. Instead, the rest-frame radio power is more sensitive to the spectral index, especially for sources at higher redshift, due to the $k$-correction term appearing in Eq. 5. The interested reader should take $P_{150}$ with more caution particularly for the radio halos in the sample at high $z$, which could be characterized by values of $\alpha>1.3$ \citep{digennaro21spectral}. \\
\indent
The quantities derived for the radio halos and relics in our sample are reported in Tables~\ref{tab:RH_sample} and \ref{tab:RR_sample}, respectively. In the tables, uncertainties on $S_{150}$ (hence, on $P_{150}$) take into account the statistical ($\sigma_{\rm stat}$), systematic ($\sigma_{\rm sys}$), and subtraction ($\sigma_{\rm sub}$) errors, which were summed in quadrature. The statistical error

\begin{equation}\label{eq:sigma_stat}
 \sigma_{\rm stat} = \rm{rms} \sqrt{\rm{N_{beam}}}
\end{equation}

\noindent
is related to the rms of the image in the integration area (and, in the case of halos, to the fitting process; see Sect.~\ref{sec:flux_halos}). The systematic error,

\begin{equation}\label{eq:sigma_sys}
 \sigma_{\rm sys} = \delta_{\rm cal} S_{\rm diffuse}
,\end{equation}

\noindent
is given by the uncertainty of the flux scale calibration, $\delta_{\rm cal}$, which is set to $10\%$ for \lotss-DR2 (see Sect.~\ref{sec:fluxscale} and \citealt{shimwell22}). The subtraction error,

\begin{equation}\label{eq:sigma_sub}
 \sigma_{\rm sub} = \xi_{\rm res} S_{\rm discrete}
,\end{equation}

\noindent
takes into account the presence of possible residuals in the diffuse emission region from the discrete sources that were subtracted in the \uv-plane. To determine the fraction of residual contaminating emission, $\xi_{\rm res}$, we visually inspected the images used for the subtraction for a subsample of clusters characterized by different values of flux density in discrete sources, and found that the following percentages,

\begin{equation}
 \xi_{\rm res} = 
 \begin{cases}
  16\% & \text{if $S_{\rm discrete} < 10$ mJy} \\
  8\% & \text{if $10$ mJy $ < S_{\rm discrete} < 100$ mJy} \\
  4\% & \text{if $100$ mJy $ < S_{\rm discrete} < 1000$ mJy} \\
  2\% & \text{if $S_{\rm discrete} > 1000$ mJy} \\
 \end{cases}
,\end{equation}

\noindent
provide a good approximation to quantify the level of contamination in our measurements.

\subsection{Radio halos}\label{sec:flux_halos}

We employed the \halofdcaE\footnote{\url{https://github.com/JortBox/Halo-FDCA}} \citep[\halofdca;][]{boxelaar21} to measure the integrated flux density from the observed radio halos. This code fits the two-dimensional surface brightness profile with a \mcmcE\ (\mcmc) method that estimates the best-fit parameters and associated uncertainties. As proposed by \citet{murgia09}, for the fitting we assume exponential profiles in the form

\begin{equation}\label{eq:exponential}
 I(r) = I_0 e^{-G(r)}\:,
\end{equation}

\noindent
where the fitted parameters are $I_0$, which is the central brightness, and $G(r),$ which is a function that determines the model morphology (\ie,\ circular, elliptical or skewed; see \citealt{boxelaar21} for more details). As done for the \lotss-DR1 cluster sample \citep{vanweeren21}, we primarily used a simple circular exponential model to fit the discrete-source-subtracted images that were obtained with a Gaussian \uv\  taper corresponding to 50 kpc at the cluster redshift. If the signal to noise in these images was low or the emission was clearly non circular, we instead used the discrete-source-subtracted images with \uv\  taper of 100 kpc or of the elliptical exponential model. In total, the circular model has four free parameters: $I_0$, the coordinates of the center ($x_0$ and $y_0$), and a single \textit{e}-folding radius ($r_1$). The elliptical model has two additional free parameters: a second \textit{e}-folding radius ($r_2$) and a rotation angle ($\phi$). \\
\indent
Prior to fitting the radio halos we carefully examined the images for contaminating extended sources that had been poorly subtracted, such as tailed radio galaxies and radio relics. We also identified regions affected by residual calibration errors. These problematic regions were  manually masked during the fitting. In order to reduce the processing time and the size of the area to manually inspect for masking, we provided as input to \halofdca\ images with a FoV reduced to approximately $1.5 \rfive \times 1.5 \rfive$. \\
\indent
A qualitative assessment of the fit quality can be done by inspecting the residual images and corner plots produced by \halofdca, which we have made publicly available for each cluster on the project website. An example of these plots for the prominent radio halo that is hosted in the cluster PSZ2 G149.75+34.68 (see Fig.~\ref{fig:example_reimaging}) is reported in Fig.~\ref{fig:a665_fdca}. We note that \halofdca\ returns the signal-to-noise ratio (S/N) and the $\chi^2_{\rm red}$ of the fit, which can also be used to assess the fit quality. However, we urge caution in interpreting these values. The S/N in \halofdca\ is defined as the integrated flux density value divided by the uncertainty introduced by the fit (which is based on the image noise and the number of data points related to the area of the halo), and hence it is not a parameter that determines the significance of the detection. The $\chi^2_{\rm red}$ instead is calculated from the difference between the fitted exponential model and the data, and high $\chi^2_{\rm red}$ values may suggest that a halo has a lot of substructure. For example, this is the case of the halo shown in Fig.~\ref{fig:a665_fdca}, where $\chi^2_{\rm red} = 7.23$. We note a positive trend between S/N and $\chi^2_{\rm red}$ for the radio halos and candidate radio halos in our sample (Fig.~\ref{fig:SNR_vs_chi2red}), indicating that the exponential model, while being a reasonable description of the data, may be affected by the presence of asymmetries and substructures that become evident only when the data are of sufficient statistical quality. \\
\indent
To demonstrate the performance of \halofdca, in Appendix~\ref{app:flux} we compare the integrated flux density obtained by integrating the flux density within a circle (or ellipse, depending on the model used in \halofdca) that roughly encompasses the $2\sigma$ contour of the radio halo ($S_{\rm 2\sigma}$) with that obtained with \halofdca\ ($S_{\rm fit}$). We generally see good agreement between the two quantities, but, as detailed in the Appendix, we found that the fitting could not be done reliably due to the low significance of the emission for 10 out of the 83 fitted radio halos and candidate halos. These sources are reported as RH*/cRH* in Table~\ref{tab:sample} and their integrated flux densities cannot be determined accurately with current data. For all halos except PSZ2 G107.10+65.32 and PSZ2 G139.18+56.37 we use as reference the \halofdca\ derived integrated flux density. For these two clusters there is significant substructure and we instead derived the integrated flux density by summing the pixels within the $2\sigma$ contour level. \\
\indent
As suggested by \citet{murgia09}, when calculating the \halofdca\ derived integrated flux densities we integrated the best-fit models up to a radius of three times the \textit{e}-folding radius. This choice leads to a flux density that is $\sim$80\% of the one that would be obtained by integrating the model up to infinity and is motivated by the fact that halos do not extend indefinitely. \\
\indent
The quantities derived for the radio halos and candidate radio halo in our sample are reported in Table~\ref{tab:RH_sample}.

\begin{figure*}
 \centering
 \includegraphics[width=\hsize,trim={0cm 0cm 0cm 0cm},clip]{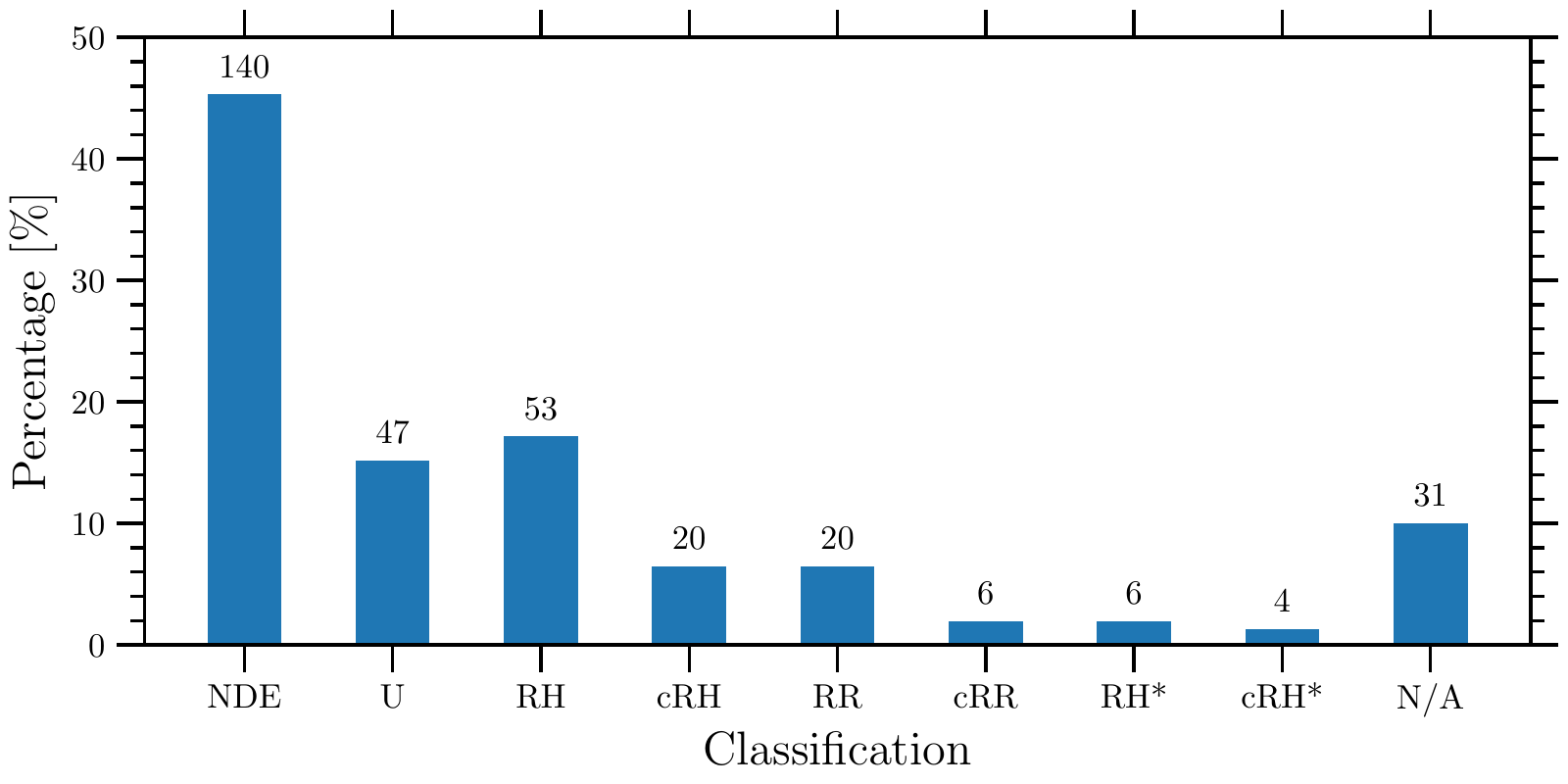}
 \caption{Summary of the number of clusters classified in our sample divided per category (NDE=no diffuse emission; U=uncertain; RH=radio halo; RR=radio relic; c=candidate; *=flux density measurement not reliable; N/A=not applicable). We note that a cluster can be classified under multiple categories, such as both RH and RR. The percentage on the $y$ axis is with respect to the total number of PSZ2 in \lotss-DR2 (309 targets).}
 \label{fig:classification_histo}
\end{figure*}

\subsection{Radio relics}

The integrated flux density of radio relics was computed in polygonal regions encompassing the $2\sigma$ contour of the diffuse emission. In the process, particular care was devoted to excluding possible artifacts due to calibration errors and/or other contaminating extended sources in the cluster. As done for halos, we primarily used the images obtained with a Gaussian \uv\  taper corresponding to 50 kpc at the cluster redshift with discrete sources removed. In a handful of relics (PSZ2 G089.52+62.34 N2, PSZ2 G091.79-27.00, PSZ2 G113.91-37.01 S, PSZ2 G166.62+42.13 E, and PSZ2 G205.90+73.76 N/S), we instead used the discrete-source-subtracted images with \uv\  taper of 100 kpc, where the integrated flux density of the relics is $>$10\% than that obtained in the 50 kpc images. From inspection of the images of PSZ2G190.61+66.46, we found that the relic was partially included in the model for source subtraction and hence that a small fraction of the relic emission was removed. Since there are no compact sources in the region of the relic, only for this target we chose to use the 50 kpc image prior to the source subtraction to measure its properties. \\
\indent
We estimated the LLS of each relic by computing the distance between the pixels above $2\sigma$ with the largest separation in the diffuse emission. Owing to the reliability of the flux density within a beam,  the error in the measured LLS corresponds to the size of the restoring beam of the image. \\
\indent
The downstream extension of radio relics, or relic width, often varies significantly across the extent of the relic. Additionally, the irregular shapes of some radio relics make it challenging to decide at which position to define and measure the width. We therefore decided to take a statistical approach to measuring the width of a relic. The straight line joining the LLS pixel pair typically lies approximately perpendicular to the width direction. This was verified by eye. Therefore, by measuring the largest separation between relic pixels along the line perpendicular to the LLS at each pixel along the LLS, we obtained an estimate of the width at all points along the relic. We then took the median of these width measurements as the characteristic relic width and adopt as error the standard deviation of the measurements, which reflects the spread of values obtained. \\
\indent
We used the coordinates of X-ray centroids as reference points for calculating the projected distance of the radio relics from the clusters. The X-ray centroid of each cluster was calculated within a region of $r_{500}$ centered at the coordinates reported in the PSZ2 catalog. Only for PSZ2 G107.10+65.32 S, we used a region of 500 kpc centered at the X-ray peak due to the PSZ2 coordinates being outside this subcluster. \\
\indent
Due to possible projection effects, the shock front associated with the relic may not be located at the sharp surface brightness edge of the relic, but at the brightest region of the relic. We therefore chose to define the position of the relic as the center of the brightest 10\% of pixels within the relic, weighted by their flux density. The distance between the relic and the centroid of the cluster X-ray emission ($D_{\rm RR-c}$) as well as the separation between double relics ($D_{\rm RR-RR}$) were computed from these coordinates. We include an additional error to account for the projection between the relic and X-ray centroid axis. Since the merger axis of clusters hosting double radio relics is approximately on the plane of the sky, the additional error corresponds to a projection of 10\deg. For all other relics we take an offset of 30\deg.  \\
\indent
The quantities derived for the radio relics and candidate radio relics in our sample are reported in Table~\ref{tab:RR_sample}.

\section{Results and discussion}\label{sec:results}

\subsection{Number and distribution of sources}\label{sec:number_sources}

\begin{figure*}
 \centering
 \includegraphics[width=\hsize,trim={0cm 0cm 0cm 0cm},clip]{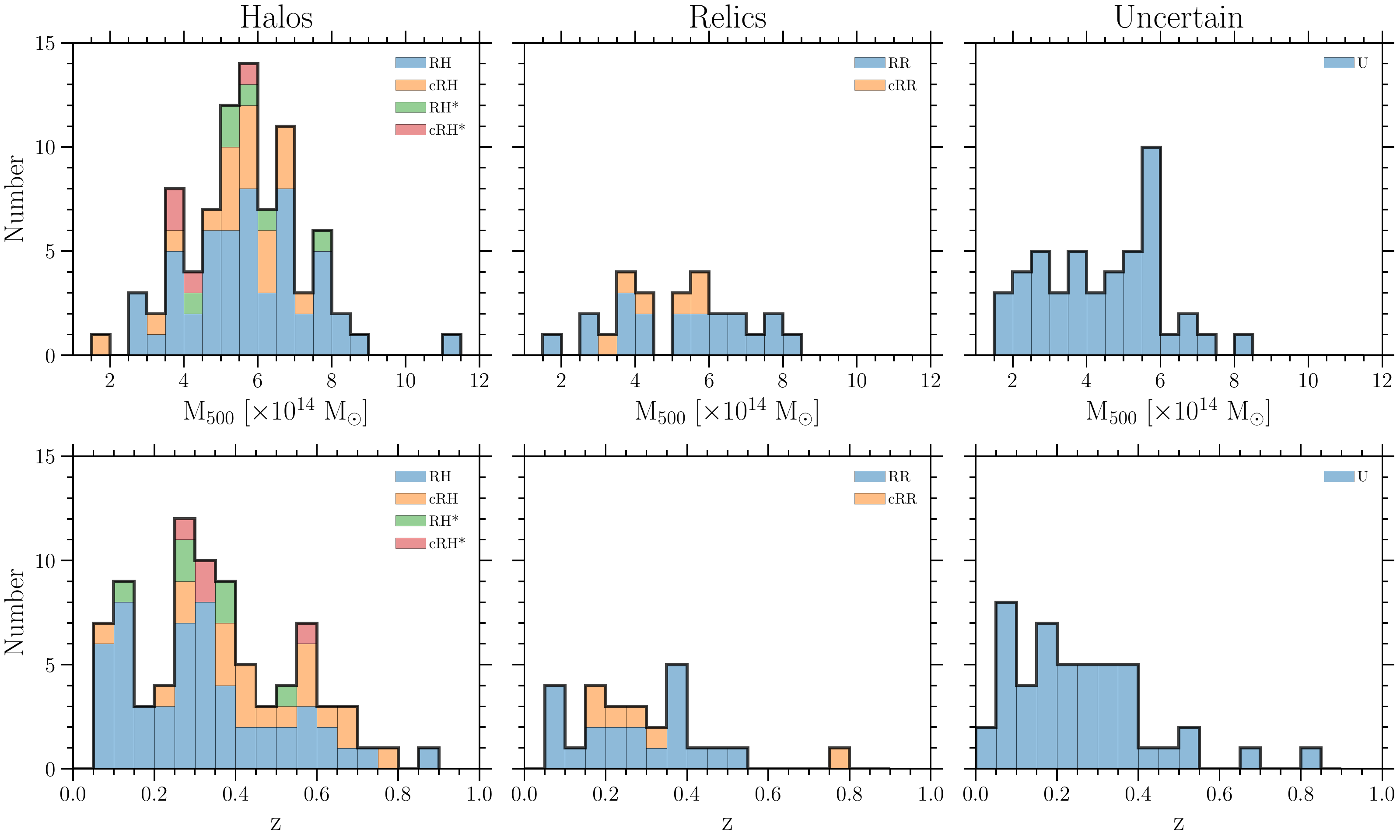}
 \caption{Histograms of the mass (\textit{top} panels) and redshift (\textit{bottom} panels) distribution for the main categories of objects classified in this work.}
 \label{fig:samples_distribution}
\end{figure*}

With 309 objects, the \lotss-DR2/PSZ2 sample represents the largest statistical sample of galaxy clusters observed with highly sensitive low-frequency observations that has been ever used to search for and study diffuse synchrotron emission in the ICM. The number of clusters divided per category as defined in Sect.~\ref{sec:classification} is summarized in Fig.~\ref{fig:classification_histo}. We found 73 clusters hosting a radio halo (\ie,\ 53 RH and 20 cRH), 26 clusters hosting one or more relics (\ie,\ 20 RR and 6 cRR), and 47 clusters with diffuse emission of uncertain origin. Additional 10 clusters are found to host a radio halo from visual inspection but since the surface brightness profile fittings were of poor quality they were reported with an asterisk in Fig.~\ref{fig:classification_histo} and Table~\ref{tab:sample} (these are the 6 RH* and 4 cRH* discussed in Appendix~\ref{app:flux}). No diffuse radio emission from the ICM was found for 140 targets while for 36 objects it was not possible to investigate the presence of diffuse emission either due to the impossibility of applying the extraction + recalibration method described in Sect.~\ref{sec:datareduction} (5 out of 36) or due to the bad image quality (31 out of 36). Notes on individual clusters (including references on previous studies employing radio observations) are reported in Appendix~\ref{app:notes}. \\
\indent
The number of clusters with newly discovered radio halos and relics in our sample is 50 (\ie,\ 50\% of the total number of halo and relic detections) and these clusters are highlighted in Table~\ref{tab:sample}. These new discoveries consist of 35 clusters hosting a radio halo (\ie,\ 21 RH and 14 cRH, namely 48\% of the number of halo detections) and 15 clusters hosting one or more radio relics (\ie,\ 9 RR and 6 cRR, namely 58\% of the number of relic detections). In addition, all the 10 halos classified with an asterisk (\ie,\ 6 RH* and 4 cRH*) except PSZ2 G118.34+68.79 \citep[see][]{vanweeren21} are reported in this work for the first time. In Fig.~\ref{fig:samples_distribution} we show the mass and redshift distribution for the clusters hosting halos, (one or more) relics, and uncertain sources in our sample. \\
\indent
We detected radio halos over the entire \mfive\ range of the PSZ2 sample we studied, which spans nearly one order of magnitude. The bulk of detections occur in the mass range $5 \times 10^{14}$ \msun\ $ \lesssim \mfive\ \lesssim 7 \times 10^{14}$ \msun. In Cassano et al. (in preparation) we discuss the occurrence of halos with the cluster mass, properly taking into account the selection effects and completeness of the PSZ2 catalog. Of particular interest is the large number of detections of halos in clusters with a mass $\mfive < 5 \times 10^{14}$ \msun\ (\ie,\ 17 RH, 4 cRH, 3 cRH*, and 1 cRH*). This mass regime is poorly studied and it has been disclosed only very recently thanks to sensitive observations with new generation instruments \citep[\eg,][for recent works]{hlavaceklarrondo18, hoang19a2146, vanweeren21, botteon19lyra, botteon21ant, botteon21a1775, duchesne21palaeontology}. We note that the candidate radio halo in PSZ2 G192.77+33.14 [$\mfive = (1.66 \pm 0.20) \times 10^{14}$ \msun], if confirmed, would be the least massive system presently known to host a radio halo. Radio halos in our sample cover also a wide redshift range, from $z=0.05$ (PSZ2 G192.77+33.14) up to $z=0.888$ (PSZ2 G160.83+81.66; see \citealt{digennaro21fast, digennaro21spectral}), with the bulk located at $0.25 \lesssim z \lesssim 0.4$. The discovery of bright radio halos in systems (partially overlapping with our sample) with $z > 0.6$ has been recently made possible thanks to \lotss\ observations \citep{cassano19, digennaro21fast}. The evolution of the radio halo properties with redshift will be discussed in Cassano et al. (in preparation). Finally, we note that in Fig.~\ref{fig:samples_distribution} we have not reported the candidate radio halo in PSZ2 G144.23-18.19, which is our only detection in a cluster without redshift and mass, while the double radio halo cluster in PSZ2 G107.10+65.32 \citep[see][]{botteon18a1758} was counted only once. \\
\indent
The radio relic sample has a broad distribution both in terms of \mfive\ and $z$ (despite being more limited in size with respect to that of halos). Unlike halos, we find only one radio relic in clusters at $z>0.6$, namely in PSZ2 G069.39+68.05 ($z=0.762$). This relic is claimed here for the first time. The least massive cluster in our sample is also that with the lowest redshift: PSZ2 G089.52+62.34 [$\mfive = (1.83 \pm 0.19) \times 10^{14}$ \msun\ at $z=0.07$], which was already reported in \citet{vanweeren21}. The total number of relic sources is 35, and the number of clusters hosting more than one relic in the ICM is 8. We only found 6 cases of classic, symmetric (\ie,\ located in diametrically opposite sides of the cluster), double radio relic systems (PSZ2 G071.21+28.86, PSZ2 G099.48+55.60, PSZ2 G113.91-37.01, PSZ2 G165.46+66.15, PSZ2 G181.06+48.47, PSZ2 G205.90+73.76).  We remark that clusters hosting more than one relic are reported only once in Fig.~\ref{fig:samples_distribution}. The statistical analysis of radio relics is presented in Jones et al. (in preparation). \\
\indent
Clusters with uncertain diffuse emission are mainly observed for $z < 0.4$ and $\mfive < 6 \times 10^{14}$ \msun. From visual inspection, we expect that a large fraction of these systems may host a radio halo (which, however, cannot be firmly claimed at the moment), while a smaller fraction may contain different kinds of sources in the ICM. Follow-up observations are required to confirm the nature of the emission observed. These clusters are not considered to derive scaling relations in Cuciti et al. (in preparation) and Jones et al. (in preparation), while for the study of the occurrence of radio halos in Cassano et al. (in preparation) we consider the two extreme cases where all or none of the uncertain sources are radio halos.

\subsection{Classification}

In this work and in the accompanying papers of the series we focus on the detection and characterization of radio halos and relics in the PSZ2 sample. These types of emissions are the most widely studied classes of diffuse synchrotron sources in the ICM, and are arguably the best defined and understood. \\
\indent
Diffuse sources in our sample were classified by visually inspecting  \lofar\ 120$-$168 MHz images at different resolutions, with and without discrete sources removed. We also made use of overlays of our radio data with optical and X-ray data when available. The most challenging aspect of the inspections was to disentangle the sources of interest (halos and relics) from other contaminating sources such as AGN and phoenixes and also from calibration artifacts. This separation relied upon the high resolution of \lofar\ and availability and high quality of the auxiliary data sets. However, in some cases it still remained very challenging to conclusively classify the emission. Indeed, low-frequency observations are very sensitive to low-energy relativistic electrons in the ICM, which even if emitted from a single compact region can occupy a significant fraction of the cluster volume due to their long lifetimes. As a consequence, the filling factor of emission associated with AGN increases at low frequency, and diffuse emission of somewhat uncertain origin and low significance can be identified almost in all clusters. This highlights the role of seed electrons in the formation of cluster-scale diffuse sources, which are an important ingredient for (re)acceleration models. \\
\indent
Although we found the application of the decision tree (Fig.~\ref{fig:decision_tree}) useful to make the classification less biased and more reproducible, we acknowledge that the visual inspection of the 309 objects in our sample is a painstaking procedure. It is clear that the strategy used in this work needs further development to properly assess significantly larger cluster samples or to eventually conduct a blind search for cluster emission. In this respect, machine-based techniques represent an appealing solution to classify the emission in large object samples \citep[\eg,][]{aniyan17, alhassan18, dominguezsanchez18, lukic18, lukic19classification, sadeghi21, vavilova21}. As part of this work, we have made public all our images and the detailed results of our decision-tree-based classification, which we hope can provide a good training set for algorithms that attempt either the full classification or to aid the automation at specific intersections in a decision-tree-type approach. Another approach would be establishing a citizen science project dedicated to galaxy clusters (\ie,\ a \lofar\ Galaxy Clusters Zoo) with the goal of distributing the classification work among citizen scientists, similarly to what is currently done by the \lofar\ Radio Galaxy Zoo\footnote{\url{http://lofargalaxyzoo.nl}} project, which has the aim of joining different components of the same radio galaxies observed by \lotss\ and identifying their optical counterparts. \\
\indent
As a final note, the unprecedented high resolution and high sensitivity at low frequencies provided by \lofar\ have the potential of probing previously unseen features of halos and relics, but also to unveil new kinds of emission in the ICM. Collecting the large variety of source morphologies in clusters in an automated way \citep[\eg,][]{gheller18, mostert21} will then help us to study the nature of the sources and the role of the surrounding environment to shape the radio emission.

\subsection{Prospects}

The fraction of identifications in each of our classifications with respect to the full \lotss-DR2 sample is shown Fig.~\ref{fig:classification_histo}. In the following, reported fractions were computed excluding the 36 targets of the sample for which we could not verify the presence of diffuse emission in the radio images (\ie,\ 5 not extracted and 31 N/A). The reported uncertainties are computed based on Poisson statistics. \\
\indent
The fractions of PSZ2 clusters in which a radio halo or one or more radio relics is claimed in this work are $19\pm9\%$ and $7\pm5\%$, respectively, and increase to $27\pm10\%$ and $10\pm6\%$ if candidates are also included. If the 10 halos classified with an asterisk for which we could not provide a reliable flux density measurement (Appendix~\ref{app:flux}) are also taken into account, the fraction of clusters with a radio halo detected rises further, to $30\pm11\%$. The fraction of clusters hosting sources with uncertain origin is $17\pm8\%$, while that of targets that do not show the presence of diffuse synchrotron emission in the ICM is $51\pm14\%$. \\
\indent
In our earlier work \citep{vanweeren21} we presented the first statistical analysis of galaxy clusters observed with \lotss\ and studied the 26 PSZ2 clusters residing in the $\sim$420 deg$^2$ covered by \lotss-DR1 (this region is contained within \lotss-DR2). We found that $73\pm15\%$ of PSZ2 clusters hosted some type of diffuse ICM related radio emission. More specifically, $62\pm15\%$ and $27\pm10\%$ PSZ2 clusters hosted a radio halo or one or more relics (including candidates), respectively. When comparing the fraction of diffuse emission in \lotss-DR1 with that of \lotss-DR2 one should take into account three important factors. Firstly, the \lotss-DR1 area is on average more sensitive than \lotss-DR2, as it covers a sky area above declination $+45\deg$, while \lotss-DR2 goes down to declination $+16\deg$, where the sensitivity of \lofar\ is reduced \citep{shimwell22}. This observational limitation makes the detection of faint ICM emission more challenging and likely biases low the fraction of extended sources in \lotss-DR2 compared to \lotss-DR1. Secondly, the area covered by \lotss-DR1 comprises a relatively small region of the sky, about 2\% of the northern sky, and may therefore be affected by cosmic variance. \lotss-DR2 instead spans about 27\% of the northern sky, enabling the possibility of performing reliable statistical studies. Thirdly, 28 PSZ2 entries in the \lotss-DR2 sample (\ie,\ 9\% of the total) lack $z$ and \mfive, and are not confirmed galaxy clusters at the moment, while all the PSZ2 detections in \lotss-DR1 are confirmed galaxy clusters. If the sources without $z$ and \mfive\ in \lotss-DR2 were \planck\ false detections, they should be rejected from the sample, resulting in an increase in the fraction of clusters hosting diffuse radio emission in the ICM (as all these 28 entries except PSZ2 G144.23-18.19 have been classified as NDE). We checked the quality flag \texttt{Q\_NEURAL} reported in the PSZ2 catalog for these 28 entries and found that 15 have \texttt{Q\_NEURAL<0.4}, which the threshold below which \citet{aghanim15} classify low reliability detections. This suggests that about half of the clusters for which we do not have $z$ and \mfive\ are likely spurious \planck\ detections. \\
\indent
The area covered by \lotss-DR2 is ideal to study the extragalactic sky as it avoids regions at low Galactic latitude and low-declination fields (\ie,\ between declination 0\deg\ and +15\deg), where the sensitivity of \lotss\ is generally a factor of 2$-$3 lower than the nominal survey noise\footnote{The sensitivity of \lotss\ observations scales with the elevation of the target as $A \times \cos(90\deg-\rm{elevation})^{-2}$, where $A=63$ \mujyb\ and the dependence on elevation is fixed according to the projected size of the \lofar\ stations \citep{shimwell22}.}. Compared to \lotss-DR1, it has a sky coverage $\sim$13 times larger (5634 deg$^2$ vs. 424 deg$^2$) and it samples a broader declination and right ascension range. This makes the new data release more representative of the quality of \lotss\ for extragalactic studies. We thus use our findings to refine the predictions on the number of PSZ2 sources that will be found to contain relics and halos at the completion of \lotss. \\
\indent
We consider that there are 835 detections in the PSZ2 catalog above a declination of 0\deg\ and assume a uniform sensitivity for \lotss\ for simplicity. We note that the presence of the Galactic plane is taken into account in the PSZ2 selection due to the lack of \planck\ detections in the zone of avoidance. The results found in \lotss-DR2 indicate that we will find $251\pm92$ clusters that host a halo and $83\pm50$ clusters that host one or more relics (including candidates), and, as in our study, approximately half of them should be new discoveries. When extrapolating the number of halos, we considered the fraction of $30\pm11\%$, which takes into account also the ten halos reported with an asterisk, as in these systems the visual inspection and classification tree led to the class of RH or cRH. If we assume that also all uncertain sources trace a halo, we obtain a conservative upper limit on the number of clusters hosting a radio halo that \lotss\ should find in PSZ2 clusters at its completion, namely $<401\pm117$. By considering the number of relic sources found in \lotss-DR2, we instead predict that at the completion of the survey \lotss\ will have observed $109\pm58$ radio relics in PSZ2 clusters. \\
\indent
In the context of the turbulent re-acceleration models for radio halos, \citet{cassano10lofar, cassano12} predicted the observation of 350$-$450 radio halos at $z \le 0.6$ in \lotss, while the number of radio relics expected to be observed in \lotss\ according to \citet{nuza12} is $\sim$2500. As these model expectations were not tailored for PSZ2 clusters, the comparison between the number of halos and relics at completion of \lotss\ and the extrapolations obtained from our analysis of the \lotss-DR2/PSZ2 sample is not straightforward. In this respect, the thoroughly statistical analysis of the results obtained for the PSZ2 clusters in \lotss-DR2 and implications on our understanding of halos and relics will be presented in the forthcoming papers of the series, where expectations will be further refined by considering: the sensitivity of present observations, the PSZ2 selection functions, and the new $P_{150}-\mfive$ correlations. Moreover, as radio halos and relics can be found in \lotss\ even in non-PSZ2 clusters, a census of diffuse radio emission in non-PSZ2 clusters in \lotss-DR2 is also currently ongoing \citep{hoang22sub}. \\
\indent
An important parameter to test the models of particle acceleration in the ICM is the spectrum of diffuse emission. For example, turbulent re-acceleration models predict the existence of a population of radio halos with steep spectra while spectral gradients in relics depend on the mechanisms of particle (re)acceleration at cluster shocks. Interestingly, we note that about half of the radio halos reported in this work are new discoveries (see Sect.~\ref{sec:number_sources}) and the turbulent re-acceleration scenario predicts that about half of the radio halos in \lotss\ should have ultra-steep spectra \citep[][]{cassano12}. If follow up observations of the radio halos in \lotss-DR2 will confirm that about 50\% of them have very steep spectra, this would be an important corroboration of such a class of models. Still, since \lotss\ is currently exploring an uncharted territory in terms of resolution and sensitivity compared to other completed surveys, it is not yet possible to perform spectral analysis for the full cluster sample. In this respect, we have planned a series of targeted observations for a selected number of objects with \meerkat, \vla,\ and \ugmrt. For a systematic study, ongoing and future sensitive radio surveys covering the northern sky, such as \lolss\ \citep{degasperin21}, \lodess\footnote{The LOFAR Decametre Sky Survey is a groundbreaking 14$-$30 MHz survey that will cover the sky above declination +20$\deg$.}, \apertifE\ (\apertif; Hess et al., in preparation), and \vlassE\ \citep[\vlass;][]{lacy20}, will supplement \lotss\ providing a multifrequency view of clusters that will be critical to investigate the statistics of the spectral properties of halos and relics as well as to properly disentangle diffuse sources from the emission of AGN, allowing us to determine the role of seed relativistic electrons in the ICM. \\
\indent
The presence of X-ray data is critical for the classification of diffuse sources in the ICM, and therefore the ongoing \erosita\ All-Sky Survey \citep[eRASS;][]{merloni12arx, predehl21} will be fundamental for providing the X-ray detection of many clusters that currently do not have pointed \xmm\ or \chandra\ observations (\ie,\ about half of the targets in our sample). This instrument will also allow the detection of low-mass and/or high-$z$ clusters \citep{pillepich12} that are currently not confirmed or missed by \planck\ due to its low sensitivity. Indeed, combining X-ray and SZ surveys is particularly efficient to find new galaxy clusters and derive their mass \citep[see, \eg,\ ComPRASS;][]{tarrio19}. The potential of the joint analysis of \lofar\ and \erosita\ observations of galaxy clusters has been demonstrated by recent papers \citep[\eg,][]{ghirardini21supercluster, pasini21arx, brienza21}. Until the eRASS data are publicly released, we are planning to follow-up a subsample of clusters with \chandra\ and/or \xmm. \\
\indent
Finally, we note that there are opportunities to further  improve the image quality with respect to that presented in this paper. This could be achieved by reapplying the extraction + self-calibration step and fine-tuning the parameters of the self-calibration \citep{vanweeren21}. The subtraction of discrete sources can also be more careful than that obtained during the analysis of this sample. Moreover, the addition of \lofar\ international baselines will help to improve the calibration of the targets affected by the sidelobes of a central bright AGN, enabling the search for diffuse emission in regions that cannot be thoroughly examined with the current calibration. In this respect, we note that a number of targets presented in this sample will be the subject of focused studies in the future (see Appendix~\ref{app:notes}).

\section{Conclusions}\label{sec:conclusions}

We have presented the largest statistical sample of galaxy clusters used to date to search for and study diffuse synchrotron sources in the ICM. We examined the 120$-$168 MHz radio emission from 309 galaxy clusters selected from PSZ2 that span a redshift and mass range of $0.016 < z < 0.9$ and $1.1 \times 10^{14}$ \msun\ $ < \mfive < 11.7 \times 10^{14}$ \msun, respectively, and have been covered by \lotss-DR2. We produced radio images with different resolutions and with or without the discrete source subtracted as well as overlays with \panstarrs\ optical images for all the targets in our sample. When available, we also used targeted \chandra\ and \xmm\ observations to compare the radio emission with that of the X-rays and derive morphological parameters of the ICM. All these images are publicly available on the project website \footnote{\url{https://lofar-surveys.org/planck_dr2.html}}. \\
\indent
We divided the diffuse synchrotron emission into the classes of halos, relics, and uncertain sources. The physical properties of halos and relics have been collected into tables (also available on the website) and are used in Cassano et al. (in preparation), Cuciti et al. (in preparation), and Jones et al. (in preparation) to discuss their statistical properties, such as occurrence and scaling relations. Overall, we found 83 clusters that host a radio halo and 26 clusters that host one or more radio relic (including candidates), of which about half are new discoveries. These numbers correspond to a detection fraction in our sample of $30\pm11\%$ and $10\pm6\%$, respectively. Based on these results, we expect to find $251\pm92$ cluster that host a halo and $83\pm50$ clusters that host at least one relic in the PSZ2 catalog at the completion of \lotss. Other searches are being made to examine different cluster samples and try to gauge how many more halos and relics can be found in \lotss\ in non-PSZ2 clusters \citep{hoang22sub}. \\
\indent
In the future, \lotss\ will benefit from the synergy of complementary radio surveys (\eg,\ \lolss, \lodess, and \apertif), which will be fundamental for studying the spectral properties of the observed sources. The all-sky survey with \erosita\ will enable a systematic comparison of the radio and X-ray properties of the PSZ2 clusters as well as the discoveries of new galaxy clusters (especially with low mass and at high $z$).

\begin{acknowledgements}
We thank the anonymous referee for constructive comments
that helped to improve the presentation of our results.
ABot and RJvW acknowledge support from the VIDI research programme with project number 639.042.729, which is financed by the Netherlands Organisation for Scientific Research (NWO).
RC, FG, MR and GB acknowledge support from INAF mainstream project `Galaxy Clusters Science with LOFAR' 1.05.01.86.05.
VC and GDG acknowledge support from the Alexander von Humboldt Foundation. 
XZ acknowledges support from Chinese Scholarship Council (CSC) and thanks Vittorio Ghirardini for useful discussions.
AS is supported by the Women In Science Excel (WISE) programme of the NWO, and acknowledges the World Premier Research Center Initiative (WPI) and the Kavli IPMU for the continued hospitality.
RJvW and CG acknowledge support from the ERC-Stg ClusterWeb 804208. 
MB and FdG acknowledge support from the Deutsche Forschungsgemeinschaft under Germany's Excellence Strategy - EXC 2121 ``Quantum Universe'' - 390833306.
DNH and ABon acknowledge support from ERC-Stg DRANOEL 714245.
MJH acknowledges from the UK Science and Technology Facilities Council [ST/V000624/1].
AI acknowledges the Italian PRIN-Miur 2017 (PI A. Cimatti).
ABon acknowledges support from MIUR FARE grant ``SMS''. 
AD acknowledges support by the BMBF Verbundforschung under the grant 05A20STA.
LOFAR \citep{vanhaarlem13} is the LOw Frequency ARray designed and constructed by ASTRON. It has observing, data processing, and data storage facilities in several countries, which are owned by various parties (each with their own funding sources), and are collectively operated by the ILT foundation under a joint scientific policy. The ILT resources have benefitted from the following recent major funding sources: CNRS-INSU, Observatoire de Paris and Universit\'{e} d'Orl\'{e}ans, France; BMBF, MIWF-NRW, MPG, Germany; Science Foundation Ireland (SFI), Department of Business, Enterprise and Innovation (DBEI), Ireland; NWO, The Netherlands; The Science and Technology Facilities Council, UK; Ministry of Science and Higher Education, Poland; Istituto Nazionale di Astrofisica (INAF), Italy. This research made use of the Dutch national e-infrastructure with support of the SURF Cooperative (e-infra 180169) and the LOFAR e-infra group, and of the LOFAR-IT computing infrastructure supported and operated by INAF, and by the Physics Dept.~of Turin University (under the agreement with Consorzio Interuniversitario per la Fisica Spaziale) at the C3S Supercomputing Centre, Italy. The J\"{u}lich LOFAR Long Term Archive and the German LOFAR network are both coordinated and operated by the J\"{u}lich Supercomputing Centre (JSC), and computing resources on the supercomputer JUWELS at JSC were provided by the Gauss Centre for Supercomputing e.V. (grant CHTB00) through the John von Neumann Institute for Computing (NIC). This research made use of the University of Hertfordshire high-performance computing facility and the LOFAR-UK computing facility located at the University of Hertfordshire and supported by STFC [ST/P000096/1]. The scientific results reported in this article are based in part on data obtained from the \chandra\ Data Archive and on observations obtained with \xmm, an ESA science mission with instruments and contributions directly funded by ESA Member States and NASA. SRON Netherlands Institute for Space Research is supported financially by the NWO. This research made use of APLpy, an open-source plotting package for Python \citep{robitaille12}, Astropy, a community-developed core Python package for Astronomy \citep{astropy13, astropy18}, Matplotlib \citep{hunter07}, NumPy \citep{harris20}, and SciPy \citep{virtanen20}.
\end{acknowledgements}

%
%

\bibliographystyle{aa}
\bibliography{library.bib}

\begin{appendix}

\onecolumn

\section{Tables}\label{app:tables}

In this appendix we collect the tables listing: the main properties of the targets in the full PSZ2/\lotss-DR2 sample (Table~\ref{tab:sample}), the X-ray morphological parameters for the clusters observed with \chandra/\xmm\ (Table~\ref{tab:cw}), and the quantities derived for the radio halos (Table~\ref{tab:RH_sample}) and radio relics (Table~\ref{tab:RR_sample}) observed. All tables are available in FITS format on the project website \url{https://lofar-surveys.org/planck_dr2.html}.

\begin{landscape}
\fontsize{6}{5}\selectfont
\begin{longtable}{lrrrrrrrrrrrl}
   \caption{Full sample of PSZ2 clusters in \lotss-DR2.} 
   \label{tab:sample} \\
   \hline
   \hline
     \multicolumn{1}{l}{Name} &
     \multicolumn{1}{c}{RA} &
     \multicolumn{1}{c}{Dec} &
     \multicolumn{1}{c}{$z$} &
     \multicolumn{1}{c}{$\mfive$} &
     \multicolumn{1}{c}{$\mfive_{\rm err}$} &
     \multicolumn{1}{c}{$\rfive$} &
     \multicolumn{1}{c}{$\rfive_{\rm err}$} &
     \multicolumn{1}{c}{IQ} &
     \multicolumn{1}{c}{rms} &
     \multicolumn{1}{c}{Classification} &
     \multicolumn{1}{c}{X-ray} &
     \multicolumn{1}{c}{Comment} \\
      &
     \multicolumn{1}{c}{[deg]} &
     \multicolumn{1}{c}{[deg]} &
      &
     \multicolumn{1}{c}{[$\times 10^{14}$ \msun]} &
     \multicolumn{1}{c}{[$\times 10^{14}$ \msun]} &
     \multicolumn{1}{c}{[kpc]} &
     \multicolumn{1}{c}{[kpc]} &
      &
     \multicolumn{1}{c}{[\mjyb]} &
      &
      &
      \\
   \hline
   \endfirsthead
   \caption{continued.}\\
   \hline
   \hline
     \multicolumn{1}{l}{Name} &
     \multicolumn{1}{c}{RA} &
     \multicolumn{1}{c}{Dec} &
     \multicolumn{1}{c}{$z$} &
     \multicolumn{1}{c}{$\mfive$} &
     \multicolumn{1}{c}{$\mfive_{\rm err}$} &
     \multicolumn{1}{c}{$\rfive$} &
     \multicolumn{1}{c}{$\rfive_{\rm err}$} &
     \multicolumn{1}{c}{IQ} &
     \multicolumn{1}{c}{rms} &
     \multicolumn{1}{c}{Classification} &
     \multicolumn{1}{c}{X-ray} &
     \multicolumn{1}{c}{Comment} \\
      &
     \multicolumn{1}{c}{[deg]} &
     \multicolumn{1}{c}{[deg]} &
      &
     \multicolumn{1}{c}{[$\times 10^{14}$ \msun]} &
     \multicolumn{1}{c}{[$\times 10^{14}$ \msun]} &
     \multicolumn{1}{c}{[kpc]} &
     \multicolumn{1}{c}{[kpc]} &
      &
     \multicolumn{1}{c}{[\mjyb]} &
      &
      &
      \\
   \hline
   \endhead
   \hline
   \endfoot
   \hline
   \endlastfoot
   PSZ2 G023.17+86.71 & 196.4731 & 26.5262 & 0.306 & 5.03 & 0.56 & 1093 & 41 & 1 & 0.093 & RH & C/X & New RH \\
PSZ2 G031.93+78.71 & 205.4709 & 26.3684 & 0.072 & 2.72 & 0.24 & 966 & 28 & 2 & 0.099 & RH, U & C/X & U emission on the SE, outside R500 \\
PSZ2 G033.81+77.18 & 207.2221 & 26.5946 & 0.062 & 4.47 & 0.14 & 1144 & 12 & 3 & 0.102 & N/A & C/X & --- \\
PSZ2 G040.58+77.12 & 207.3622 & 28.0946 & 0.075 & 2.63 & 0.22 & 955 & 26 & 2 & 0.112 & RH & C/X & New RH \\
PSZ2 G045.13+67.78 & 217.9958 & 29.5567 & 0.219 & 4.83 & 0.45 & 1113 & 35 & 1 & 0.105 & NDE & --- & --- \\
PSZ2 G045.87+57.70 & 229.5874 & 29.4696 & 0.611 & 7.03 & 0.68 & 1085 & 35 & 1 & 0.110 & RH & X & --- \\
PSZ2 G046.88+56.48 & 231.0469 & 29.9129 & 0.115 & 5.31 & 0.23 & 1191 & 17 & 1 & 0.124 & RH & C/X & --- \\
PSZ2 G048.10+57.16 & 230.3161 & 30.6278 & 0.078 & 3.59 & 0.21 & 1058 & 21 & 1 & 0.101 & RH, RR & C/X & --- \\
PSZ2 G048.75+53.18 & 234.9669 & 30.6960 & 0.098 & 2.53 & 0.31 & 935 & 39 & 1 & 0.124 & NDE & C & --- \\
PSZ2 G049.18+65.05 & 221.1194 & 31.2334 & 0.234 & 4.73 & 0.49 & 1099 & 38 & 1 & 0.096 & NDE & C & --- \\
PSZ2 G049.32+44.37 & 245.1404 & 29.8941 & 0.097 & 3.67 & 0.26 & 1059 & 25 & 2 & 0.116 & RH & C/X & New RH \\
PSZ2 G050.46+67.54 & 218.1680 & 31.5879 & 0.131 & 2.92 & 0.34 & 971 & 38 & 1 & 0.098 & NDE & C & --- \\
PSZ2 G052.08+46.13 & 243.4439 & 32.0814 &  &  &  &  &  & 1 & 0.093 & NDE & --- & --- \\
PSZ2 G053.53+59.52 & 227.5546 & 33.4915 & 0.113 & 5.85 & 0.23 & 1231 & 16 & 1 & 0.093 & RH, U & C/X & U emissions in the outskirts \\
PSZ2 G053.80+36.49 & 254.9008 & 31.7432 &  &  &  &  &  & 1 & 0.115 & NDE & --- & --- \\
PSZ2 G054.85+54.30 & 233.8283 & 34.3446 & 0.235 & 4.39 & 0.46 & 1072 & 38 & 2 & 0.099 & NDE & --- & --- \\
PSZ2 G054.99+53.41 & 234.9018 & 34.4313 & 0.229 & 5.73 & 0.38 & 1174 & 26 & 3 & 0.122 & N/A & C/X & --- \\
PSZ2 G055.59+31.85 & 260.6150 & 32.1354 & 0.224 & 7.78 & 0.31 & 1302 & 17 & 1 & 0.106 & RH & C/X & --- \\
PSZ2 G055.80+32.90 & 259.4624 & 32.5608 & 0.105 & 2.58 & 0.31 & 939 & 37 & 1 & 0.124 & NDE & --- & --- \\
PSZ2 G056.14+28.06 & 265.0748 & 31.6030 & 0.426 & 5.53 & 0.57 & 1077 & 37 & 1 & 0.133 & NDE & --- & --- \\
PSZ2 G056.38+23.36 & 270.3845 & 30.3962 &  &  &  &  &  & 2 & 0.226 & NDE & --- & --- \\
PSZ2 G056.62+88.42 & 194.4923 & 27.7528 & 0.045 & 3.30 & 0.07 & 1039 & 7 & -1 & -1.000 & --- & --- & Toward the direction of the Coma cluster \\
PSZ2 G056.77+36.32 & 255.6663 & 34.0511 & 0.095 & 4.38 & 0.20 & 1124 & 17 & 1 & 0.123 & RH & C/X & New RH \\
PSZ2 G057.61+34.93 & 257.4672 & 34.4613 & 0.080 & 3.73 & 0.19 & 1071 & 18 & 2 & 0.140 & RR, U & C/X & U emission in the center \\
PSZ2 G057.73+51.58 & 237.1413 & 36.1034 & 0.238 & 5.59 & 0.51 & 1160 & 36 & 1 & 0.113 & NDE & --- & --- \\
PSZ2 G057.78+52.32 & 236.2149 & 36.1219 & 0.065 & 2.38 & 0.22 & 926 & 28 & 1 & 0.124 & NDE & C/X & --- \\
PSZ2 G057.80+88.00 & 194.9118 & 27.9537 & 0.023 & 7.17 & 0.09 & 1356 & 6 & -1 & -1.000 & --- & --- & Coma cluster \\
PSZ2 G057.92+27.64 & 266.0678 & 32.9986 & 0.076 & 2.68 & 0.21 & 961 & 25 & 2 & 0.118 & NDE & C/X & --- \\
PSZ2 G058.29+18.55 & 276.3360 & 30.4320 & 0.065 & 3.88 & 0.17 & 1091 & 16 & 3 & 0.555 & N/A & C/X & --- \\
PSZ2 G058.31+41.96 & 249.0508 & 36.0554 &  &  &  &  &  & 1 & 0.095 & NDE & --- & --- \\
PSZ2 G059.18+32.91 & 260.2029 & 35.3248 & 0.383 & 5.21 & 0.56 & 1074 & 39 & 1 & 0.145 & NDE & --- & --- \\
PSZ2 G059.29+44.49 & 245.9901 & 36.9729 & 0.343 & 5.76 & 0.69 & 1128 & 45 & 1 & 0.123 & NDE & --- & --- \\
PSZ2 G059.47+33.06 & 260.0786 & 35.6002 & 0.387 & 6.09 & 0.58 & 1129 & 36 & 2 & 0.170 & U & C/X & --- \\
PSZ2 G059.52+16.23 & 279.3393 & 30.6642 & 0.284 & 4.37 & 0.55 & 1052 & 45 & 3 & 2.503 & N/A & --- & Redshift updated from Streblyanska et al. (2019) \\
PSZ2 G059.76+14.59 & 281.1949 & 30.2468 & 0.303 & 4.80 & 0.51 & 1077 & 38 & 3 & 0.457 & N/A & --- & Redshift updated from Streblyanska et al. (2019) \\
PSZ2 G060.10+15.59 & 280.2784 & 30.9322 & 0.190 & 4.67 & 0.47 & 1112 & 37 & -1 & -1.000 & --- & --- & --- \\
PSZ2 G060.16+64.50 & 221.0737 & 35.9376 & 0.361 & 5.00 & 0.63 & 1068 & 45 & 1 & 0.088 & NDE & --- & --- \\
PSZ2 G060.55+27.00 & 267.5745 & 35.0761 & 0.171 & 3.48 & 0.41 & 1015 & 40 & 2 & 0.147 & NDE & C/X & --- \\
PSZ2 G061.75+88.11 & 194.7270 & 28.0231 & 0.044 & 3.40 & 0.63 & 1051 & 66 & -1 & -1.000 & --- & --- & Toward the direction of the Coma cluster \\
PSZ2 G062.94+43.69 & 247.1565 & 39.5602 & 0.030 & 2.87 & 0.12 & 997 & 14 & 3 & 0.123 & N/A & C/X & --- \\
PSZ2 G063.38+53.44 & 234.5073 & 39.4147 & 0.422 & 6.24 & 0.60 & 1123 & 36 & 1 & 0.111 & cRH & --- & New cRH \\
PSZ2 G065.28+44.53 & 246.0751 & 41.2470 & 0.182 & 3.90 & 0.48 & 1050 & 44 & 1 & 0.099 & NDE & C & --- \\
PSZ2 G065.45+78.10 & 204.8180 & 33.0102 & 0.273 & 4.07 & 0.53 & 1031 & 46 & 1 & 0.111 & NDE & --- & --- \\
PSZ2 G065.79+41.80 & 249.7168 & 41.5990 & 0.336 & 5.22 & 0.59 & 1094 & 42 & 1 & 0.094 & NDE & --- & --- \\
PSZ2 G065.89+31.59 & 263.2231 & 40.6419 & 0.133 & 2.74 & 0.36 & 949 & 42 & 1 & 0.119 & NDE & --- & --- \\
PSZ2 G066.26+20.82 & 276.8511 & 38.2587 & 0.278 & 4.13 & 0.50 & 1034 & 42 & 2 & 0.194 & NDE & --- & --- \\
PSZ2 G066.34+26.14 & 270.3078 & 39.8546 & 0.622 & 6.10 & 0.67 & 1031 & 38 & 2 & 0.110 & cRH & --- & New cRH. Redshift updated from Aguado-Barahona et al. (2019) \\
PSZ2 G066.41+27.03 & 269.2019 & 40.1327 & 0.576 & 7.70 & 0.53 & 1134 & 26 & 1 & 0.118 & RH & C/X & New RH \\
PSZ2 G066.68+68.44 & 215.4320 & 37.2821 & 0.163 & 3.79 & 0.34 & 1047 & 32 & 1 & 0.083 & NDE & C/X & --- \\
PSZ2 G066.85+22.48 & 275.0416 & 39.2985 & 0.190 & 3.14 & 0.41 & 974 & 43 & 3 & 0.354 & N/A & --- & Redshift updated from Aguado-Barahona et al. (2019) \\
PSZ2 G067.17+67.46 & 216.5032 & 37.8160 & 0.171 & 7.24 & 0.26 & 1295 & 16 & 2 & 0.192 & U & C/X & --- \\
PSZ2 G067.52+34.75 & 259.3134 & 42.4618 & 0.175 & 4.54 & 0.34 & 1107 & 28 & 2 & 0.112 & NDE & C/X & --- \\
PSZ2 G068.36+81.81 & 200.6946 & 31.6565 & 0.308 & 6.60 & 0.43 & 1195 & 26 & 2 & 0.163 & NDE & X & --- \\
PSZ2 G069.39+68.05 & 215.4044 & 38.3613 & 0.762 & 5.69 & 0.77 & 952 & 43 & 2 & 0.183 & cRH, cRR & --- & New cRH and cRR. Redshift updated from Buddendiek et al. (2015) \\
PSZ2 G070.89+49.26 & 239.1791 & 44.6533 & 0.610 & 6.46 & 0.69 & 1055 & 38 & 1 & 0.087 & NDE & X & --- \\
PSZ2 G071.21+28.86 & 268.0130 & 44.6695 & 0.366 & 6.75 & 0.45 & 1178 & 26 & 1 & 0.101 & RH, RR & X & Double RR \\
PSZ2 G071.39+59.54 & 225.3359 & 42.3464 & 0.292 & 5.87 & 0.42 & 1157 & 27 & 2 & 0.084 & RH* & C/X & New RH* \\
PSZ2 G071.63+29.78 & 266.8257 & 45.1899 & 0.157 & 4.13 & 0.29 & 1080 & 25 & 1 & 0.101 & NDE & C/X & Possible revived fossil plasma in the S \\
PSZ2 G072.62+41.46 & 250.0941 & 46.7103 & 0.228 & 11.69 & 0.26 & 1490 & 11 & 3 & 0.111 & N/A & C/X & --- \\
PSZ2 G073.31+67.52 & 215.1737 & 39.9245 & 0.609 & 6.74 & 0.59 & 1071 & 31 & 3 & 0.475 & N/A & C/X & --- \\
PSZ2 G073.97-27.82 & 328.4233 & 17.6986 & 0.233 & 9.80 & 0.28 & 1402 & 13 & 2 & 0.216 & NDE & C/X & --- \\
PSZ2 G074.37+71.11 & 210.9114 & 38.4697 & 0.485 & 5.75 & 0.58 & 1067 & 36 & 3 & 0.169 & N/A & C & --- \\
PSZ2 G075.08+19.83 & 281.6991 & 45.7759 & 0.200 & 3.45 & 0.38 & 1002 & 37 & -1 & -1.000 & --- & --- & Redshift updated from Streblyanska et al. (2019) \\
PSZ2 G076.55+60.29 & 223.0302 & 44.4554 & 0.287 & 4.13 & 0.48 & 1031 & 40 & 1 & 0.067 & RH* & C & New RH*. Redshift updated from Streblyanska et al. (2018) \\
PSZ2 G077.90-26.63 & 330.2281 & 20.9701 & 0.147 & 5.06 & 0.26 & 1160 & 20 & 2 & 0.216 & RH* & C/X & New RH* \\
PSZ2 G080.16+57.65 & 225.2369 & 47.2874 & 0.088 & 2.51 & 0.21 & 937 & 26 & 1 & 0.127 & RR & C/X & --- \\
PSZ2 G080.41-33.24 & 336.5396 & 17.3767 & 0.107 & 3.77 & 0.27 & 1066 & 26 & 1 & 0.139 & NDE & C/X & Revived fossil plasma \\
PSZ2 G080.55-24.82 & 330.8550 & 23.9105 & 0.266 & 4.28 & 0.53 & 1051 & 43 & 1 & 0.135 & NDE & --- & --- \\
PSZ2 G080.64+64.31 & 216.8313 & 44.1308 & 0.502 & 5.23 & 0.59 & 1026 & 39 & 1 & 0.133 & NDE & C & --- \\
PSZ2 G080.70+48.31 & 238.2988 & 51.1483 & 0.235 & 3.20 & 0.41 & 965 & 41 & 1 & 0.068 & cRH & --- & New cRH \\
PSZ2 G081.02+50.57 & 234.8082 & 50.6137 & 0.501 & 4.69 & 0.54 & 991 & 38 & 1 & 0.061 & RH & X & New RH \\
PSZ2 G081.72+70.15 & 210.0995 & 41.0129 & 0.250 & 4.16 & 0.44 & 1047 & 37 & 1 & 0.097 & U & C & Two uncertain sources located to the W with respect to the bulk of the X-ray emission \\
PSZ2 G083.14+66.57 & 213.4445 & 43.6523 & 0.089 & 2.07 & 0.26 & 878 & 36 & 2 & 0.077 & NDE & --- & --- \\
PSZ2 G083.29-31.03 & 337.1296 & 20.6221 & 0.412 & 8.27 & 0.44 & 1239 & 22 & 1 & 0.118 & RH & C/X & New RH \\
PSZ2 G083.86+85.09 & 196.4623 & 30.8931 & 0.183 & 4.66 & 0.33 & 1114 & 27 & 1 & 0.079 & NDE & C/X & --- \\
PSZ2 G084.10+58.72 & 222.3158 & 48.5589 & 0.731 & 5.40 & 0.62 & 947 & 36 & 1 & 0.073 & RH & C/X & --- \\
PSZ2 G084.13-35.41 & 340.4853 & 17.5233 & 0.314 & 5.50 & 0.58 & 1122 & 40 & 2 & 0.127 & RH, U & X & New RH. U emission to the S of the RH \\
PSZ2 G084.69+42.28 & 246.7659 & 55.4802 & 0.130 & 2.70 & 0.26 & 946 & 31 & 1 & 0.068 & NDE & X & --- \\
PSZ2 G085.23+39.42 & 251.6292 & 56.4674 & 0.257 & 3.88 & 0.39 & 1021 & 35 & 1 & 0.083 & cRH & --- & New cRH \\
PSZ2 G086.28+74.76 & 204.4773 & 38.8708 & 0.699 & 5.61 & 0.71 & 972 & 41 & 3 & 0.091 & N/A & --- & Redshift updated from Streblyanska et al. (2018) \\
PSZ2 G086.43-24.95 & 335.5578 & 27.1430 & 0.231 & 3.81 & 0.50 & 1024 & 46 & 1 & 0.127 & NDE & --- & --- \\
PSZ2 G086.54-26.67 & 336.7802 & 25.8145 & 0.165 & 3.45 & 0.39 & 1015 & 39 & 1 & 0.120 & NDE & C & --- \\
PSZ2 G086.58+73.11 & 205.8750 & 40.1328 & 0.222 & 3.83 & 0.50 & 1029 & 45 & 1 & 0.073 & cRR, U & --- & New cRR. U emission in the center \\
PSZ2 G086.93+53.18 & 228.4787 & 52.7984 & 0.675 & 5.45 & 0.51 & 971 & 30 & 1 & 0.101 & RH & C/X & --- \\
PSZ2 G087.39-34.58 & 342.3068 & 19.7139 & 0.760 & 6.06 & 0.63 & 973 & 34 & 2 & 0.184 & NDE & --- & Redshift updated from Streblyanska et al. (2018) \\
PSZ2 G087.39+50.92 & 231.5690 & 54.1431 & 0.748 & 5.16 & 0.57 & 927 & 34 & 3 & 0.098 & N/A & X & --- \\
PSZ2 G087.44-21.56 & 334.0988 & 30.4251 & 0.258 & 4.15 & 0.51 & 1043 & 43 & 1 & 0.120 & NDE & --- & --- \\
PSZ2 G088.53+41.18 & 247.3887 & 58.5338 & 0.133 & 2.56 & 0.34 & 929 & 42 & 1 & 0.076 & NDE & --- & Complex radio emission associated to the brightest optical galaxy \\
PSZ2 G088.98+55.07 & 224.7806 & 52.8020 & 0.702 & 4.92 & 0.62 & 929 & 39 & 3 & 0.105 & N/A & C/X & --- \\
PSZ2 G089.39+69.36 & 208.4259 & 43.4835 & 0.680 & 5.75 & 0.72 & 987 & 41 & 1 & 0.067 & cRH & --- & --- \\
PSZ2 G089.52+62.34 & 215.5299 & 48.4769 & 0.070 & 1.83 & 0.19 & 847 & 30 & 2 & 0.078 & RR & C & Two RR located in the N \\
PSZ2 G091.27-38.62 & 347.3346 & 17.8953 & 0.105 & 3.14 & 0.39 & 1003 & 42 & 2 & 0.146 & NDE & --- & Redshift updated from Barrena et al. (2018) \\
PSZ2 G091.79-27.00 & 341.3603 & 28.1444 & 0.362 & 6.31 & 0.48 & 1154 & 29 & 1 & 0.104 & RR & X & New RR \\
PSZ2 G092.11-33.73 & 345.4059 & 22.4833 &  &  &  &  &  & 2 & 0.161 & NDE & --- & --- \\
PSZ2 G092.46-35.22 & 346.4589 & 21.3244 &  &  &  &  &  & 1 & 0.141 & NDE & --- & --- \\
PSZ2 G092.69+59.92 & 216.6347 & 51.2521 & 0.462 & 4.79 & 0.60 & 1013 & 43 & 1 & 0.078 & NDE & C/X & Redshift updated from Aguado-Barahona et al. (2019) \\
PSZ2 G092.71+73.46 & 203.8014 & 41.0004 & 0.228 & 8.13 & 0.27 & 1320 & 14 & 3 & 0.075 & N/A & C/X & --- \\
PSZ2 G093.04-32.38 & 345.4276 & 24.0403 & 0.512 & 6.34 & 0.72 & 1090 & 42 & 2 & 0.200 & NDE & --- & Redshift updated from Barrena et al. (2018) \\
PSZ2 G093.71-30.90 & 345.1839 & 25.6114 &  &  &  &  &  & 3 & 0.350 & N/A & --- & --- \\
PSZ2 G093.94-38.82 & 349.3656 & 18.7010 & 0.042 & 2.22 & 0.17 & 912 & 23 & 1 & 0.192 & U & C/X & --- \\
PSZ2 G094.44+36.13 & 255.2898 & 64.2391 & 0.453 & 4.09 & 0.43 & 964 & 34 & 1 & 0.078 & NDE & C/X & --- \\
PSZ2 G094.56+51.03 & 227.0614 & 57.8918 & 0.539 & 5.87 & 0.44 & 1051 & 26 & 3 & 0.086 & N/A & X & --- \\
PSZ2 G094.61-41.24 & 350.9896 & 16.7655 & 0.042 & 1.76 & 0.23 & 845 & 37 & 2 & 0.128 & U & C/X & --- \\
PSZ2 G095.00-37.14 & 349.3564 & 20.5736 &  &  &  &  &  & 1 & 0.144 & NDE & --- & --- \\
PSZ2 G095.22+67.41 & 207.8526 & 46.3557 & 0.062 & 1.50 & 0.21 & 795 & 38 & 1 & 0.084 & U & X & --- \\
PSZ2 G095.29+44.13 & 238.1346 & 62.0545 & 0.331 & 4.03 & 0.35 & 1006 & 29 & 2 & 0.103 & cRH* & --- & New cRH* \\
PSZ2 G096.14+56.24 & 218.8388 & 55.1414 & 0.140 & 2.77 & 0.25 & 951 & 28 & 2 & 0.086 & NDE & --- & --- \\
PSZ2 G096.43-20.89 & 342.0362 & 35.5560 & 0.350 & 5.39 & 0.51 & 1100 & 34 & 2 & 0.122 & cRR & --- & New cRR. Redshift updated from Boada et al. (2019) \\
PSZ2 G096.83+52.49 & 223.2488 & 58.0366 & 0.318 & 4.92 & 0.37 & 1080 & 27 & 1 & 0.073 & RH & C & --- \\
PSZ2 G097.15+39.20 & 246.9034 & 65.3963 & 0.206 & 2.94 & 0.32 & 948 & 34 & 1 & 0.088 & NDE & --- & --- \\
PSZ2 G097.52+51.70 & 223.8162 & 58.8752 & 0.700 & 5.18 & 0.51 & 946 & 31 & 2 & 0.105 & U & X & --- \\
PSZ2 G097.72+38.12 & 248.9776 & 66.2023 & 0.171 & 6.59 & 0.16 & 1255 & 10 & 1 & 0.076 & RH & C/X & --- \\
PSZ2 G097.78+46.95 & 231.0525 & 62.0227 & 0.338 & 3.43 & 0.39 & 950 & 36 & 2 & 0.104 & NDE & --- & --- \\
PSZ2 G098.30-41.15 & 353.5972 & 18.0003 & 0.436 & 7.05 & 0.61 & 1164 & 34 & 1 & 0.100 & cRH & --- & New cRH \\
PSZ2 G098.38+77.22 & 199.6059 & 38.5853 & 0.780 & 6.62 & 0.71 & 994 & 36 & 1 & 0.102 & NDE & --- & Redshift updated from Boada et al. (2019) \\
PSZ2 G098.39+57.68 & 215.5698 & 54.9011 &  &  &  &  &  & 1 & 0.091 & NDE & --- & --- \\
PSZ2 G098.44+56.59 & 216.7791 & 55.7497 & 0.132 & 2.83 & 0.27 & 960 & 30 & 1 & 0.090 & NDE & --- & --- \\
PSZ2 G098.62+51.76 & 222.8271 & 59.3306 & 0.298 & 3.35 & 0.48 & 958 & 47 & 2 & 0.075 & NDE & --- & Redshift updated from Streblyanska et al. (2018) \\
PSZ2 G098.75-28.63 & 348.4410 & 29.6117 &  &  &  &  &  & 2 & 0.113 & NDE & --- & --- \\
PSZ2 G099.24+42.54 & 238.0378 & 65.3405 & 0.223 & 2.80 & 0.39 & 927 & 43 & 1 & 0.104 & U & --- & U emissions in the center and outskirts \\
PSZ2 G099.48+37.72 & 248.6714 & 67.6423 & 0.167 & 2.51 & 0.28 & 911 & 34 & 1 & 0.104 & U & --- & --- \\
PSZ2 G099.48+55.60 & 217.1485 & 56.8766 & 0.105 & 2.81 & 0.21 & 967 & 25 & 1 & 0.098 & RR & C/X & New RR. Double RR \\
PSZ2 G099.55+34.23 & 257.6109 & 68.7321 & 0.310 & 3.61 & 0.38 & 977 & 34 & 1 & 0.112 & cRH*, U & --- & New cRH*. U emission in the W. Redshift updated from Aguado-Barahona et al. (2019) \\
PSZ2 G099.86+58.45 & 213.6782 & 54.7836 & 0.630 & 6.85 & 0.49 & 1067 & 25 & 2 & 0.075 & RH & C/X & --- \\
PSZ2 G100.14+41.67 & 239.0257 & 66.3545 & 0.234 & 4.04 & 0.28 & 1044 & 24 & 3 & 0.315 & N/A & C & --- \\
PSZ2 G100.22-29.64 & 350.2543 & 29.2136 & 0.485 & 7.64 & 0.59 & 1173 & 30 & 2 & 0.135 & NDE & --- & --- \\
PSZ2 G100.22+33.81 & 258.4193 & 69.3734 & 0.598 & 4.61 & 0.47 & 947 & 33 & 2 & 0.169 & NDE & --- & Redshift updated from Streblyanska et al. (2019) \\
PSZ2 G100.45-38.42 & 354.1311 & 21.1519 & 0.057 & 1.84 & 0.27 & 853 & 42 & 1 & 0.143 & NDE & C/X & Revived fossil plasma \\
PSZ2 G100.96-24.07 & 348.3483 & 34.5796 & 0.424 & 5.70 & 0.73 & 1089 & 47 & 2 & 0.119 & cRH & --- & --- \\
PSZ2 G101.52-29.98 & 351.5963 & 29.3258 & 0.227 & 4.88 & 0.52 & 1114 & 39 & 1 & 0.114 & NDE & --- & --- \\
PSZ2 G102.90-31.04 & 353.3020 & 28.7678 & 0.592 & 6.73 & 0.72 & 1078 & 39 & 2 & 0.111 & NDE & --- & --- \\
PSZ2 G103.40-32.99 & 354.5043 & 27.0818 & 0.031 & 1.51 & 0.13 & 805 & 23 & 3 & 0.164 & N/A & X & --- \\
PSZ2 G104.15-38.85 & 357.2018 & 21.7235 &  &  &  &  &  & 1 & 0.095 & NDE & --- & --- \\
PSZ2 G105.55+77.21 & 197.7544 & 39.2463 & 0.072 & 2.20 & 0.22 & 900 & 30 & 1 & 0.140 & U & X & --- \\
PSZ2 G105.76+54.73 & 212.5897 & 59.6801 & 0.316 & 4.41 & 0.45 & 1042 & 36 & 1 & 0.149 & NDE & --- & --- \\
PSZ2 G105.82-38.36 & 358.3931 & 22.5789 &  &  &  &  &  & 1 & 0.098 & NDE & --- & --- \\
PSZ2 G106.41+50.82 & 216.3149 & 63.1860 & 0.139 & 2.89 & 0.25 & 964 & 27 & 1 & 0.073 & U & C/X & --- \\
PSZ2 G106.61+66.71 & 202.6355 & 49.1529 & 0.331 & 4.67 & 0.56 & 1056 & 42 & 1 & 0.055 & RH & C & --- \\
PSZ2 G107.10+65.32 & 203.1777 & 50.5183 & 0.280 & 8.22 & 0.28 & 1300 & 15 & 1 & 0.075 & RH, RR & C/X & Double RH. Radio bridge. RR in A1758S. Revived fossil plasma \\
PSZ2 G107.11-39.50 & 359.7704 & 21.7518 & 0.533 & 6.93 & 0.67 & 1114 & 36 & 1 & 0.102 & U & --- & --- \\
PSZ2 G107.39-31.48 & 357.6390 & 29.5399 & 0.150 & 3.45 & 0.45 & 1019 & 45 & 1 & 0.104 & NDE & --- & --- \\
PSZ2 G107.67-39.78 & 0.3019 & 21.5932 & 0.415 & 5.73 & 0.75 & 1095 & 48 & 2 & 0.101 & U & --- & --- \\
PSZ2 G108.24+58.11 & 207.2301 & 57.3469 & 0.322 & 3.85 & 0.53 & 994 & 46 & 1 & 0.183 & U & --- & --- \\
PSZ2 G108.27+48.66 & 216.7773 & 65.6528 & 0.674 & 4.93 & 0.49 & 940 & 32 & 1 & 0.134 & cRH & --- & New cRH \\
PSZ2 G109.14-28.02 & 358.2865 & 33.2800 & 0.457 & 6.92 & 0.62 & 1147 & 34 & 3 & 0.096 & N/A & --- & --- \\
PSZ2 G109.22-44.01 & 2.5556 & 17.7559 & 0.173 & 3.26 & 0.48 & 993 & 49 & 2 & 0.190 & cRR & --- & New cRR \\
PSZ2 G109.97+52.84 & 209.9356 & 62.5279 & 0.326 & 4.81 & 0.38 & 1069 & 28 & 1 & 0.067 & RH & C & New RH \\
PSZ2 G111.75+70.37 & 198.2642 & 46.2812 & 0.183 & 4.34 & 0.33 & 1088 & 27 & 1 & 0.076 & RH, RR & C/X & RR in the NE \\
PSZ2 G112.07-39.86 & 3.8730 & 22.2454 &  &  &  &  &  & 2 & 0.135 & NDE & --- & --- \\
PSZ2 G112.35-32.86 & 2.6884 & 29.1673 & 0.348 & 5.50 & 0.68 & 1108 & 46 & 2 & 0.166 & NDE & X & --- \\
PSZ2 G112.48+56.99 & 203.9932 & 59.2210 & 0.070 & 2.99 & 0.15 & 998 & 17 & 1 & 0.058 & RH & C & New RH \\
PSZ2 G112.54+59.53 & 202.4759 & 56.8123 & 0.830 & 5.76 & 0.66 & 930 & 36 & 1 & 0.071 & NDE & --- & Redshift updated from Zohren et al. (2019) \\
PSZ2 G113.27+48.39 & 209.7444 & 67.4431 &  &  &  &  &  & 1 & 0.072 & NDE & --- & --- \\
PSZ2 G113.29-29.69 & 2.9363 & 32.4325 & 0.107 & 3.71 & 0.27 & 1060 & 25 & 2 & 0.114 & U & C/X & --- \\
PSZ2 G113.91-37.01 & 4.9004 & 25.2965 & 0.371 & 7.58 & 0.55 & 1223 & 30 & 1 & 0.122 & RH, RR & C/X & New RH and RR. Double RR \\
PSZ2 G114.14+58.96 & 201.3065 & 57.5834 & 0.116 & 2.17 & 0.30 & 884 & 41 & 2 & 0.069 & NDE & --- & Possible revived fossil plasma \\
PSZ2 G114.31+64.89 & 198.7658 & 51.8329 & 0.284 & 6.76 & 0.37 & 1216 & 22 & 1 & 0.071 & RH & C/X & --- \\
PSZ2 G114.39-20.98 & 2.2522 & 41.1782 & 0.154 & 3.81 & 0.35 & 1053 & 32 & 1 & 0.187 & NDE & --- & --- \\
PSZ2 G114.79-33.71 & 5.1456 & 28.6657 & 0.094 & 3.82 & 0.22 & 1075 & 21 & 2 & 0.133 & NDE & C/X & --- \\
PSZ2 G114.83+57.25 & 201.4457 & 59.3305 & 0.170 & 3.27 & 0.30 & 995 & 31 & 1 & 0.076 & NDE & --- & --- \\
PSZ2 G114.90-34.35 & 5.3526 & 28.0504 & 0.095 & 2.47 & 0.32 & 929 & 41 & 2 & 0.140 & U & --- & --- \\
PSZ2 G114.99+70.36 & 196.7286 & 46.5257 & 0.226 & 5.70 & 0.35 & 1174 & 24 & 2 & 0.086 & U & C & Complex radio emission \\
PSZ2 G115.58-44.56 & 7.3620 & 17.9947 & 0.170 & 4.32 & 0.40 & 1091 & 34 & 1 & 0.100 & NDE & --- & --- \\
PSZ2 G115.67-27.57 & 5.0172 & 34.8529 & 0.457 & 5.31 & 0.75 & 1050 & 50 & 1 & 0.100 & cRH & --- & --- \\
PSZ2 G116.32-36.33 & 6.9294 & 26.2373 & 0.367 & 5.31 & 0.63 & 1087 & 43 & 3 & 0.128 & N/A & C/X & --- \\
PSZ2 G116.50-44.47 & 8.0345 & 18.1536 & 0.396 & 7.61 & 0.60 & 1213 & 32 & 1 & 0.111 & RH*, RR & X & New RH* and RR \\
PSZ2 G118.34+68.79 & 195.3484 & 48.2419 & 0.255 & 3.77 & 0.49 & 1012 & 44 & 1 & 0.071 & cRH* & --- & --- \\
PSZ2 G118.49+48.17 & 201.0202 & 68.6690 & 0.320 & 3.68 & 0.41 & 980 & 37 & 1 & 0.077 & NDE & --- & Redshift updated from Aguado-Barahona et al. (2019) \\
PSZ2 G118.92+52.38 & 198.5514 & 64.5643 & 0.220 & 4.39 & 0.29 & 1078 & 24 & 1 & 0.059 & U & --- & --- \\
PSZ2 G119.92+59.12 & 195.7642 & 57.9334 & 0.196 & 3.44 & 0.33 & 1002 & 32 & 1 & 0.072 & U & --- & --- \\
PSZ2 G120.08-44.41 & 10.7148 & 18.4071 & 0.267 & 4.65 & 0.64 & 1080 & 49 & 2 & 0.218 & NDE & --- & --- \\
PSZ2 G121.03+57.02 & 194.9266 & 60.0741 & 0.344 & 5.69 & 0.43 & 1123 & 28 & 1 & 0.089 & RR, U & C & New RR. U emission in the center \\
PSZ2 G121.13+49.64 & 195.8869 & 67.4415 & 0.221 & 3.17 & 0.36 & 967 & 37 & 1 & 0.091 & NDE & X & --- \\
PSZ2 G121.77+51.75 & 194.5723 & 65.3600 & 0.233 & 3.98 & 0.41 & 1038 & 35 & 1 & 0.056 & NDE & --- & --- \\
PSZ2 G121.87-45.97 & 12.0892 & 16.8876 &  &  &  &  &  & 2 & 0.237 & NDE & --- & --- \\
PSZ2 G122.30+54.52 & 193.6463 & 62.5963 & 0.318 & 4.53 & 0.45 & 1050 & 35 & 1 & 0.094 & NDE & --- & --- \\
PSZ2 G122.47-38.41 & 12.4689 & 24.4527 & 0.082 & 2.52 & 0.31 & 939 & 38 & 2 & 0.115 & U & --- & --- \\
PSZ2 G122.89-36.82 & 12.8246 & 26.0510 & 0.346 & 5.57 & 0.73 & 1114 & 49 & 1 & 0.083 & NDE & --- & --- \\
PSZ2 G123.00-35.52 & 12.9231 & 27.3461 & 0.380 & 6.47 & 0.65 & 1156 & 39 & 2 & 0.077 & RH* & X & New RH* \\
PSZ2 G123.66+67.25 & 192.4223 & 49.8718 & 0.284 & 4.38 & 0.51 & 1053 & 41 & 1 & 0.064 & NDE & C & --- \\
PSZ2 G124.20-36.48 & 14.0018 & 26.3784 & 0.197 & 7.65 & 0.36 & 1308 & 20 & 3 & 0.113 & N/A & C/X & --- \\
PSZ2 G125.30-27.99 & 15.4081 & 34.8346 & 0.223 & 5.02 & 0.55 & 1126 & 41 & 1 & 0.111 & U & --- & --- \\
PSZ2 G125.71+53.86 & 189.2251 & 63.1843 & 0.302 & 6.55 & 0.33 & 1195 & 20 & 3 & 0.072 & N/A & C/X & --- \\
PSZ2 G126.20-33.17 & 16.0136 & 29.6172 & 0.358 & 5.45 & 0.74 & 1101 & 50 & 1 & 0.123 & NDE & --- & --- \\
PSZ2 G126.28+65.62 & 190.6409 & 51.4447 & 0.820 & 5.54 & 0.66 & 922 & 37 & 1 & 0.055 & U & --- & Redshift updated from Burenin et al. (2018) \\
PSZ2 G126.57+51.61 & 187.4359 & 65.3561 & 0.815 & 6.47 & 0.61 & 973 & 31 & 1 & 0.066 & NDE & C & Redshift updated from Burenin et al. (2018) \\
PSZ2 G126.61-37.63 & 16.0784 & 25.1450 & 0.166 & 3.84 & 0.50 & 1050 & 46 & 3 & 0.169 & N/A & X & --- \\
PSZ2 G126.72-21.03 & 17.6030 & 41.6924 & 0.220 & 4.11 & 0.57 & 1055 & 49 & 1 & 0.110 & NDE & --- & Redshift updated from Burenin et al. (2018) \\
PSZ2 G127.44-34.74 & 17.0569 & 27.9786 & 0.249 & 4.90 & 0.71 & 1106 & 54 & 2 & 0.131 & NDE & --- & --- \\
PSZ2 G127.50-30.52 & 17.5164 & 32.1829 & 0.353 & 5.29 & 0.67 & 1092 & 46 & 1 & 0.107 & NDE & X & --- \\
PSZ2 G128.15-24.71 & 18.8768 & 37.9135 & 0.263 & 4.48 & 0.60 & 1068 & 48 & 1 & 0.069 & NDE & --- & Redshift updated from Streblyanska et al. (2019) \\
PSZ2 G129.99-22.42 & 21.3891 & 39.9803 &  &  &  &  &  & 1 & 0.081 & NDE & --- & --- \\
PSZ2 G130.25-26.50 & 20.9524 & 35.9025 & 0.216 & 4.49 & 0.50 & 1088 & 40 & 2 & 0.102 & NDE & --- & --- \\
PSZ2 G131.27-25.82 & 22.2049 & 36.4447 &  &  &  &  &  & 1 & 0.141 & NDE & --- & --- \\
PSZ2 G132.54-42.16 & 20.4346 & 20.1398 & 0.194 & 3.99 & 0.51 & 1054 & 45 & 1 & 0.152 & NDE & X & --- \\
PSZ2 G133.59+50.68 & 176.7040 & 65.0891 & 0.529 & 5.16 & 0.57 & 1011 & 37 & 1 & 0.076 & RH* & X & New RH* \\
PSZ2 G133.60+69.04 & 187.2235 & 47.6123 & 0.254 & 5.88 & 0.40 & 1173 & 27 & 1 & 0.069 & RH & C & Possible revived fossil plasma \\
PSZ2 G133.92-42.73 & 21.4010 & 19.4054 & 0.636 & 7.24 & 0.96 & 1085 & 48 & 1 & 0.116 & NDE & --- & Redshift updated from Burenin (2017) \\
PSZ2 G134.26-44.28 & 21.3587 & 17.8353 &  &  &  &  &  & 1 & 0.152 & NDE & --- & --- \\
PSZ2 G134.59+53.38 & 177.7969 & 62.3578 & 0.345 & 4.45 & 0.54 & 1034 & 42 & 2 & 0.080 & NDE & --- & --- \\
PSZ2 G134.70+48.91 & 173.3085 & 66.3869 & 0.116 & 3.40 & 0.22 & 1027 & 22 & 2 & 0.107 & NDE & C/X & --- \\
PSZ2 G135.06+54.39 & 178.0903 & 61.3191 & 0.317 & 5.41 & 0.42 & 1115 & 29 & 1 & 0.065 & NDE & --- & --- \\
PSZ2 G135.17+65.43 & 184.8205 & 50.9217 & 0.544 & 6.01 & 0.60 & 1057 & 35 & 2 & 0.071 & RH, RR & C & --- \\
PSZ2 G135.19+57.88 & 180.5387 & 58.0593 & 0.103 & 2.21 & 0.23 & 893 & 31 & 3 & 0.133 & N/A & C & --- \\
PSZ2 G136.31+54.67 & 176.9598 & 60.7656 & 0.477 & 5.98 & 0.51 & 1084 & 31 & 1 & 0.069 & NDE & --- & Redshift updated from Streblyanska et al. (2019) \\
PSZ2 G136.33-44.53 & 22.8257 & 17.3003 &  &  &  &  &  & 1 & 0.197 & NDE & --- & --- \\
PSZ2 G136.64-25.03 & 28.2988 & 36.1888 & 0.016 & 1.12 & 0.11 & 732 & 24 & 2 & 0.125 & NDE & C/X & --- \\
PSZ2 G136.92+59.46 & 180.0852 & 56.2524 & 0.065 & 1.81 & 0.16 & 846 & 25 & 2 & 0.103 & U & X & --- \\
PSZ2 G137.24+53.93 & 175.2772 & 61.1942 & 0.470 & 7.00 & 0.48 & 1146 & 26 & 1 & 0.063 & NDE & --- & Redshift updated from Boada et al. (2019) \\
PSZ2 G137.58+53.88 & 174.8622 & 61.1503 &  &  &  &  &  & 1 & 0.084 & NDE & --- & --- \\
PSZ2 G137.74-27.08 & 28.7835 & 33.9443 & 0.087 & 2.83 & 0.28 & 975 & 32 & 2 & 0.123 & NDE & X & Possible revived fossil plasma \\
PSZ2 G138.32-39.82 & 25.5157 & 21.5330 & 0.280 & 5.98 & 0.56 & 1169 & 37 & 1 & 0.138 & RH & C & --- \\
PSZ2 G139.00+50.92 & 170.0451 & 63.2596 & 0.784 & 5.90 & 0.70 & 955 & 38 & 1 & 0.077 & NDE & --- & Redshift updated from Aguado-Barahona et al. (2019) \\
PSZ2 G139.18+56.37 & 175.5928 & 58.5227 & 0.322 & 6.87 & 0.38 & 1205 & 22 & 1 & 0.068 & RH & C/X & --- \\
PSZ2 G139.72-17.13 & 34.9806 & 42.8317 & 0.155 & 3.61 & 0.44 & 1033 & 42 & 1 & 0.153 & NDE & --- & Redshift updated from Streblyanska et al. (2019) \\
PSZ2 G141.05-32.61 & 30.0905 & 27.8208 & 0.554 & 6.92 & 0.77 & 1104 & 41 & 2 & 0.124 & cRH & --- & New cRH \\
PSZ2 G141.98+69.31 & 183.2387 & 46.3648 & 0.713 & 5.29 & 0.68 & 948 & 40 & 1 & 0.087 & NDE & --- & Redshift updated from Aguado-Barahona et al. (2019) \\
PSZ2 G143.26+65.24 & 179.8427 & 49.7742 & 0.363 & 7.65 & 0.43 & 1230 & 23 & 1 & 0.061 & RH & C/X & --- \\
PSZ2 G143.44+53.66 & 168.7489 & 59.4513 & 0.366 & 5.31 & 0.57 & 1088 & 39 & 2 & 0.067 & U & --- & Possible revived fossil plasma \\
PSZ2 G143.62+42.61 & 150.8432 & 67.1371 & 0.206 & 4.95 & 0.34 & 1127 & 26 & 1 & 0.087 & NDE & --- & --- \\
PSZ2 G144.23-18.19 & 39.7214 & 40.1886 &  &  &  &  &  & 1 & 0.096 & cRH & --- & New cRH \\
PSZ2 G144.33+62.85 & 177.3062 & 51.6086 & 0.132 & 2.66 & 0.35 & 940 & 42 & 1 & 0.063 & NDE & --- & --- \\
PSZ2 G144.84-35.16 & 32.4230 & 24.3554 &  &  &  &  &  & 1 & 0.092 & NDE & --- & --- \\
PSZ2 G144.99-24.64 & 37.1891 & 34.0270 & 0.190 & 4.25 & 0.50 & 1077 & 42 & 2 & 0.097 & cRR, U & --- & New cRR. U emission located in the center of the images \\
PSZ2 G145.25+50.84 & 163.3572 & 60.8620 &  &  &  &  &  & 1 & 0.073 & NDE & --- & --- \\
PSZ2 G145.65+59.30 & 173.1514 & 54.2102 & 0.347 & 4.73 & 0.61 & 1054 & 46 & 2 & 0.076 & U & X & --- \\
PSZ2 G146.13+40.97 & 144.7845 & 66.4371 & 0.342 & 4.70 & 0.65 & 1054 & 49 & 1 & 0.076 & NDE & --- & Redshift updated from Aguado-Barahona et al. (2019) \\
PSZ2 G146.82+40.97 & 144.1849 & 65.9752 & 0.259 & 4.49 & 0.27 & 1071 & 21 & 1 & 0.070 & NDE & --- & --- \\
PSZ2 G147.17+42.67 & 147.4525 & 64.9221 & 0.460 & 5.65 & 0.72 & 1071 & 46 & 1 & 0.078 & NDE & --- & Redshift updated from Aguado-Barahona et al. (2019) \\
PSZ2 G147.88+53.24 & 164.3924 & 58.0187 & 0.600 & 6.47 & 0.60 & 1060 & 33 & 1 & 0.065 & cRH & --- & --- \\
PSZ2 G148.36+75.23 & 184.6185 & 40.2211 & 0.304 & 4.75 & 0.50 & 1073 & 38 & 1 & 0.075 & RH & C & New RH \\
PSZ2 G149.22+54.18 & 164.6039 & 56.7926 & 0.137 & 5.87 & 0.22 & 1222 & 15 & 2 & 0.083 & RH & C & --- \\
PSZ2 G149.75+34.68 & 127.7147 & 65.8623 & 0.182 & 8.86 & 0.32 & 1381 & 17 & 1 & 0.069 & RH & C/X & --- \\
PSZ2 G150.24+48.72 & 155.8514 & 59.8085 & 0.205 & 3.56 & 0.42 & 1010 & 40 & 1 & 0.066 & NDE & --- & --- \\
PSZ2 G150.56+46.67 & 152.1830 & 60.7778 & 0.398 & 5.21 & 0.62 & 1068 & 42 & 1 & 0.061 & cRH & --- & --- \\
PSZ2 G150.56+58.32 & 168.7952 & 53.3275 & 0.470 & 7.55 & 0.51 & 1175 & 26 & 2 & 0.075 & RH & C/X & --- \\
PSZ2 G151.19+48.27 & 154.4084 & 59.5510 & 0.289 & 5.08 & 0.47 & 1103 & 34 & 2 & 0.059 & RH, RR & C/X & --- \\
PSZ2 G151.62+54.78 & 163.7128 & 55.3610 & 0.486 & 5.37 & 0.72 & 1042 & 47 & 3 & 0.155 & N/A & --- & --- \\
PSZ2 G152.33+81.28 & 187.6699 & 34.6290 & 0.333 & 5.36 & 0.55 & 1105 & 38 & 2 & 0.084 & NDE & --- & Possible revived fossil plasma \\
PSZ2 G152.40+75.00 & 183.3147 & 39.8577 & 0.453 & 5.14 & 0.70 & 1040 & 47 & 1 & 0.074 & NDE & --- & Redshift updated from Streblyanska et al. (2018) \\
PSZ2 G152.47+42.11 & 142.4647 & 61.6579 & 0.900 & 6.58 & 0.77 & 946 & 37 & 1 & 0.096 & NDE & --- & Redshift updated from Aguado-Barahona et al. (2019) \\
PSZ2 G153.29+36.56 & 130.6776 & 62.6758 & 0.650 & 6.32 & 1.27 & 1031 & 68 & 1 & 0.077 & NDE & --- & --- \\
PSZ2 G153.56+36.82 & 131.1332 & 62.4117 &  &  &  &  &  & 1 & 0.068 & NDE & --- & --- \\
PSZ2 G153.57+36.26 & 129.9422 & 62.5148 & 0.132 & 3.37 & 0.39 & 1018 & 39 & 1 & 0.077 & NDE & --- & --- \\
PSZ2 G153.68+36.96 & 131.3866 & 62.2867 &  &  &  &  &  & 1 & 0.070 & NDE & --- & --- \\
PSZ2 G153.80+33.79 & 124.5597 & 62.6820 &  &  &  &  &  & 1 & 0.070 & NDE & --- & --- \\
PSZ2 G154.13+40.19 & 137.8076 & 61.1324 & 0.290 & 5.47 & 0.45 & 1131 & 31 & 1 & 0.087 & cRH & --- & New cRH \\
PSZ2 G155.80+70.40 & 178.4833 & 42.8600 & 0.333 & 4.42 & 0.56 & 1036 & 44 & 1 & 0.071 & NDE & --- & Possible revived fossil plasma \\
PSZ2 G156.26+59.64 & 167.1024 & 50.2811 & 0.588 & 6.77 & 0.60 & 1081 & 32 & 2 & 0.096 & cRH & --- & --- \\
PSZ2 G157.63+78.02 & 184.3872 & 36.7167 & 0.377 & 5.56 & 0.57 & 1100 & 38 & 2 & 0.072 & U & --- & --- \\
PSZ2 G159.86+42.57 & 139.8791 & 56.3460 & 0.270 & 4.61 & 0.57 & 1076 & 44 & 1 & 0.080 & NDE & --- & --- \\
PSZ2 G160.83+81.66 & 186.7250 & 33.5703 & 0.888 & 5.70 & 0.66 & 906 & 35 & 1 & 0.087 & RH & C/X & --- \\
PSZ2 G160.94+44.85 & 143.3681 & 54.9523 &  &  &  &  &  & 1 & 0.082 & NDE & --- & --- \\
PSZ2 G163.61+34.30 & 124.7099 & 54.5333 & 0.596 & 5.76 & 0.74 & 1021 & 44 & 2 & 0.079 & cRH* & --- & New cRH* \\
PSZ2 G163.69+53.52 & 155.5975 & 50.1197 & 0.158 & 4.73 & 0.31 & 1130 & 24 & 2 & 0.090 & U & C & --- \\
PSZ2 G163.87+48.54 & 148.2005 & 51.8883 & 0.214 & 3.62 & 0.43 & 1013 & 41 & 1 & 0.059 & NDE & C & --- \\
PSZ2 G164.65+46.37 & 144.5898 & 52.0336 & 0.342 & 6.01 & 0.55 & 1144 & 35 & 1 & 0.070 & RH & C & New RH \\
PSZ2 G165.06+54.13 & 155.9404 & 49.1327 & 0.144 & 4.94 & 0.28 & 1152 & 22 & 1 & 0.078 & RH & C & --- \\
PSZ2 G165.46+66.15 & 170.9186 & 43.0135 & 0.196 & 3.70 & 0.42 & 1027 & 39 & 1 & 0.108 & RR & C & Double RR \\
PSZ2 G165.68+44.01 & 140.5859 & 51.8876 & 0.210 & 3.76 & 0.50 & 1027 & 46 & 1 & 0.072 & NDE & --- & --- \\
PSZ2 G165.76+31.15 & 119.5415 & 52.5986 & 0.259 & 4.43 & 0.66 & 1066 & 54 & 2 & 0.131 & NDE & --- & --- \\
PSZ2 G165.95+41.01 & 135.7227 & 52.2021 & 0.062 & 1.79 & 0.26 & 843 & 40 & 1 & 0.079 & NDE & C/X & --- \\
PSZ2 G166.09+43.38 & 139.4836 & 51.7221 & 0.217 & 6.85 & 0.32 & 1251 & 20 & 1 & 0.074 & RH & C/X & --- \\
PSZ2 G166.62+42.13 & 137.3948 & 51.5489 & 0.232 & 5.34 & 0.40 & 1145 & 28 & 1 & 0.081 & RH, RR & C & RH contaminated by RR emission. Multiple RR \\
PSZ2 G168.33+69.73 & 174.0505 & 40.1032 & 0.288 & 4.79 & 0.50 & 1083 & 37 & 1 & 0.102 & U & C & Possible revived fossil plasma \\
PSZ2 G169.62+33.84 & 124.1776 & 49.5502 & 0.347 & 5.92 & 0.63 & 1136 & 40 & 3 & 0.125 & N/A & --- & --- \\
PSZ2 G170.26+73.90 & 178.0167 & 37.2537 & 0.165 & 3.12 & 0.42 & 981 & 45 & 2 & 0.094 & U & --- & --- \\
PSZ2 G170.98+39.45 & 132.7602 & 48.5104 & 0.538 & 8.08 & 0.68 & 1170 & 33 & 2 & 0.092 & U & C & --- \\
PSZ2 G172.63+35.15 & 126.3699 & 47.1532 & 0.127 & 3.90 & 0.34 & 1071 & 31 & 1 & 0.180 & RH & C & New RH \\
PSZ2 G172.74+65.30 & 167.9029 & 40.8574 & 0.079 & 2.45 & 0.21 & 932 & 27 & 2 & 0.105 & U & C/X & Possible revived fossil plasma \\
PSZ2 G175.60+35.47 & 127.0828 & 44.7549 & 0.145 & 3.08 & 0.40 & 983 & 43 & 2 & 0.109 & U & C & --- \\
PSZ2 G176.27+37.54 & 130.0378 & 44.3618 & 0.567 & 6.14 & 0.86 & 1056 & 50 & 2 & 0.091 & RH & C & --- \\
PSZ2 G177.03+32.64 & 123.3526 & 43.2651 & 0.511 & 6.07 & 0.76 & 1075 & 45 & 1 & 0.120 & NDE & --- & Redshift updated from Aguado-Barahona et al. (2019) \\
PSZ2 G178.00+42.32 & 136.6840 & 43.1361 & 0.237 & 3.97 & 0.54 & 1036 & 47 & 1 & 0.073 & NDE & --- & --- \\
PSZ2 G178.94+56.00 & 154.9481 & 41.0098 & 0.092 & 2.23 & 0.30 & 899 & 40 & 1 & 0.074 & NDE & --- & --- \\
PSZ2 G179.09+60.12 & 160.1864 & 39.9592 & 0.137 & 3.84 & 0.33 & 1061 & 30 & 1 & 0.093 & RH & C/X & New RH \\
PSZ2 G180.60+76.65 & 179.3110 & 33.6110 & 0.214 & 6.30 & 0.34 & 1218 & 22 & 2 & 0.094 & NDE & C & --- \\
PSZ2 G180.88+31.04 & 121.9739 & 39.7755 & 0.365 & 5.99 & 0.70 & 1133 & 44 & 3 & 0.112 & N/A & X & --- \\
PSZ2 G181.06+48.47 & 144.8501 & 40.7576 & 0.240 & 4.23 & 0.49 & 1057 & 41 & 1 & 0.066 & RR & C & New RR. Double RR \\
PSZ2 G182.59+55.83 & 154.2574 & 39.0280 & 0.206 & 5.83 & 0.34 & 1191 & 23 & 3 & 0.095 & N/A & C/X & --- \\
PSZ2 G183.30+34.98 & 127.4335 & 38.4548 & 0.392 & 6.56 & 0.64 & 1156 & 38 & 2 & 0.097 & cRH, U & --- & New cRH. U emisssion located in the E \\
PSZ2 G183.90+42.99 & 137.6963 & 38.8112 & 0.561 & 6.95 & 0.74 & 1102 & 39 & 1 & 0.096 & RH & X & --- \\
PSZ2 G184.24+43.69 & 138.6019 & 38.6019 & 0.397 & 5.41 & 0.63 & 1082 & 42 & 1 & 0.077 & NDE & --- & --- \\
PSZ2 G184.68+28.91 & 120.2399 & 36.0972 & 0.288 & 5.50 & 0.52 & 1134 & 36 & 1 & 0.088 & RH & C/X & New RH \\
PSZ2 G185.08+34.02 & 126.5397 & 36.8505 & 0.365 & 5.41 & 0.66 & 1095 & 44 & 1 & 0.111 & NDE & --- & --- \\
PSZ2 G186.37+37.26 & 130.7485 & 36.3528 & 0.282 & 11.00 & 0.37 & 1431 & 16 & 1 & 0.107 & RH & C/X & --- \\
PSZ2 G186.61+62.94 & 162.6646 & 35.8429 & 0.509 & 5.68 & 0.65 & 1052 & 40 & 1 & 0.067 & cRH & --- & New cRH \\
PSZ2 G186.99+38.65 & 132.5385 & 36.0766 & 0.378 & 6.84 & 0.52 & 1178 & 30 & 2 & 0.092 & RH, RR & C & New RH and RR \\
PSZ2 G187.53+21.92 & 113.0646 & 31.6261 & 0.171 & 5.17 & 0.39 & 1158 & 29 & 2 & 0.160 & U & C/X & --- \\
PSZ2 G187.74+20.66 & 111.7531 & 31.0220 & 0.193 & 4.29 & 0.50 & 1080 & 42 & 3 & 0.218 & N/A & --- & --- \\
PSZ2 G189.23+20.55 & 112.1961 & 29.6730 & 0.398 & 5.46 & 0.74 & 1085 & 50 & 2 & 0.147 & NDE & --- & --- \\
PSZ2 G189.31+59.24 & 157.9544 & 35.0406 & 0.126 & 3.24 & 0.31 & 1007 & 32 & 1 & 0.103 & RH & C & New RH. Revived fossil plasma \\
PSZ2 G190.61+66.46 & 166.5376 & 33.5720 & 0.488 & 5.55 & 0.65 & 1053 & 41 & 1 & 0.077 & RH, RR & C & New RH and RR \\
PSZ2 G191.57+58.88 & 157.4152 & 33.9099 &  &  &  &  &  & 1 & 0.107 & NDE & --- & --- \\
PSZ2 G192.18+56.12 & 154.0911 & 33.6604 & 0.124 & 3.62 & 0.30 & 1045 & 29 & 1 & 0.089 & RH & C/X & New RH \\
PSZ2 G192.77+33.14 & 127.1718 & 30.4022 & 0.050 & 1.66 & 0.20 & 825 & 34 & 2 & 0.116 & cRH & --- & New cRH \\
PSZ2 G192.90+29.63 & 123.3039 & 29.3652 & 0.360 & 5.17 & 0.76 & 1080 & 53 & 1 & 0.119 & cRH & --- & New cRH \\
PSZ2 G193.63+54.85 & 152.5729 & 32.8307 & 0.292 & 5.19 & 0.52 & 1110 & 37 & 2 & 0.134 & U & X & --- \\
PSZ2 G194.98+54.12 & 151.7502 & 32.0212 & 0.375 & 5.66 & 0.66 & 1108 & 43 & 2 & 0.120 & U & C & --- \\
PSZ2 G195.24+29.34 & 123.6579 & 27.3366 & 0.284 & 4.94 & 0.62 & 1095 & 46 & 2 & 0.113 & U & --- & --- \\
PSZ2 G195.60+44.06 & 140.0993 & 30.5038 & 0.295 & 6.13 & 0.48 & 1172 & 31 & 2 & 0.105 & NDE & C/X & --- \\
PSZ2 G197.13+33.46 & 128.6143 & 26.9731 & 0.456 & 5.87 & 0.69 & 1086 & 43 & 1 & 0.109 & U & --- & --- \\
PSZ2 G198.46+46.01 & 142.7118 & 28.8476 & 0.299 & 5.55 & 0.54 & 1132 & 37 & 1 & 0.088 & cRR, U & --- & New cRR \\
PSZ2 G199.61+53.41 & 151.2012 & 29.2311 & 0.371 & 5.54 & 0.59 & 1101 & 39 & 1 & 0.095 & NDE & --- & --- \\
PSZ2 G199.75+46.59 & 143.5387 & 28.0764 & 0.553 & 5.91 & 0.76 & 1048 & 45 & 2 & 0.098 & NDE & --- & --- \\
PSZ2 G200.06+77.22 & 178.5581 & 29.2431 & 0.361 & 4.62 & 0.61 & 1041 & 46 & 3 & 0.130 & N/A & --- & --- \\
PSZ2 G202.66+66.98 & 166.8142 & 28.7965 & 0.483 & 5.28 & 0.70 & 1038 & 46 & 1 & 0.115 & NDE & --- & Redshift updated from Streblyanska et al. (2018) \\
PSZ2 G203.22+66.40 & 166.1721 & 28.5492 & 0.580 & 5.89 & 0.71 & 1036 & 42 & 3 & 0.130 & N/A & --- & --- \\
PSZ2 G205.90+73.76 & 174.5534 & 27.9203 & 0.447 & 7.39 & 0.55 & 1177 & 29 & 1 & 0.101 & RH, RR & C & New RH and RR. Double RR \\
 \end{longtable}
 \tablefoot{Column 1: PSZ2 name; Cols. 2 and 3: coordinates; Col. 4: redshift; Cols. 5 and 6: mass and its error; Cols. 7 and 8: radius and its error; Cols. 9 and 10: image quality (IQ; see Sect.~\ref{sec:imaging}) and rms noise of the reference \lofar\ image (-1 denotes the targets that were not extracted and self-calibrated); Col. 11: classification of the radio emission (see Sect.~\ref{sec:classification}); Col. 12: presence of archival X-ray observations (C=\chandra, X=\xmm); Col. 13: comments. Columns 1-6 are provided by the PSZ2 catalog \citep{planck16xxvii}.}
\end{landscape}
\FloatBarrier

\setlength{\LTcapwidth}{\textwidth} 
\footnotesize
 \begin{longtable}{lcrrrrr}
   \caption{Subsample of galaxy clusters with \chandra/\xmm\ data available for which we computed the X-ray morphological parameters.} 
   \label{tab:cw} \\
   \hline
   \hline
   \multicolumn{1}{l}{Name} &
   \multicolumn{1}{c}{Subcluster} &
   \multicolumn{1}{c}{$c$} &
   \multicolumn{1}{c}{$c_{\rm err}$} &
   \multicolumn{1}{c}{$w$} &
   \multicolumn{1}{c}{$w_{\rm err}$} &
   \multicolumn{1}{c}{X-ray} \\
   \hline
   \endfirsthead
   \caption{continued.}\\
   \hline
   \hline
   \multicolumn{1}{l}{Name} &
   \multicolumn{1}{c}{Subcluster} &
   \multicolumn{1}{c}{$c$} &
   \multicolumn{1}{c}{$c_{\rm err}$} &
   \multicolumn{1}{c}{$w$} &
   \multicolumn{1}{c}{$w_{\rm err}$} &
   \multicolumn{1}{c}{X-ray} \\
   \hline
   \endhead
   \hline
   \endfoot
   \hline
   \endlastfoot
   PSZ2 G023.17+86.71 &  & 1.23e-01 & 1.00e-02 & 2.17e-02 & 2.65e-03 & C+X \\
PSZ2 G031.93+78.71 &  & 2.14e-01 & 1.53e-03 & 2.83e-02 & 2.39e-04 & X \\
PSZ2 G033.81+77.18 &  & 4.26e-01 & 1.77e-03 & 8.67e-03 & 2.43e-03 & C+X \\
PSZ2 G040.58+77.12 &  & 2.27e-01 & 6.40e-03 & 6.11e-03 & 6.80e-04 & C+X \\
PSZ2 G045.87+57.70 &  & 2.54e-01 & 5.10e-03 & 2.18e-02 & 7.02e-04 & X \\
PSZ2 G046.88+56.48 &  & 8.24e-02 & 4.15e-03 & 2.34e-02 & 1.75e-03 & C+X \\
PSZ2 G048.10+57.16 &  & 8.80e-02 & 2.87e-03 & 5.91e-02 & 7.59e-03 & C+X \\
PSZ2 G048.75+53.18 &  & 3.40e-01 & 8.41e-03 & 6.49e-03 & 1.22e-03 & C \\
PSZ2 G049.18+65.05 &  & 2.87e-01 & 1.71e-02 & 7.97e-03 & 3.10e-03 & C \\
PSZ2 G049.32+44.37 &  & 1.84e-01 & 6.00e-03 & 1.09e-02 & 1.10e-03 & C+X \\
PSZ2 G050.46+67.54 &  & 3.61e-01 & 5.14e-03 & 1.86e-03 & 5.16e-04 & C \\
PSZ2 G053.53+59.52 &  & 1.39e-01 & 2.72e-03 & 1.32e-02 & 3.49e-03 & C+X \\
PSZ2 G054.99+53.41 &  & 1.44e-01 & 1.09e-02 & 1.65e-02 & 3.19e-03 & C+X \\
PSZ2 G055.59+31.85 &  & 3.00e-01 & 1.49e-02 & 5.65e-03 & 3.84e-03 & C+X \\
PSZ2 G056.77+36.32 &  & 3.03e-01 & 1.05e-02 & 3.88e-03 & 1.78e-03 & C+X \\
PSZ2 G057.61+34.93 &  & 1.08e-01 & 4.11e-03 & 1.40e-02 & 1.02e-03 & C+X \\
PSZ2 G057.78+52.32 & E & 2.26e-01 & 1.83e-03 & 6.11e-03 & 4.55e-04 & X \\
PSZ2 G057.78+52.32 & W & 2.25e-01 & 5.74e-03 & 1.61e-02 & 1.44e-03 & X \\
PSZ2 G057.92+27.64 &  & 4.52e-01 & 1.96e-02 & 5.48e-03 & 3.93e-03 & C+X \\
PSZ2 G058.29+18.55 & E & 1.25e-01 & 1.53e-02 & 3.63e-02 & 2.22e-02 & C+X \\
PSZ2 G058.29+18.55 & W & 4.22e-01 & 3.26e-03 & 1.34e-02 & 5.43e-04 & X \\
PSZ2 G059.47+33.06 &  & 3.66e-01 & 3.98e-02 & 1.36e-02 & 9.59e-04 & C+X \\
PSZ2 G060.55+27.00 &  & 4.22e-01 & 1.05e-02 & 4.26e-03 & 8.72e-04 & C+X \\
PSZ2 G062.94+43.69 &  & 4.36e-01 & 3.40e-04 & 3.00e-03 & 1.95e-05 & X \\
PSZ2 G065.28+44.53 &  & 2.11e-01 & 6.56e-03 & 3.44e-02 & 1.78e-03 & C \\
PSZ2 G066.41+27.03 &  & 8.83e-02 & 8.98e-03 & 2.75e-02 & 2.28e-02 & C+X \\
PSZ2 G066.68+68.44 &  & 3.42e-01 & 8.91e-03 & 8.70e-03 & 3.25e-03 & C+X \\
PSZ2 G067.17+67.46 &  & 2.23e-01 & 8.98e-03 & 4.29e-02 & 1.61e-03 & C+X \\
PSZ2 G067.52+34.75 &  & 3.89e-01 & 2.42e-03 & 4.10e-03 & 2.71e-04 & X \\
PSZ2 G068.36+81.81 &  & 1.37e-01 & 3.05e-03 & 2.69e-02 & 8.34e-04 & X \\
PSZ2 G070.89+49.26 &  & 1.36e-01 & 4.25e-03 & 2.06e-02 & 1.20e-03 & X \\
PSZ2 G071.21+28.86 &  & 6.38e-02 & 4.45e-03 & 1.29e-02 & 2.08e-03 & X \\
PSZ2 G071.39+59.54 &  & 1.53e-01 & 1.51e-02 & 1.65e-02 & 3.03e-03 & C+X \\
PSZ2 G071.63+29.78 &  & 8.24e-02 & 3.34e-03 & 2.45e-02 & 1.38e-02 & C+X \\
PSZ2 G072.62+41.46 &  & 1.30e-01 & 7.29e-03 & 2.68e-02 & 5.08e-03 & C+X \\
PSZ2 G073.31+67.52 &  & 1.56e-01 & 1.42e-02 & 1.58e-02 & 2.81e-03 & C+X \\
PSZ2 G073.97-27.82 &  & 2.77e-01 & 5.97e-03 & 1.11e-02 & 5.59e-04 & C+X \\
PSZ2 G074.37+71.11 &  & 1.43e-01 & 2.00e-02 & 2.82e-02 & 6.10e-03 & C \\
PSZ2 G076.55+60.29 &  & 2.38e-01 & 1.70e-02 & 2.87e-02 & 3.84e-03 & C \\
PSZ2 G077.90-26.63 &  & 2.19e-01 & 7.61e-03 & 1.80e-02 & 1.05e-03 & C+X \\
PSZ2 G080.16+57.65 &  & 1.30e-01 & 1.12e-02 & 3.27e-02 & 2.05e-03 & C+X \\
PSZ2 G080.41-33.24 &  & 1.98e-01 & 1.72e-02 & 5.85e-02 & 1.31e-02 & C+X \\
PSZ2 G080.64+64.31 &  & 4.53e-01 & 1.24e-02 & 6.22e-03 & 1.50e-03 & C \\
PSZ2 G081.02+50.57 &  & 1.49e-01 & 5.22e-03 & 3.77e-02 & 1.37e-03 & X \\
PSZ2 G081.72+70.15 &  & 1.21e-01 & 1.80e-02 & 1.83e-02 & 5.38e-03 & C \\
PSZ2 G083.29-31.03 &  & 1.77e-01 & 1.27e-02 & 2.97e-02 & 1.07e-02 & C+X \\
PSZ2 G083.86+85.09 &  & 1.89e-01 & 9.11e-03 & 3.33e-02 & 4.09e-03 & C+X \\
PSZ2 G084.10+58.72 &  & 1.77e-01 & 2.79e-02 & 2.04e-02 & 9.05e-03 & C+X \\
PSZ2 G084.13-35.41 &  & 9.53e-02 & 5.82e-03 & 3.79e-02 & 2.14e-03 & X \\
PSZ2 G084.69+42.28 &  & 2.70e-01 & 4.30e-03 & 1.29e-02 & 5.93e-04 & X \\
PSZ2 G086.54-26.67 &  & 3.04e-01 & 6.40e-03 & 5.42e-03 & 8.78e-04 & C \\
PSZ2 G086.93+53.18 &  & 1.26e-01 & 1.86e-02 & 1.86e-02 & 3.51e-03 & C+X \\
PSZ2 G087.39+50.92 &  & 2.13e-01 & 1.16e-02 & 2.34e-02 & 2.13e-03 & X \\
PSZ2 G088.98+55.07 &  & 2.94e-01 & 1.20e-01 & 6.44e-02 & 1.62e-02 & C+X \\
PSZ2 G089.52+62.34 &  & 1.13e-01 & 9.41e-03 & 3.20e-02 & 2.26e-03 & C \\
PSZ2 G091.79-27.00 &  & 7.29e-02 & 6.16e-03 & 4.54e-02 & 2.52e-03 & X \\
PSZ2 G092.69+59.92 &  & 1.29e-01 & 4.45e-02 & 6.63e-02 & 4.53e-02 & C+X \\
PSZ2 G092.71+73.46 &  & 1.54e-01 & 5.69e-03 & 1.44e-02 & 2.21e-03 & C+X \\
PSZ2 G093.94-38.82 & EN & 2.14e-01 & 2.36e-03 & 4.07e-02 & 5.69e-04 & X \\
PSZ2 G093.94-38.82 & ES & 1.93e-01 & 2.30e-03 & 3.18e-02 & 5.75e-04 & X \\
PSZ2 G093.94-38.82 & W & 3.29e-01 & 3.09e-03 & 1.68e-02 & 5.35e-04 & X \\
PSZ2 G094.44+36.13 &  & 2.83e-01 & 2.85e-02 & 1.33e-02 & 4.92e-03 & C+X \\
PSZ2 G094.56+51.03 &  & 1.02e-01 & 4.14e-03 & 5.69e-02 & 1.67e-03 & X \\
PSZ2 G094.61-41.24 &  & 3.23e-01 & 1.26e-03 & 7.82e-03 & 2.16e-04 & X \\
PSZ2 G095.22+67.41 &  & 1.25e-01 & 2.28e-03 & 2.06e-02 & 8.51e-04 & X \\
PSZ2 G096.83+52.49 &  & 2.09e-01 & 3.63e-03 & 8.69e-03 & 8.71e-04 & C \\
PSZ2 G097.52+51.70 &  & 2.17e-01 & 8.06e-03 & 1.92e-02 & 1.10e-03 & X \\
PSZ2 G097.72+38.12 &  & 1.70e-01 & 6.56e-03 & 3.21e-02 & 7.91e-03 & C+X \\
PSZ2 G099.48+55.60 &  & 8.41e-02 & 4.91e-03 & 2.55e-02 & 3.21e-03 & C+X \\
PSZ2 G099.86+58.45 &  & 1.33e-01 & 1.04e-02 & 2.15e-02 & 5.58e-03 & C+X \\
PSZ2 G100.14+41.67 &  & 2.50e-01 & 3.02e-03 & 5.67e-02 & 5.84e-04 & C \\
PSZ2 G100.45-38.42 &  & 4.11e-01 & 1.50e-03 & 2.64e-03 & 1.35e-04 & X \\
PSZ2 G103.40-32.99 &  & 1.08e-01 & 1.21e-03 & 5.21e-03 & 5.06e-04 & X \\
PSZ2 G105.55+77.21 &  & 1.82e-01 & 2.31e-03 & 2.52e-02 & 6.40e-04 & X \\
PSZ2 G106.41+50.82 &  & 3.49e-01 & 2.12e-02 & 1.90e-02 & 1.22e-03 & C+X \\
PSZ2 G106.61+66.71 &  & 1.40e-01 & 3.20e-02 & 5.08e-02 & 7.99e-03 & C \\
PSZ2 G107.10+65.32 & N & 1.07e-01 & 5.96e-03 & 8.61e-02 & 1.15e-03 & C+X \\
PSZ2 G107.10+65.32 & S & 1.42e-01 & 6.93e-03 & 3.60e-02 & 2.18e-03 & C+X \\
PSZ2 G109.97+52.84 &  & 3.34e-01 & 5.04e-03 & 8.21e-03 & 8.83e-04 & C \\
PSZ2 G111.75+70.37 &  & 9.16e-02 & 6.56e-03 & 5.71e-02 & 3.19e-03 & C+X \\
PSZ2 G112.35-32.86 &  & 2.63e-01 & 9.71e-03 & 1.35e-02 & 1.41e-03 & X \\
PSZ2 G112.48+56.99 &  & 1.74e-01 & 4.69e-03 & 4.60e-03 & 9.61e-04 & C \\
PSZ2 G113.29-29.69 &  & 1.69e-01 & 1.02e-02 & 1.33e-02 & 5.02e-03 & C+X \\
PSZ2 G113.91-37.01 &  & 1.57e-01 & 1.65e-02 & 4.60e-02 & 2.54e-03 & C+X \\
PSZ2 G114.31+64.89 &  & 1.66e-01 & 2.70e-02 & 1.28e-02 & 1.92e-03 & C+X \\
PSZ2 G114.79-33.71 &  & 1.53e-01 & 9.18e-03 & 7.39e-03 & 4.54e-03 & C+X \\
PSZ2 G114.99+70.36 &  & 1.46e-01 & 5.90e-03 & 1.72e-02 & 1.80e-03 & C \\
PSZ2 G116.32-36.33 & N & 1.50e-01 & 1.70e-02 & 1.24e-02 & 4.73e-03 & C+X \\
PSZ2 G116.32-36.33 & S & 2.97e-01 & 1.26e-02 & 9.13e-03 & 1.50e-03 & X \\
PSZ2 G116.50-44.47 &  & 1.30e-01 & 6.75e-03 & 5.60e-02 & 2.37e-03 & X \\
PSZ2 G121.03+57.02 &  & 9.77e-02 & 7.83e-03 & 1.10e-01 & 3.74e-03 & C \\
PSZ2 G121.13+49.64 &  & 9.88e-02 & 5.27e-03 & 3.28e-02 & 1.94e-03 & X \\
PSZ2 G123.00-35.52 &  & 1.56e-01 & 5.28e-03 & 2.39e-02 & 1.08e-03 & X \\
PSZ2 G123.66+67.25 &  & 2.50e-01 & 3.05e-02 & 1.59e-02 & 5.27e-03 & C \\
PSZ2 G124.20-36.48 & N & 3.06e-01 & 3.56e-03 & 5.41e-02 & 8.91e-04 & C+X \\
PSZ2 G124.20-36.48 & S & 1.01e-01 & 1.04e-02 & 1.90e-02 & 6.16e-03 & C+X \\
PSZ2 G125.71+53.86 &  & 1.96e-01 & 1.68e-02 & 1.04e-02 & 3.56e-03 & C+X \\
PSZ2 G126.61-37.63 &  & 1.70e-01 & 5.69e-03 & 8.77e-03 & 1.09e-03 & X \\
PSZ2 G127.50-30.52 &  & 1.16e-01 & 7.32e-03 & 1.39e-02 & 2.04e-03 & X \\
PSZ2 G132.54-42.16 &  & 2.11e-01 & 8.94e-03 & 2.55e-03 & 1.58e-03 & X \\
PSZ2 G133.59+50.68 &  & 9.27e-02 & 5.08e-03 & 1.94e-02 & 2.18e-03 & X \\
PSZ2 G133.60+69.04 &  & 8.67e-02 & 8.92e-03 & 3.80e-02 & 3.49e-03 & C \\
PSZ2 G134.70+48.91 &  & 2.52e-01 & 2.79e-02 & 5.34e-03 & 1.98e-03 & C+X \\
PSZ2 G135.17+65.43 &  & 1.05e-01 & 1.87e-02 & 4.72e-02 & 7.64e-03 & C \\
PSZ2 G135.19+57.88 &  & 1.66e-01 & 9.31e-03 & 1.33e-02 & 2.56e-03 & C \\
PSZ2 G136.92+59.46 &  & 9.37e-02 & 2.33e-03 & 8.87e-02 & 1.22e-03 & X \\
PSZ2 G137.74-27.08 &  & 1.46e-01 & 2.35e-03 & 4.31e-02 & 6.91e-04 & X \\
PSZ2 G138.32-39.82 &  & 1.98e-01 & 6.58e-03 & 1.32e-02 & 1.33e-03 & C \\
PSZ2 G139.18+56.37 &  & 8.60e-02 & 5.90e-03 & 4.70e-02 & 8.49e-03 & C+X \\
PSZ2 G143.26+65.24 &  & 1.42e-01 & 2.62e-02 & 2.46e-02 & 1.49e-03 & C+X \\
PSZ2 G145.65+59.30 &  & 1.44e-01 & 6.56e-03 & 1.20e-02 & 1.45e-03 & X \\
PSZ2 G148.36+75.23 &  & 2.06e-01 & 9.02e-03 & 5.27e-02 & 2.43e-03 & C \\
PSZ2 G149.22+54.18 &  & 1.36e-01 & 3.42e-03 & 3.72e-03 & 8.17e-04 & C \\
PSZ2 G149.75+34.68 &  & 1.72e-01 & 3.39e-03 & 6.13e-02 & 3.66e-03 & C+X \\
PSZ2 G150.56+58.32 &  & 1.33e-01 & 1.62e-02 & 3.16e-02 & 1.77e-02 & C+X \\
PSZ2 G151.19+48.27 &  & 7.73e-02 & 8.77e-03 & 2.41e-02 & 1.18e-02 & C+X \\
PSZ2 G160.83+81.66 &  & 2.78e-01 & 3.08e-02 & 1.72e-02 & 3.92e-03 & C+X \\
PSZ2 G163.69+53.52 &  & 1.98e-01 & 5.88e-03 & 8.31e-03 & 1.41e-03 & C \\
PSZ2 G163.87+48.54 &  & 4.61e-01 & 3.51e-03 & 1.61e-03 & 3.40e-04 & C \\
PSZ2 G164.65+46.37 &  & 2.46e-01 & 1.00e-02 & 6.05e-02 & 2.10e-03 & C \\
PSZ2 G165.06+54.13 &  & 1.88e-01 & 5.22e-03 & 1.77e-02 & 1.53e-03 & C \\
PSZ2 G165.46+66.15 &  & 6.96e-02 & 5.22e-03 & 3.31e-02 & 3.13e-03 & C \\
PSZ2 G166.09+43.38 &  & 1.84e-01 & 6.80e-03 & 1.83e-02 & 5.71e-03 & C+X \\
PSZ2 G166.62+42.13 &  & 6.85e-02 & 5.52e-03 & 3.48e-02 & 3.03e-03 & C \\
PSZ2 G168.33+69.73 &  & 2.64e-01 & 3.00e-02 & 1.91e-02 & 4.29e-03 & C \\
PSZ2 G170.98+39.45 &  & 1.14e-01 & 1.61e-02 & 2.69e-02 & 6.51e-03 & C \\
PSZ2 G172.63+35.15 &  & 1.84e-01 & 8.75e-03 & 2.01e-02 & 2.14e-03 & C \\
PSZ2 G172.74+65.30 &  & 2.18e-01 & 1.10e-02 & 2.44e-02 & 1.69e-02 & C+X \\
PSZ2 G175.60+35.47 &  & 2.66e-01 & 1.13e-02 & 1.05e-02 & 2.00e-03 & C \\
PSZ2 G176.27+37.54 &  & 2.43e-01 & 1.69e-02 & 1.90e-02 & 3.87e-03 & C \\
PSZ2 G179.09+60.12 &  & 5.15e-01 & 6.11e-03 & 6.54e-03 & 2.17e-03 & C+X \\
PSZ2 G180.60+76.65 &  & 2.89e-01 & 6.39e-03 & 2.41e-03 & 5.32e-04 & C \\
PSZ2 G180.88+31.04 &  & 1.01e-01 & 1.15e-02 & 1.83e-02 & 4.22e-03 & X \\
PSZ2 G181.06+48.47 &  & 1.41e-01 & 1.05e-02 & 6.95e-02 & 3.04e-03 & C \\
PSZ2 G182.59+55.83 &  & 2.86e-01 & 1.26e-02 & 5.95e-03 & 9.26e-04 & C+X \\
PSZ2 G183.90+42.99 &  & 1.56e-01 & 4.77e-03 & 1.82e-02 & 1.13e-03 & X \\
PSZ2 G184.68+28.91 &  & 2.93e-01 & 1.47e-02 & 7.89e-03 & 2.99e-03 & C+X \\
PSZ2 G186.37+37.26 &  & 1.47e-01 & 8.43e-03 & 9.89e-03 & 5.32e-03 & C+X \\
PSZ2 G186.99+38.65 &  & 1.99e-01 & 8.38e-03 & 3.85e-02 & 2.11e-03 & C \\
PSZ2 G187.53+21.92 &  & 3.04e-01 & 1.59e-02 & 7.20e-03 & 5.27e-03 & C+X \\
PSZ2 G189.31+59.24 &  & 2.45e-01 & 3.64e-03 & 4.76e-02 & 7.56e-04 & C \\
PSZ2 G190.61+66.46 &  & 1.05e-01 & 1.61e-02 & 2.87e-02 & 5.77e-03 & C \\
PSZ2 G192.18+56.12 &  & 1.72e-01 & 7.12e-03 & 1.70e-02 & 1.12e-02 & C+X \\
PSZ2 G193.63+54.85 &  & 1.67e-01 & 7.38e-03 & 5.62e-02 & 1.86e-03 & X \\
PSZ2 G194.98+54.12 &  & 1.84e-01 & 1.37e-02 & 6.07e-02 & 3.54e-03 & C \\
PSZ2 G195.60+44.06 & E1 & 9.39e-02 & 5.86e-03 & 1.94e-02 & 2.35e-03 & X \\
PSZ2 G195.60+44.06 & E2 & 1.23e-01 & 7.51e-03 & 4.63e-02 & 1.97e-02 & C+X \\
PSZ2 G195.60+44.06 & W1 & 2.83e-01 & 7.56e-03 & 8.57e-03 & 9.76e-04 & X \\
PSZ2 G195.60+44.06 & W2 & 9.70e-02 & 2.07e-03 & 4.79e-02 & 7.30e-04 & X \\
PSZ2 G205.90+73.76 &  & 2.12e-01 & 1.79e-02 & 1.35e-02 & 3.23e-03 & C \\
 \end{longtable}
 \tablefoot{Col. 1: PSZ2 name; Col. 2: position of the subcluster; Cols. 3 and 4: concentration parameter and its error; Cols. 5 and 6: centroid shift and its error; Col. 7: instrument used (C=\chandra, X=\xmm).}

\newpage

\begin{table}
 \centering
 \caption{Sample of radio halos and candidate radio halos found in this work.}
 \label{tab:RH_sample}
\resizebox{\textwidth}{!}{
 \begin{tabular}{lrrrrrrrrrrrrlrrrl}
   \hline
   \hline
  \multicolumn{1}{l}{Name} &
  \multicolumn{1}{c}{$S_{150}$ ($2\sigma$)} &
  \multicolumn{1}{c}{${S_{150}}_{\rm err}$ ($2\sigma$)} &
  \multicolumn{1}{c}{$S_{150}$ (fit)} &
  \multicolumn{1}{c}{${S_{150}}_{\rm err}$ (fit)} &
  \multicolumn{1}{c}{$P_{150}$} &
  \multicolumn{1}{c}{${P_{150}}_{\rm err}$} &
  \multicolumn{1}{c}{$I_0$} &
  \multicolumn{1}{c}{${I_0}_{\rm err}$} &
  \multicolumn{1}{c}{$r_1$} &
  \multicolumn{1}{c}{${r_1}_{\rm err}$} &
  \multicolumn{1}{c}{$r_2$} &
  \multicolumn{1}{c}{${r_2}_{\rm err}$} &
  \multicolumn{1}{c}{Model} &
  \multicolumn{1}{c}{S/N} &
  \multicolumn{1}{c}{rms} &
  \multicolumn{1}{c}{$\chi^2_{\rm red}$} &
  \multicolumn{1}{c}{Comment} \\
     &
  \multicolumn{1}{c}{[mJy]} &
  \multicolumn{1}{c}{[mJy]} &
  \multicolumn{1}{c}{[mJy]} &
  \multicolumn{1}{c}{[mJy]} &
  \multicolumn{1}{c}{[\whz]} &
  \multicolumn{1}{c}{[\whz]} &
  \multicolumn{1}{c}{[\mujyarcsecsq]} &
  \multicolumn{1}{c}{[\mujyarcsecsq]} &
  \multicolumn{1}{c}{[kpc]} &
  \multicolumn{1}{c}{[kpc]} &
  \multicolumn{1}{c}{[kpc]} &
  \multicolumn{1}{c}{[kpc]} &
   &
   &
  \multicolumn{1}{c}{[\mjyb]} &
   &
   \\
   \hline
   PSZ2 G023.17+86.71 & 25.73 & 3.12 & 29.84 & 3.62 & 9.76e+24 & 1.18e+24 & 5.79 & 0.34 & 232.8 & 11.7 & 95.1 & 6.4 & rotated\_ellipse & 19.81 & 0.347 & 1.18 & --- \\
PSZ2 G031.93+78.71 & 189.84 & 78.34 & 176.45 & 78.03 & 2.28e+24 & 1.01e+24 & 11.39 & 0.20 & 78.7 & 1.1 &  &  & circle & 66.26 & 0.168 & 3.27 & --- \\
PSZ2 G040.58+77.12 & 50.33 & 6.76 & 42.92 & 7.82 & 6.04e+23 & 1.10e+23 & 0.67 & 0.10 & 165.3 & 21.0 &  &  & circle & 7.01 & 0.082 & 2.46 & --- \\
PSZ2 G045.87+57.70 & 10.78 & 2.59 & 11.75 & 2.79 & 2.11e+25 & 5.03e+24 & 14.18 & 1.90 & 89.0 & 9.6 &  &  & circle & 9.66 & 0.877 & 1.11 & --- \\
PSZ2 G046.88+56.48 & 177.19 & 20.99 & 236.45 & 27.77 & 8.36e+24 & 9.81e+23 & 1.57 & 0.07 & 369.9 & 13.6 &  &  & circle & 22.46 & 0.131 & 0.70 & --- \\
PSZ2 G048.10+57.16 & 300.00 & 39.36 & 354.88 & 44.67 & 5.43e+24 & 6.84e+23 & 1.58 & 0.05 & 463.4 & 15.6 & 220.7 & 6.3 & rotated\_ellipse & 32.61 & 0.160 & 0.87 & --- \\
PSZ2 G049.32+44.37 & 45.87 & 7.71 & 70.64 & 10.61 & 1.73e+24 & 2.59e+23 & 0.61 & 0.05 & 281.2 & 19.7 &  &  & circle & 12.13 & 0.099 & 0.79 & --- \\
PSZ2 G053.53+59.52 & 519.08 & 57.38 & 387.17 & 45.84 & 1.32e+25 & 1.56e+24 & 11.66 & 0.11 & 171.5 & 1.4 &  &  & circle & 104.71 & 0.141 & 13.68 & --- \\
PSZ2 G055.59+31.85 & 30.25 & 5.21 & 33.57 & 5.46 & 5.29e+24 & 8.60e+23 & 17.48 & 0.71 & 72.3 & 2.3 &  &  & circle & 30.72 & 0.433 & 1.26 & --- \\
PSZ2 G056.77+36.32 & 79.79 & 26.53 & 83.39 & 26.97 & 1.95e+24 & 6.30e+23 & 1.10 & 0.10 & 223.3 & 15.5 &  &  & circle & 14.15 & 0.099 & 0.79 & --- \\
PSZ2 G063.38+53.44 & 27.53 & 5.02 & 29.40 & 5.19 & 2.09e+25 & 3.70e+24 & 10.76 & 0.58 & 133.0 & 5.8 &  &  & circle & 20.38 & 0.713 & 1.08 & --- \\
PSZ2 G066.34+26.14 & 98.35 & 12.89 & 82.29 & 11.71 & 1.55e+26 & 2.20e+25 & 22.02 & 0.70 & 190.5 & 5.1 &  &  & circle & 32.77 & 1.304 & 1.60 & --- \\
PSZ2 G066.41+27.03 & 89.88 & 10.47 & 111.34 & 12.61 & 1.72e+26 & 1.95e+25 & 15.50 & 0.42 & 374.5 & 11.0 & 173.6 & 5.1 & rotated\_ellipse & 36.36 & 1.096 & 1.05 & --- \\
PSZ2 G069.39+68.05 & 14.60 & 2.46 & 18.49 & 4.42 & 5.86e+25 & 1.40e+25 & 4.73 & 1.01 & 211.9 & 38.5 &  &  & circle & 4.64 & 0.826 & 0.71 & --- \\
PSZ2 G071.21+28.86 & 45.99 & 7.74 & 67.70 & 10.82 & 3.41e+25 & 5.44e+24 & 1.53 & 0.13 & 490.3 & 35.6 &  &  & circle & 11.08 & 0.260 & 1.32 & --- \\
PSZ2 G080.70+48.31 & 8.18 & 2.89 & 6.39 & 2.86 & 1.12e+24 & 5.03e+23 & 4.58 & 0.65 & 64.1 & 6.5 &  &  & circle & 8.75 & 0.324 & 0.96 & --- \\
PSZ2 G081.02+50.57 & 5.39 & 1.08 & 10.59 & 2.09 & 1.15e+25 & 2.28e+24 & 1.45 & 0.33 & 238.3 & 36.7 &  &  & circle & 6.46 & 0.175 & 0.87 & --- \\
PSZ2 G083.29-31.03 & 53.22 & 5.88 & 54.71 & 6.00 & 3.67e+25 & 4.03e+24 & 6.71 & 0.28 & 226.5 & 7.8 &  &  & circle & 26.25 & 0.341 & 1.65 & --- \\
PSZ2 G084.10+58.72 & 3.80 & 0.85 & 3.97 & 0.96 & 1.13e+25 & 2.74e+24 & 6.03 & 1.79 & 85.5 & 17.5 &  &  & circle & 6.49 & 0.320 & 1.05 & --- \\
PSZ2 G084.13-35.41 & 51.57 & 13.14 & 56.55 & 13.48 & 1.97e+25 & 4.69e+24 & 6.16 & 0.31 & 201.8 & 8.4 &  &  & circle & 20.03 & 0.565 & 1.17 & --- \\
PSZ2 G085.23+39.42 & 55.35 & 7.81 & 68.62 & 8.84 & 1.49e+25 & 1.92e+24 & 5.12 & 0.18 & 211.5 & 4.8 &  &  & circle & 40.12 & 0.262 & 1.80 & --- \\
PSZ2 G086.93+53.18 & 8.91 & 1.69 & 11.50 & 2.20 & 2.67e+25 & 5.09e+24 & 3.29 & 0.54 & 190.7 & 25.2 &  &  & circle & 7.29 & 0.486 & 1.02 & --- \\
PSZ2 G089.39+69.36 & 11.77 & 2.17 & 10.93 & 2.14 & 2.58e+25 & 5.05e+24 & 10.47 & 0.79 & 104.6 & 6.2 &  &  & circle & 16.23 & 0.532 & 0.83 & --- \\
PSZ2 G096.83+52.49 & 39.63 & 7.45 & 45.21 & 7.80 & 1.62e+25 & 2.80e+24 & 7.92 & 0.23 & 211.0 & 5.9 & 122.2 & 3.9 & rotated\_ellipse & 36.70 & 0.417 & 1.53 & --- \\
PSZ2 G097.72+38.12 & 152.51 & 16.07 & 140.24 & 14.94 & 1.19e+25 & 1.27e+24 & 7.44 & 0.14 & 183.2 & 2.9 &  &  & circle & 55.12 & 0.276 & 1.26 & --- \\
PSZ2 G098.30-41.15 & 46.52 & 7.17 & 42.00 & 6.87 & 3.24e+25 & 5.31e+24 & 40.22 & 1.10 & 83.8 & 1.7 &  &  & circle & 44.54 & 0.746 & 2.24 & --- \\
PSZ2 G099.86+58.45 & 21.92 & 4.82 & 19.92 & 4.74 & 3.87e+25 & 9.21e+24 & 7.36 & 0.49 & 163.2 & 8.6 &  &  & circle & 15.69 & 0.687 & 1.10 & --- \\
PSZ2 G100.96-24.07 & 17.18 & 7.03 & 18.45 & 7.24 & 1.33e+25 & 5.21e+24 & 5.82 & 0.75 & 143.7 & 14.6 &  &  & circle & 8.78 & 0.544 & 0.95 & --- \\
PSZ2 G106.61+66.71 & 19.45 & 2.41 & 17.91 & 2.27 & 7.07e+24 & 8.98e+23 & 12.82 & 0.52 & 81.7 & 2.5 &  &  & circle & 32.54 & 0.257 & 1.97 & --- \\
PSZ2 G107.10+65.32 & 270.18 & 36.21 & 154.77 & 29.94 & 7.16e+25 & 9.60e+24 & 22.66 & 1.44 & 251.1 & 16.0 & 102.9 & 7.0 & rotated\_ellipse & 17.21 & 0.253 & 2.87 & $P_{150}$ derived from $S_{150}$ ($2\sigma$) \\
PSZ2 G108.27+48.66 & 10.26 & 3.30 & 9.39 & 3.23 & 2.17e+25 & 7.47e+24 & 6.42 & 0.84 & 123.3 & 12.0 &  &  & circle & 9.63 & 0.563 & 0.94 & --- \\
PSZ2 G109.97+52.84 & 13.61 & 1.91 & 16.25 & 2.58 & 6.18e+24 & 9.80e+23 & 1.39 & 0.16 & 233.5 & 25.0 &  &  & circle & 9.07 & 0.127 & 1.95 & --- \\
PSZ2 G111.75+70.37 & 28.12 & 20.02 & 29.28 & 20.05 & 2.90e+24 & 1.99e+24 & 2.53 & 0.14 & 218.1 & 12.9 & 105.7 & 6.5 & rotated\_ellipse & 20.34 & 0.149 & 1.49 & --- \\
PSZ2 G112.48+56.99 & 41.72 & 7.00 & 55.16 & 8.36 & 6.71e+23 & 1.02e+23 & 0.80 & 0.05 & 160.7 & 9.3 &  &  & circle & 16.05 & 0.071 & 1.00 & --- \\
PSZ2 G113.91-37.01 & 78.03 & 8.22 & 122.78 & 13.64 & 6.38e+25 & 7.09e+24 & 5.09 & 0.23 & 365.3 & 15.0 &  &  & circle & 20.92 & 0.786 & 1.47 & --- \\
PSZ2 G114.31+64.89 & 41.10 & 11.72 & 66.61 & 13.27 & 1.83e+25 & 3.64e+24 & 12.40 & 0.59 & 143.9 & 6.6 &  &  & circle & 19.50 & 0.369 & 1.67 & --- \\
PSZ2 G115.67-27.57 & 4.61 & 0.84 & 4.53 & 0.74 & 3.93e+24 & 6.39e+23 & 5.00 & 1.33 & 80.3 & 14.0 &  &  & circle & 7.86 & 0.287 & 0.93 & --- \\
PSZ2 G133.60+69.04 & 145.00 & 18.87 & 162.80 & 20.30 & 3.43e+25 & 4.28e+24 & 7.07 & 0.17 & 377.1 & 5.2 & 200.9 & 4.3 & rotated\_ellipse & 74.85 & 0.248 & 3.13 & --- \\
PSZ2 G135.17+65.43 & 35.76 & 6.89 & 40.76 & 7.35 & 5.46e+25 & 9.84e+24 & 6.43 & 0.50 & 232.7 & 12.1 &  &  & circle & 19.65 & 0.576 & 1.24 & --- \\
PSZ2 G138.32-39.82 & 28.07 & 4.29 & 31.00 & 4.80 & 8.22e+24 & 1.27e+24 & 2.89 & 0.24 & 201.7 & 14.3 &  &  & circle & 12.23 & 0.419 & 1.07 & --- \\
PSZ2 G139.18+56.37 & 398.78 & 49.86 & 288.44 & 41.53 & 1.47e+26 & 1.84e+25 & 45.75 & 0.25 & 170.2 & 0.8 &  &  & circle & 195.52 & 0.369 & 20.10 & $P_{150}$ derived from $S_{150}$ ($2\sigma$) \\
PSZ2 G141.05-32.61 & 66.30 & 8.94 & 71.57 & 9.78 & 1.00e+26 & 1.37e+25 & 6.30 & 0.32 & 314.6 & 12.8 &  &  & circle & 19.40 & 0.525 & 1.68 & --- \\
PSZ2 G143.26+65.24 & 33.72 & 4.79 & 34.02 & 5.12 & 1.68e+25 & 2.53e+24 & 4.30 & 0.27 & 206.5 & 11.7 &  &  & circle & 16.23 & 0.389 & 1.07 & --- \\
PSZ2 G144.23-18.19 & 17.33 & 3.12 & 23.80 & 3.96 & 2.10e+24 & 3.79e+23 & 1.80 & 0.18 & 174.2 & 15.0 &  &  & circle & 11.02 & 0.214 & 1.05 & $P_{150}$ and $r_1$ computed assuming $z=0.2$ \\
PSZ2 G147.88+53.24 & 12.90 & 3.63 & 21.86 & 4.27 & 3.75e+25 & 7.33e+24 & 5.87 & 0.38 & 187.0 & 10.9 &  &  & circle & 14.18 & 0.595 & 1.03 & --- \\
PSZ2 G148.36+75.23 & 20.97 & 2.33 & 18.86 & 2.16 & 6.07e+24 & 6.95e+23 & 8.57 & 0.58 & 96.6 & 5.1 &  &  & circle & 18.46 & 0.343 & 1.00 & --- \\
PSZ2 G149.22+54.18 & 332.74 & 52.81 & 315.77 & 51.81 & 1.64e+25 & 2.69e+24 & 6.42 & 0.11 & 246.1 & 3.6 &  &  & circle & 56.88 & 0.234 & 2.07 & --- \\
PSZ2 G149.75+34.68 & 711.34 & 74.64 & 574.12 & 61.67 & 5.62e+25 & 6.04e+24 & 12.23 & 0.08 & 303.9 & 1.6 &  &  & circle & 157.82 & 0.184 & 7.23 & --- \\
PSZ2 G150.56+46.67 & 13.87 & 1.58 & 16.18 & 1.88 & 9.98e+24 & 1.16e+24 & 4.17 & 0.26 & 153.0 & 7.6 &  &  & circle & 17.50 & 0.388 & 0.96 & --- \\
PSZ2 G150.56+58.32 & 54.22 & 11.09 & 60.83 & 11.53 & 5.65e+25 & 1.07e+25 & 10.66 & 0.37 & 315.3 & 11.2 & 132.6 & 4.8 & rotated\_ellipse & 29.55 & 0.607 & 1.22 & --- \\
PSZ2 G151.19+48.27 & 38.62 & 7.69 & 47.05 & 9.01 & 1.34e+25 & 2.57e+24 & 1.66 & 0.13 & 335.5 & 23.7 &  &  & circle & 11.15 & 0.202 & 0.95 & --- \\
PSZ2 G154.13+40.19 & 13.50 & 1.82 & 13.20 & 1.83 & 3.80e+24 & 5.26e+23 & 9.04 & 0.64 & 76.2 & 4.1 &  &  & circle & 17.60 & 0.422 & 0.89 & --- \\
PSZ2 G156.26+59.64 & 6.59 & 1.65 & 7.81 & 1.97 & 1.27e+25 & 3.22e+24 & 9.74 & 1.90 & 85.9 & 13.7 &  &  & circle & 6.48 & 0.948 & 0.94 & --- \\
PSZ2 G160.83+81.66 & 11.67 & 2.70 & 10.56 & 2.67 & 4.98e+25 & 1.26e+25 & 15.90 & 1.26 & 91.7 & 5.5 &  &  & circle & 16.69 & 0.647 & 1.21 & --- \\
PSZ2 G164.65+46.37 & 23.52 & 5.28 & 22.40 & 5.29 & 9.56e+24 & 2.26e+24 & 5.87 & 0.33 & 217.1 & 12.0 & 87.5 & 5.4 & rotated\_ellipse & 18.86 & 0.413 & 0.89 & --- \\
PSZ2 G165.06+54.13 & 43.38 & 9.18 & 60.61 & 10.76 & 3.51e+24 & 6.23e+23 & 0.88 & 0.06 & 305.0 & 18.4 &  &  & circle & 14.18 & 0.108 & 0.76 & --- \\
PSZ2 G166.09+43.38 & 142.10 & 15.44 & 112.78 & 12.81 & 1.65e+25 & 1.87e+24 & 7.19 & 0.16 & 202.0 & 3.9 &  &  & circle & 46.32 & 0.210 & 1.39 & --- \\
PSZ2 G166.62+42.13 &  &  &  &  &  &  &  &  &  &  &  &  & --- &  &  &  & Cannot be disentangled from RR emission \\
PSZ2 G172.63+35.15 & 56.64 & 7.29 & 68.09 & 9.27 & 2.99e+24 & 4.07e+23 & 1.79 & 0.16 & 202.5 & 15.1 &  &  & circle & 12.26 & 0.172 & 1.02 & --- \\
PSZ2 G176.27+37.54 & 4.73 & 1.11 & 6.47 & 2.12 & 9.62e+24 & 3.16e+24 & 2.96 & 1.05 & 139.6 & 40.7 &  &  & circle & 3.47 & 0.375 & 1.10 & --- \\
PSZ2 G179.09+60.12 & 25.21 & 6.99 & 36.51 & 8.04 & 1.89e+24 & 4.17e+23 & 2.55 & 0.22 & 133.0 & 10.4 &  &  & circle & 11.31 & 0.293 & 1.27 & --- \\
PSZ2 G183.30+34.98 & 24.35 & 7.35 & 32.35 & 8.01 & 1.92e+25 & 4.76e+24 & 5.65 & 0.49 & 184.1 & 13.8 &  &  & circle & 11.63 & 0.761 & 0.92 & --- \\
PSZ2 G183.90+42.99 & 57.09 & 5.92 & 52.12 & 5.40 & 7.55e+25 & 7.81e+24 & 16.35 & 0.39 & 167.6 & 3.1 &  &  & circle & 45.83 & 0.606 & 1.48 & --- \\
PSZ2 G184.68+28.91 & 7.03 & 1.23 & 6.86 & 1.41 & 1.94e+24 & 3.98e+23 & 5.40 & 0.93 & 70.7 & 9.8 &  &  & circle & 7.17 & 0.473 & 0.64 & --- \\
PSZ2 G186.37+37.26 & 88.01 & 9.11 & 85.31 & 9.04 & 2.30e+25 & 2.44e+24 & 12.40 & 0.43 & 162.2 & 4.8 &  &  & circle & 30.41 & 0.533 & 1.10 & --- \\
PSZ2 G186.61+62.94 & 3.10 & 1.10 & 3.21 & 1.11 & 3.64e+24 & 1.26e+24 & 2.85 & 0.82 & 94.9 & 18.6 &  &  & circle & 6.61 & 0.191 & 0.53 & --- \\
PSZ2 G186.99+38.65 & 54.29 & 7.14 & 55.06 & 7.28 & 3.00e+25 & 3.96e+24 & 27.68 & 0.90 & 106.1 & 2.8 &  &  & circle & 35.46 & 0.625 & 1.30 & --- \\
PSZ2 G189.31+59.24 & 334.87 & 68.03 & 252.77 & 64.40 & 1.09e+25 & 2.78e+24 & 25.80 & 0.39 & 102.3 & 1.2 &  &  & circle & 78.08 & 0.189 & 4.63 & --- \\
PSZ2 G190.61+66.46 & 17.61 & 2.44 & 35.59 & 4.35 & 3.63e+25 & 4.44e+24 & 3.22 & 0.23 & 290.5 & 15.0 &  &  & circle & 16.68 & 0.252 & 1.85 & --- \\
PSZ2 G192.18+56.12 & 14.92 & 2.52 & 19.45 & 3.59 & 8.10e+23 & 1.50e+23 & 0.66 & 0.10 & 174.9 & 22.1 &  &  & circle & 7.24 & 0.102 & 1.01 & --- \\
PSZ2 G192.77+33.14 & 108.18 & 14.63 & 118.30 & 18.72 & 7.10e+23 & 1.12e+23 & 0.54 & 0.06 & 211.7 & 20.8 &  &  & circle & 9.37 & 0.062 & 0.75 & --- \\
PSZ2 G192.90+29.63 & 21.71 & 3.05 & 33.96 & 5.14 & 1.64e+25 & 2.48e+24 & 2.91 & 0.29 & 249.5 & 21.1 &  &  & circle & 9.53 & 0.412 & 0.88 & --- \\
PSZ2 G205.90+73.76 & 45.01 & 6.79 & 36.67 & 6.28 & 3.01e+25 & 5.15e+24 & 34.30 & 1.53 & 103.3 & 4.7 & 71.7 & 3.1 & rotated\_ellipse & 27.79 & 0.763 & 0.72 & --- \\
   \hline
 \end{tabular}
}
\tablefoot{Col. 1: PSZ2 name; Cols. 2 and 3: flux density integrated within the $2\sigma$ region and its error; Cols. 4 and 5: flux density obtained from the surface brightness profile fitting and its error; Cols. 6 and 7: radio power and its error; Cols. 8 and 9: best-fit central brightness and its error; Cols. 10-13: best-fit first and second \textit{e}-folding radii and their errors; Col. 14: model adopted for the fitting; Col. 15: S/N of the radio halo; Col. 16: rms noise of the image used to fit the surface brightness profile; Col. 17: $\chi^2_{\rm red}$ of the fit; Col. 18: comments.}
\end{table}

\begin{table}[h]
 \centering
 \caption{Sample of radio relics and candidate radio relics found in this work.}
 \label{tab:RR_sample}
\resizebox{\textwidth}{!}{
 \begin{tabular}{lcrrrrrrrrrrrrrrr}
   \hline
   \hline
  \multicolumn{1}{l}{Name} &
  \multicolumn{1}{c}{Position} &
  \multicolumn{1}{c}{RA$_{\rm RR}$} &
  \multicolumn{1}{c}{DEC$_{\rm RR}$} &
  \multicolumn{1}{c}{$S_{150}$} &
  \multicolumn{1}{c}{${S_{150}}_{\rm err}$} &
  \multicolumn{1}{c}{$P_{150}$} &
  \multicolumn{1}{c}{${P_{150}}_{\rm err}$} &
  \multicolumn{1}{c}{LLS} &
  \multicolumn{1}{c}{LLS$_{\rm err}$} &
  \multicolumn{1}{c}{Width} &
  \multicolumn{1}{c}{Width$_{\rm err}$} &
  \multicolumn{1}{c}{Average SB} &
  \multicolumn{1}{c}{$D_{\rm RR-c}$} &
  \multicolumn{1}{c}{${D_{\rm RR-c}}_{\rm err}$} &
  \multicolumn{1}{c}{$D_{\rm RR-RR}$} &
  \multicolumn{1}{c}{${D_{\rm RR-RR}}_{\rm err}$} \\  
   &
   &
  \multicolumn{1}{c}{[deg]} & 
  \multicolumn{1}{c}{[deg]} & 
  \multicolumn{1}{c}{[mJy]} &
  \multicolumn{1}{c}{[mJy]} &
  \multicolumn{1}{c}{[\whz]} &
  \multicolumn{1}{c}{[\whz]} &
  \multicolumn{1}{c}{[kpc]} &
  \multicolumn{1}{c}{[kpc]} &  
  \multicolumn{1}{c}{[kpc]} &
  \multicolumn{1}{c}{[kpc]} &  
  \multicolumn{1}{c}{[\mujyb]} &
  \multicolumn{1}{c}{[kpc]} &
  \multicolumn{1}{c}{[kpc]} &
  \multicolumn{1}{c}{[kpc]} &
  \multicolumn{1}{c}{[kpc]} \\
   \hline
   PSZ2 G048.10+57.16 &  & 230.0114 & 30.5267 & 264.28 & 26.83 & 3.96e+24 & 4.02e+23 & 998 & 35 & 395 & 182 & 2940 & 1515 & 206 &  &  \\
PSZ2 G057.61+34.93 &  & 257.7043 & 34.5640 & 426.57 & 42.90 & 6.74e+24 & 6.77e+23 & 1361 & 44 & 450 & 167 & 5579 & 1269 & 176 &  &  \\
PSZ2 G069.39+68.05 &  & 215.4091 & 38.3299 & 28.47 & 4.58 & 7.61e+25 & 1.22e+25 & 1489 & 56 & 109 & 153 & 1237 &  &  &  &  \\
PSZ2 G071.21+28.86 & N & 268.0816 & 44.7043 & 518.11 & 51.83 & 2.37e+26 & 2.37e+25 & 1254 & 39 & 291 & 149 & 6205 & 998 & 42 & 1941 & 39 \\
PSZ2 G071.21+28.86 & S & 267.9735 & 44.6313 & 179.66 & 17.99 & 8.23e+25 & 8.24e+24 & 813 & 39 & 136 & 86 & 5134 & 965 & 42 & 1941 & 39 \\
PSZ2 G080.16+57.65 &  & 225.5644 & 47.1942 & 37.63 & 4.78 & 7.27e+23 & 9.24e+22 & 1073 & 50 & 173 & 86 & 1667 & 1418 & 196 &  &  \\
PSZ2 G086.58+73.11 &  & 205.9581 & 39.9924 & 4.28 & 0.66 & 6.22e+23 & 9.53e+22 & 394 & 44 & 121 & 42 & 584 &  &  &  &  \\
PSZ2 G089.52+62.34 & N2 & 215.8002 & 48.7878 & 92.49 & 9.94 & 1.10e+24 & 1.19e+23 & 1213 & 35 & 296 & 200 & 2742 & 1721 & 233 &  &  \\
PSZ2 G089.52+62.34 & N1 & 215.5888 & 48.5719 & 50.64 & 6.63 & 6.04e+23 & 7.90e+22 & 273 & 35 & 107 & 39 & 5147 & 484 & 74 &  &  \\
PSZ2 G091.79-27.00 &  & 341.3831 & 28.2727 & 16.48 & 2.55 & 7.36e+24 & 1.14e+24 & 1535 & 41 & 164 & 96 & 1176 & 2510 & 339 &  &  \\
PSZ2 G096.43-20.89 &  & 341.9545 & 35.5495 & 33.56 & 4.16 & 1.38e+25 & 1.71e+24 & 1482 & 53 & 279 & 173 & 839 &  &  &  &  \\
PSZ2 G099.48+55.60 & N & 216.8435 & 57.0057 & 108.16 & 11.44 & 3.04e+24 & 3.22e+23 & 1711 & 45 & 354 & 173 & 1398 & 1366 & 49 & 2405 & 45 \\
PSZ2 G099.48+55.60 & S & 217.2878 & 56.7582 & 69.38 & 7.49 & 1.95e+24 & 2.11e+23 & 1090 & 45 & 229 & 160 & 1609 & 1068 & 48 & 2405 & 45 \\
PSZ2 G107.10+65.32 &  & 203.2562 & 50.4015 & 21.49 & 2.25 & 5.29e+24 & 5.53e+23 & 538 & 50 & 266 & 113 & 1258 & 1211 & 170 &  &  \\
PSZ2 G109.22-44.01 &  & 2.6228 & 17.7278 & 148.71 & 15.75 & 1.24e+25 & 1.31e+24 & 1168 & 40 & 211 & 100 & 3216 &  &  &  &  \\
PSZ2 G111.75+70.37 &  & 198.3618 & 46.3271 & 120.37 & 12.07 & 1.14e+25 & 1.14e+24 & 874 & 56 & 317 & 93 & 4520 & 898 & 133 &  &  \\
PSZ2 G113.91-37.01 & N & 4.9365 & 25.3543 & 164.61 & 16.57 & 7.78e+25 & 7.84e+24 & 1172 & 43 & 412 & 155 & 2033 & 1210 & 47 & 2605 & 43 \\
PSZ2 G113.91-37.01 & S & 4.8569 & 25.2329 & 40.41 & 4.43 & 1.91e+25 & 2.10e+24 & 1481 & 43 & 296 & 128 & 1361 & 1464 & 49 & 2605 & 43 \\
PSZ2 G116.50-44.47 &  & 8.0645 & 18.1766 & 13.00 & 1.54 & 7.17e+24 & 8.48e+23 & 555 & 47 & 118 & 79 & 946 & 890 & 128 &  &  \\
PSZ2 G121.03+57.02 &  & 194.9253 & 60.0416 & 12.26 & 1.57 & 4.86e+24 & 6.21e+23 & 542 & 44 & 141 & 45 & 759 & 570 & 88 &  &  \\
PSZ2 G135.17+65.43 &  & 184.7975 & 50.8694 & 6.61 & 0.93 & 7.78e+24 & 1.10e+24 & 654 & 54 & 125 & 63 & 552 & 1110 & 158 &  &  \\
PSZ2 G144.99-24.64 &  & 37.2926 & 33.9827 & 10.45 & 2.12 & 1.07e+24 & 2.18e+23 & 773 & 48 & 153 & 82 & 749 &  &  &  &  \\
PSZ2 G151.19+48.27 &  & 154.4696 & 59.5328 & 19.24 & 2.05 & 5.09e+24 & 5.44e+23 & 812 & 43 & 122 & 90 & 816 & 682 & 101 &  &  \\
PSZ2 G165.46+66.15 & N & 170.8381 & 43.1768 & 89.07 & 9.43 & 9.79e+24 & 1.04e+24 & 1069 & 49 & 339 & 120 & 2270 & 843 & 51 & 2115 & 49 \\
PSZ2 G165.46+66.15 & S & 170.9279 & 43.0079 & 171.63 & 17.51 & 1.89e+25 & 1.92e+24 & 1638 & 49 & 385 & 188 & 2571 & 1387 & 53 & 2115 & 49 \\
PSZ2 G166.62+42.13 & W & 137.2336 & 51.5893 & 454.39 & 45.53 & 7.29e+25 & 7.30e+24 & 1901 & 39 & 718 & 238 & 2529 & 1462 & 200 &  &  \\
PSZ2 G166.62+42.13 & N & 137.4405 & 51.6072 & 32.37 & 3.67 & 5.19e+24 & 5.88e+23 & 932 & 39 & 364 & 163 & 675 & 917 & 129 &  &  \\
PSZ2 G166.62+42.13 & E & 137.5390 & 51.5332 & 12.21 & 6.52 & 1.96e+24 & 1.05e+24 & 1100 & 39 & 148 & 71 & 1080 & 1204 & 166 &  &  \\
PSZ2 G181.06+48.47 & N & 144.9131 & 40.8741 & 28.22 & 2.98 & 4.89e+24 & 5.17e+23 & 1394 & 39 & 135 & 67 & 925 & 1660 & 47 & 2694 & 39 \\
PSZ2 G181.06+48.47 & S & 144.8281 & 40.6879 & 66.76 & 6.83 & 1.16e+25 & 1.18e+24 & 1660 & 39 & 207 & 122 & 1083 & 1035 & 42 & 2694 & 39 \\
PSZ2 G186.99+38.65 &  & 132.5838 & 36.1125 & 12.42 & 1.58 & 6.14e+24 & 7.80e+23 & 723 & 46 & 136 & 73 & 705 & 1030 & 145 &  &  \\
PSZ2 G190.61+66.46 &  & 166.5720 & 33.5517 & 60.49 & 6.43 & 5.48e+25 & 5.82e+24 & 573 & 48 & 184 & 180 & 2331 & 865 & 126 &  &  \\
PSZ2 G198.46+46.01 &  & 142.7257 & 28.7635 & 41.68 & 4.64 & 1.19e+25 & 1.33e+24 & 1494 & 43 & 247 & 172 & 694 &  &  &  &  \\
PSZ2 G205.90+73.76 & N & 174.5111 & 27.9806 & 5.37 & 0.97 & 3.95e+24 & 7.12e+23 & 664 & 48 & 125 & 42 & 791 & 1490 & 53 & 2958 & 48 \\
PSZ2 G205.90+73.76 & S & 174.5439 & 27.8404 & 7.55 & 1.12 & 5.55e+24 & 8.26e+23 & 655 & 48 & 155 & 58 & 1028 & 1540 & 54 & 2958 & 48 \\
   \hline
 \end{tabular}
}
\tablefoot{Col. 1: PSZ2 name; Col. 2: position of the relic with respect to the cluster; Cols. 3 and 4: coordinates of the radio relic; Cols. 5 and 6: flux density; Cols. 7 and 8: radio power and its error; Cols. 9 and 10: largest-linear size and its error; Cols. 11 and 12: width and its error; Col. 13: average surface brightness; Cols. 14 and 15: distance between radio relic and X-ray centroid and its error; Cols. 16 and 17: distance between double radio relics and its error.}
\end{table}   
\FloatBarrier

\section{Image gallery}\label{app:collage}

Figure~\ref{fig:dr2_collection} shows a collection of \lofar\ images of the targets in our \lotss-DR2/PSZ2 sample.

\begin{figure}
  \centering
  \includegraphics[width=\hsize,trim={0cm 0cm 0cm 0cm},clip,page=1]{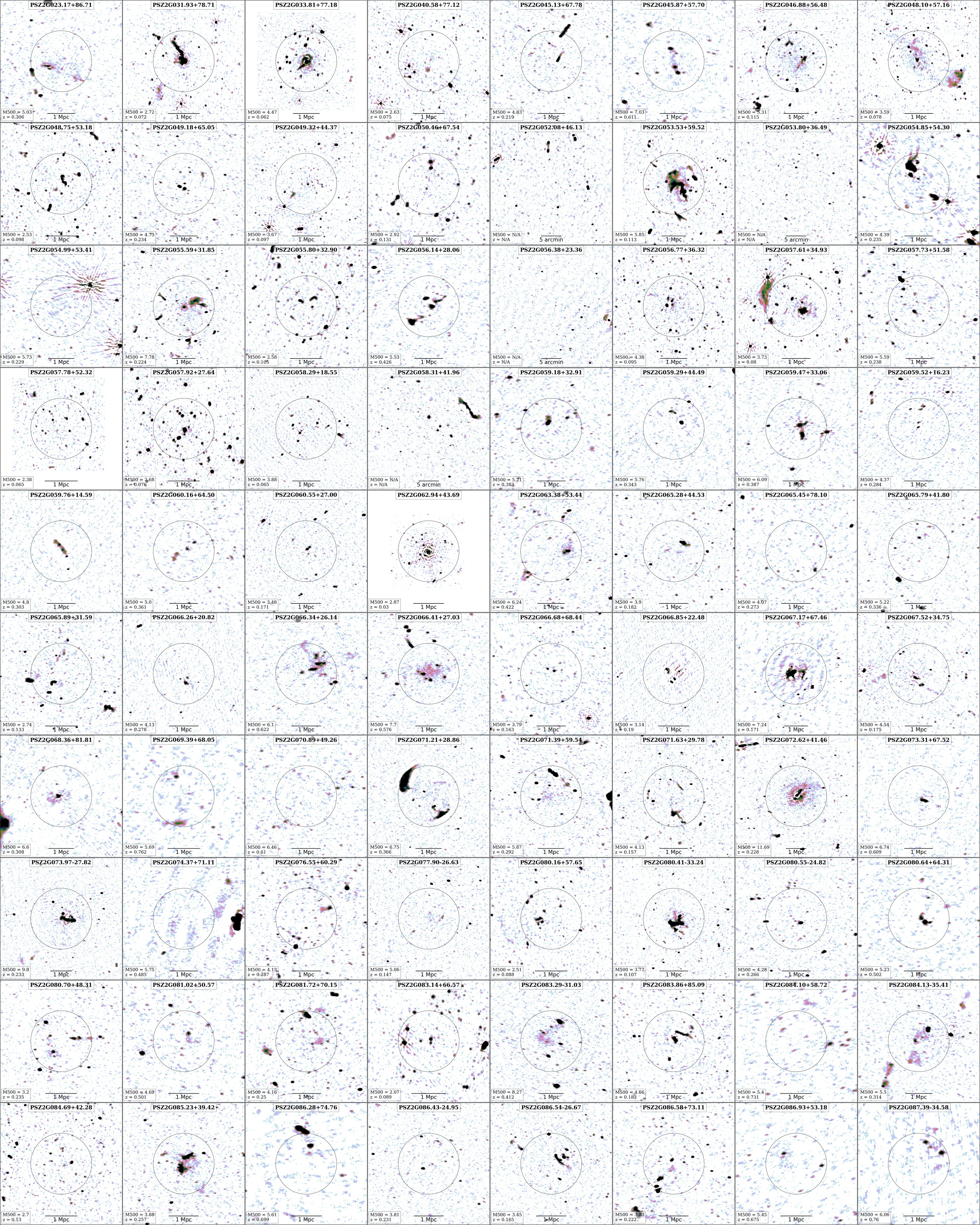}
  \caption{\lofar\ image gallery of the targets in our sample, excluding the five objects for which it was not possible to apply the extraction + recalibration scheme (see end of Sect.~\ref{sec:datareduction}). The collage is available at full resolution on the project website, \url{https://lofar-surveys.org/planck_dr2.html}.}
  \label{fig:dr2_collection}
\end{figure}

\begin{figure}
\addtocounter{figure}{-1}
\centering
  \includegraphics[width=\hsize,trim={0cm 0cm 0cm 0cm},clip]{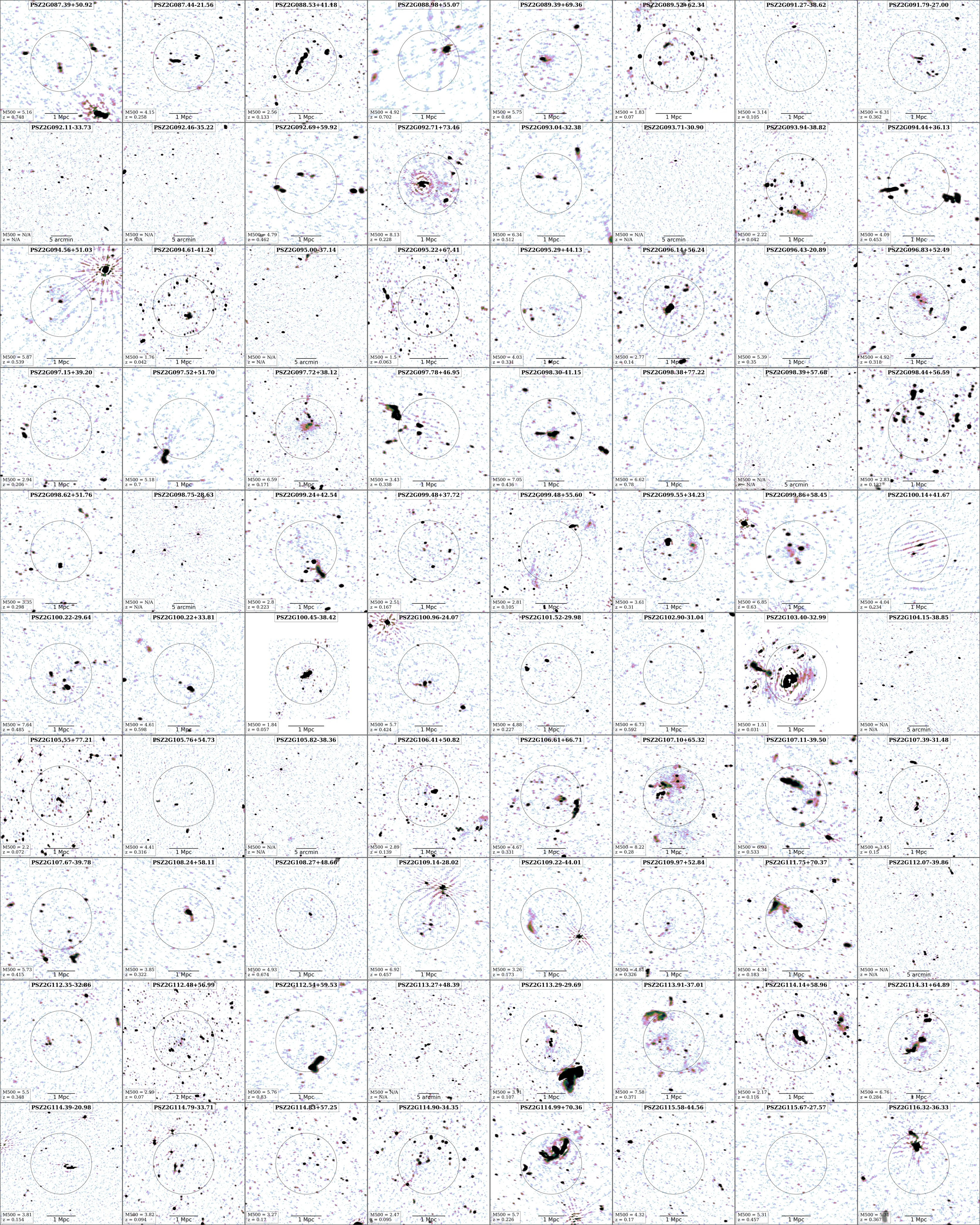}
  \caption{continued.}
\end{figure}

\begin{figure}
\addtocounter{figure}{-1}
\centering
  \includegraphics[width=\hsize,trim={0cm 0cm 0cm 0cm},clip]{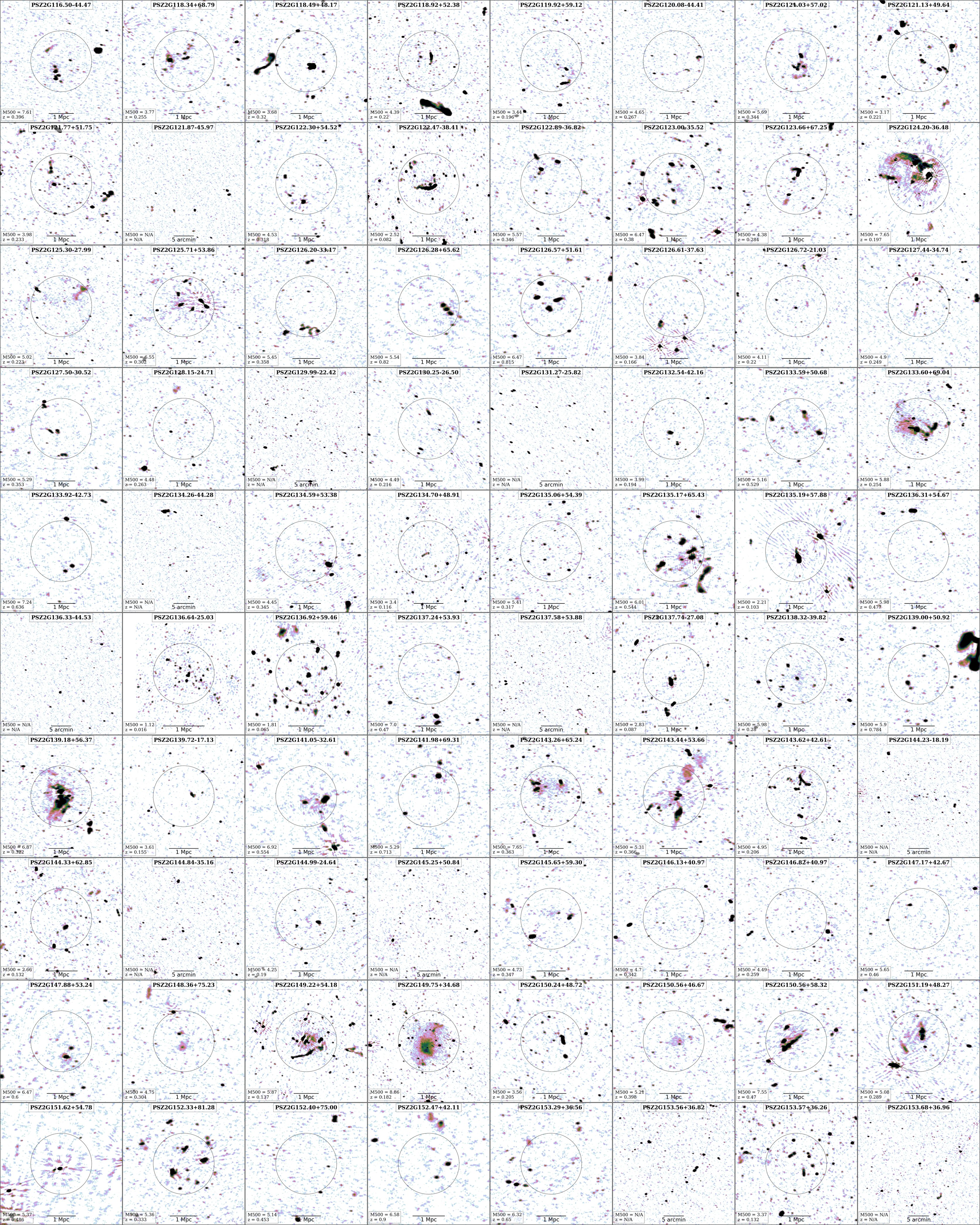}
  \caption{continued.}
\end{figure}

\begin{figure}
\addtocounter{figure}{-1}
\centering
  \includegraphics[width=\hsize,trim={0cm 0cm 0cm 0cm},clip]{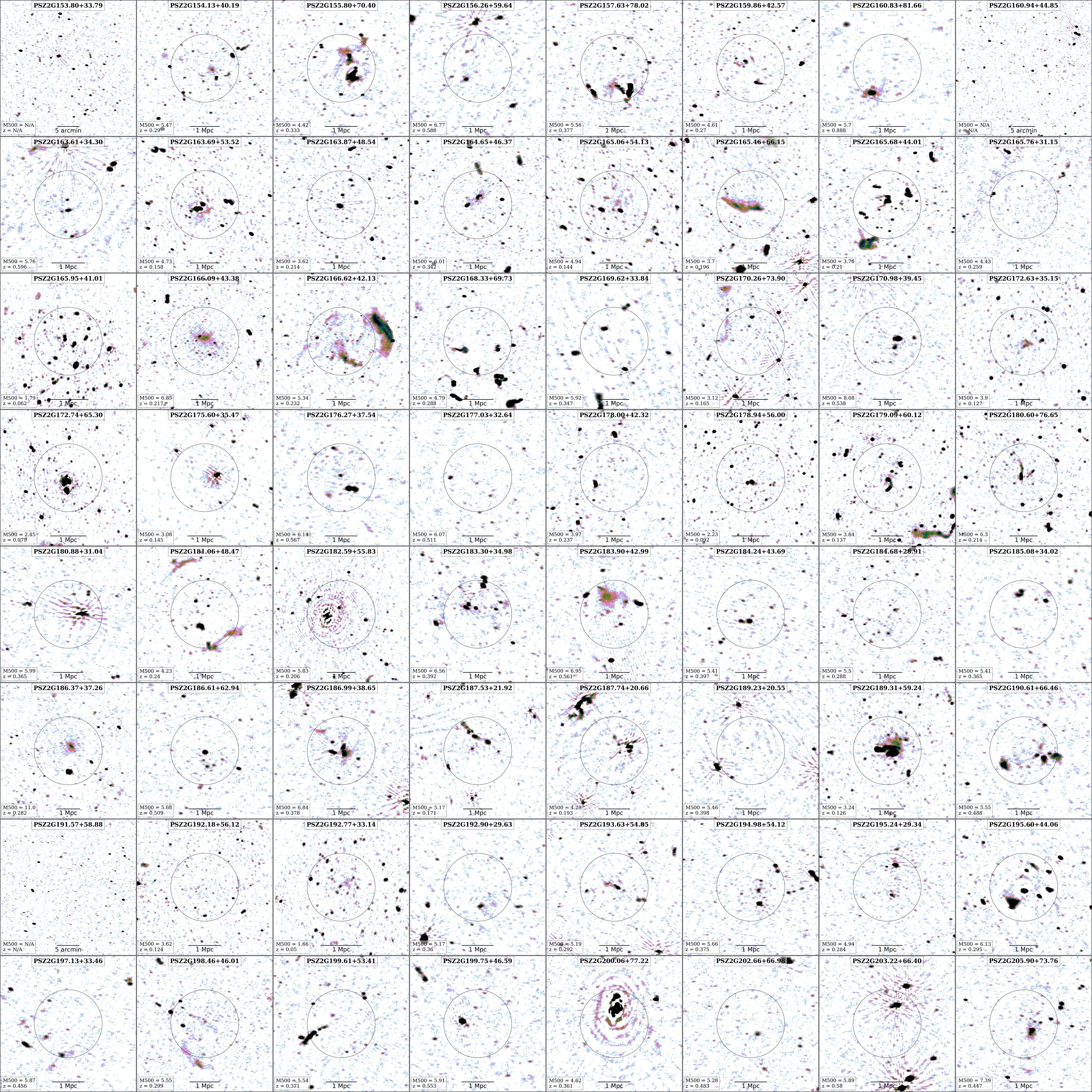}
  \caption{continued.}
\end{figure}
\FloatBarrier

\section{Comparison between flux density measurements for radio halos}\label{app:flux}

\begin{figure}
  \centering
  \includegraphics[width=.48\hsize,trim={0cm 0cm 0cm 0cm},clip,valign=c]{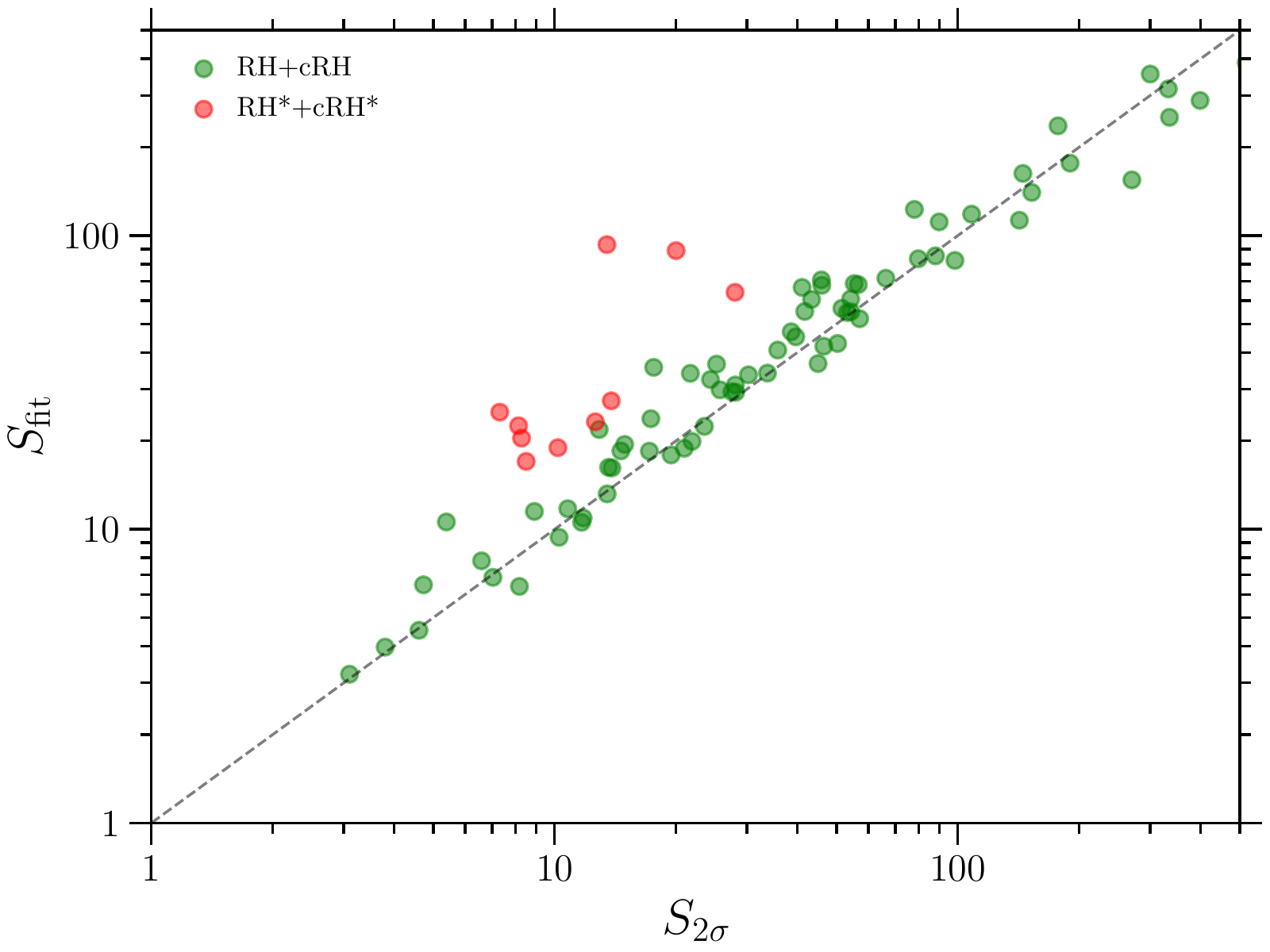}
  \includegraphics[width=.48\hsize,trim={0cm 0cm 0cm 0cm},clip,valign=c]{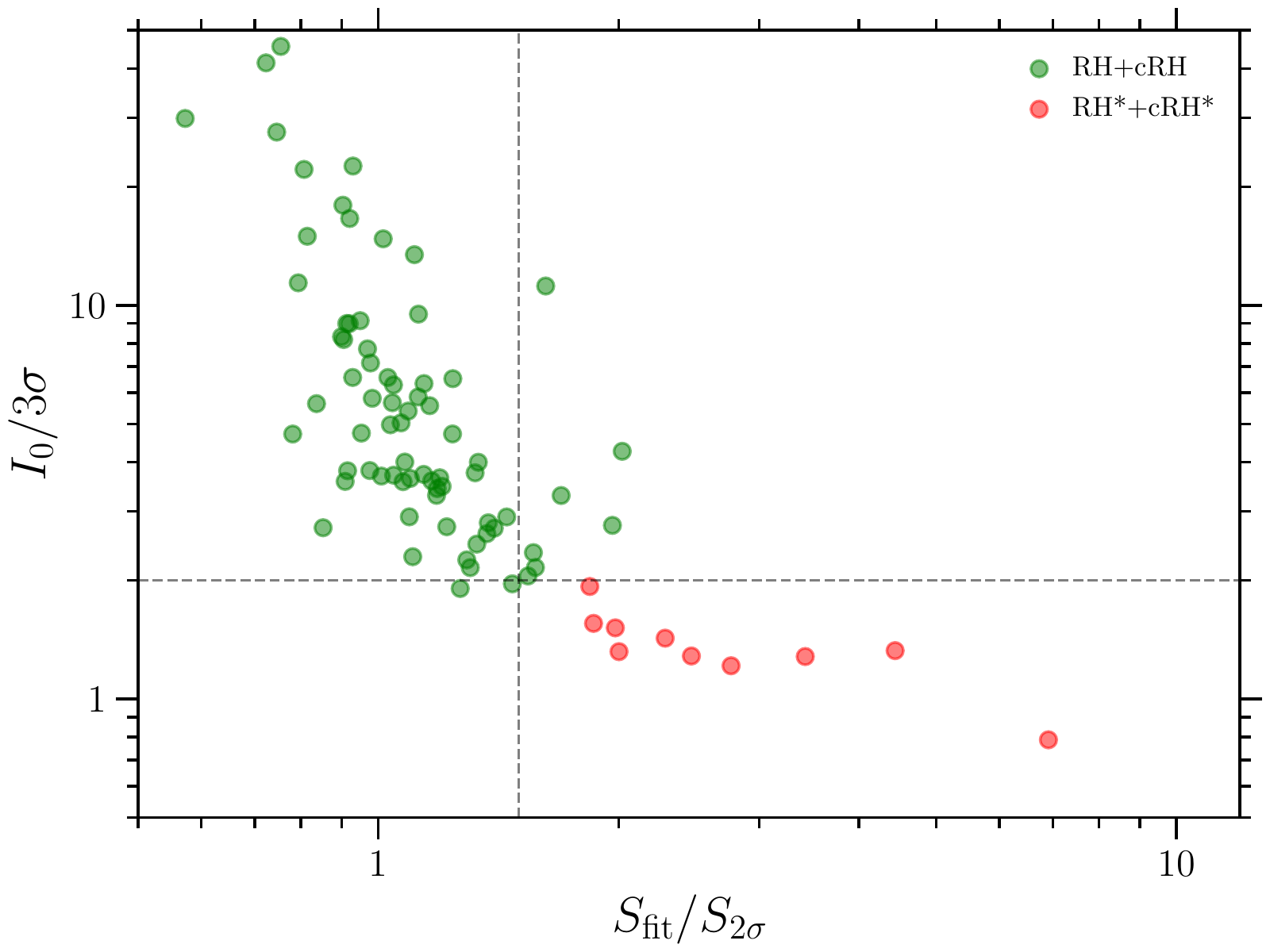}
  \caption{Flux density measurements for radio halos. \textit{Left}: Comparison between the flux density derived with \halofdca\ ($S_{\rm fit}$) and that obtained manually within the $2\sigma$ contour ($S_{\rm 2\sigma}$). The dashed line indicates the linear correlation as a reference. \textit{Right}: Diagnostic plot used to determine when \halofdca\ does not provide a reliable flux density measurement due to the low significance of the radio halo. The dashed lines indicate the thresholds of $S_{\rm fit}/S_{\rm 2\sigma} > 1.5$ and $I_0/3\sigma < 2$ that were used to identify the sources that were classified as radio halos or candidate radio halos with the decision tree of Fig.~\ref{fig:decision_tree}, but whose flux density was not considered reliable. These sources are reported in red and in the manuscript are referred to as RH* and cRH*.}
  \label{fig:fit_vs_2sigma}
\end{figure}

Here we compare the flux density for the radio halos obtained by fitting their surface brightness profile with \halofdca\ ($S_{\rm fit}$) and by manual integration of a circular/elliptical region encompassing the $2\sigma$ level contour ($S_{\rm 2\sigma}$). In both cases, regions affected by residual calibration errors and/or contaminating extended sources were masked out and the flux density beneath them was extrapolated assuming the best-fit model (in the case of \halofdca) or the average surface brightness of the halo in the non-masked region (in the case of the manual measurement). The left panel of Fig.~\ref{fig:fit_vs_2sigma} shows that there is a general agreement between the two measurements, although a few outliers are present. \\
\indent
We inspected the radio images and results of the fitting for all clusters and found that the outliers may indicate cases in which \halofdca\ provides a unreliable fit because of the low significance of the radio halo. To verify this, we produce the plot shown in the right panel of Fig.~\ref{fig:fit_vs_2sigma}, which shows the ratio between the best-fit value of the central surface brightness $I_0$ and the $3\sigma$ noise of the map as a function of the $S_{\rm fit}/S_{\rm 2\sigma}$ ratio. We observe a trend between the plotted quantities, noting indeed that the halos where the two flux density measurements are in most disagreement (large $S_{\rm fit}/S_{\rm 2\sigma}$ values) are typically those with lowest $I_0/3\sigma$. Our interpretation is that when the radio halo is observed with low significance, \halofdca\ does not converge on the diffuse emission but it provides a nonphysical best fit with a low $I_0$ and a large $r_1$, basically fitting the image noise. Importantly, these nonphysical best-fit models have still good S/N and $\chi^2_{\rm red}$ values and corner plots as the fit is still converged; thus, they cannot be simply rejected by using these tools. We therefore used the right panel of Fig.~\ref{fig:fit_vs_2sigma} as diagnostic plot, adopting the arbitrary thresholds of $S_{\rm fit}/S_{\rm 2\sigma} > 1.5$ and $I_0/3\sigma < 2$ to identify the bad cases. These radio halos and candidate radio halos are marked with an asterisk in Table~\ref{tab:sample} (\ie,\ RH*/cRH*) and represent 10 out of the 83 fitted halos. They are collected in Table~\ref{tab:asteriskRH_sample}. Deeper \lofar\ observations are required to reliably determine the integrated flux density of the halo emission in these clusters.

\begin{table}
 \centering
 \caption{Sample of radio halos and candidate radio halos that were classified with an asterisk (the RH* and cRH* in Table~\ref{tab:sample}) because of their low significance. The integrated flux density of these sources cannot be determined accurately with current data.}
 \label{tab:asteriskRH_sample}
\resizebox{\textwidth}{!}{
 \begin{tabular}{lrrrrrrrrrrlrrr}
   \hline
   \hline
  \multicolumn{1}{l}{Name} &
  \multicolumn{1}{c}{$S_{150}$ ($2\sigma$)} &
  \multicolumn{1}{c}{${S_{150}}_{\rm err}$ ($2\sigma$)} &
  \multicolumn{1}{c}{$S_{150}$ (fit)} &
  \multicolumn{1}{c}{${S_{150}}_{\rm err}$ (fit)} &
  \multicolumn{1}{c}{$P_{150}$} &
  \multicolumn{1}{c}{${P_{150}}_{\rm err}$} &
  \multicolumn{1}{c}{$I_0$} &
  \multicolumn{1}{c}{${I_0}_{\rm err}$} &
  \multicolumn{1}{c}{$r_1$} &
  \multicolumn{1}{c}{${r_1}_{\rm err}$} &
  \multicolumn{1}{c}{Model} &
  \multicolumn{1}{c}{S/N} &
  \multicolumn{1}{c}{rms} &
  \multicolumn{1}{c}{$\chi^2_{\rm red}$} \\
   &
  \multicolumn{1}{c}{[mJy]} &
  \multicolumn{1}{c}{[mJy]} &
  \multicolumn{1}{c}{[mJy]} &
  \multicolumn{1}{c}{[mJy]} &
  \multicolumn{1}{c}{[\whz]} &
  \multicolumn{1}{c}{[\whz]} &
  \multicolumn{1}{c}{[\mujyarcsecsq]} &
  \multicolumn{1}{c}{[\mujyarcsecsq]} &
  \multicolumn{1}{c}{[kpc]} &
  \multicolumn{1}{c}{[kpc]} &
   &
   &
  \multicolumn{1}{c}{[\mjyb]} &
   \\
   \hline
   PSZ2 G071.39+59.54 & 20.01 & 2.30 & 88.87 & 11.90 & 2.60e+25 & 3.48e+24 & 1.27 & 0.08 & 529.2 & 36.0 & circle & 11.25 & 0.319 & 1.15 \\
PSZ2 G076.55+60.29 & 8.14 & 1.67 & 22.52 & 5.72 & 6.33e+24 & 1.61e+24 & 0.43 & 0.07 & 453.8 & 82.3 & circle & 4.40 & 0.118 & 0.76 \\
PSZ2 G077.90-26.63 & 28.02 & 4.04 & 64.10 & 10.75 & 3.89e+24 & 6.52e+23 & 1.88 & 0.24 & 218.1 & 26.5 & circle & 7.53 & 0.439 & 1.03 \\
PSZ2 G095.29+44.13 & 13.82 & 2.08 & 27.38 & 4.36 & 1.08e+25 & 1.72e+24 & 1.21 & 0.13 & 328.7 & 32.8 & circle & 8.48 & 0.266 & 0.89 \\
PSZ2 G099.55+34.23 & 13.48 & 5.97 & 93.17 & 19.09 & 3.14e+25 & 6.44e+24 & 0.56 & 0.07 & 851.0 & 112.6 & circle & 5.95 & 0.237 & 0.84 \\
PSZ2 G116.50-44.47 & 12.62 & 6.73 & 23.23 & 8.29 & 1.42e+25 & 5.05e+24 & 3.94 & 0.78 & 187.8 & 36.0 & circle & 5.11 & 0.680 & 1.08 \\
PSZ2 G118.34+68.79 & 7.31 & 2.48 & 25.05 & 4.47 & 5.33e+24 & 9.52e+23 & 0.80 & 0.09 & 321.5 & 32.4 & circle & 8.58 & 0.208 & 1.34 \\
PSZ2 G123.00-35.52 & 8.28 & 1.86 & 20.43 & 4.27 & 1.13e+25 & 2.35e+24 & 2.23 & 0.35 & 228.4 & 34.4 & circle & 5.93 & 0.578 & 1.08 \\
PSZ2 G133.59+50.68 & 10.19 & 1.45 & 18.96 & 3.35 & 2.37e+25 & 4.18e+24 & 1.20 & 0.15 & 362.8 & 43.6 & circle & 7.07 & 0.257 & 0.92 \\
PSZ2 G163.61+34.30 & 8.50 & 2.19 & 17.02 & 3.71 & 2.87e+25 & 6.25e+24 & 1.50 & 0.24 & 324.7 & 44.9 & circle & 6.21 & 0.379 & 0.79 \\
   \hline
 \end{tabular}
}
\tablefoot{Col. 1: PSZ2 name; Cols. 2 and 3: flux density integrated within the $2\sigma$ region and its error; Cols. 4 and 5: flux density obtained from the surface brightness profile fitting and its error; Cols. 6 and 7: radio power and its error; Cols. 8 and 9: best-fit central brightness and its error; Cols. 10 and 11: best-fit \textit{e}-folding radius and its error; Col. 12: model adopted for the fitting; Col. 13: S/N of the radio halo; Col. 14: rms noise of the image used to fit the surface brightness profile; Col. 15: $\chi^2_{\rm red}$ of the fit.}
\end{table}   
\FloatBarrier

\section{Notes on individual clusters}\label{app:notes}

In the following, we make specific remarks for some clusters of our sample.

\paragraph{PSZ2 G023.17+86.71} The radio halo is elongated in the E-W direction and it is claimed in this paper for the first time.

\paragraph{PSZ2 G031.93+78.71} (A1775) It hosts a prominent head-tail radio galaxy that has been studied in the past \citep{owen97, giovannini00, giacintucci07, ternidegregory17}. The presence in the cluster center of a radio halo bounded by a cold front was recently claimed by \citet{botteon21a1775} using \lotss\ data, who referred to this emission as a ``slingshot'' radio halo. Toward the SE, a diffuse source elongated along the N-S direction and with uncertain nature is also observed.

\paragraph{PSZ2 G033.81+77.18} (A1795) Past observations with the \vla\ and \gmrt\ found the presence of diffuse emission in the cluster center, which may be related to a mini-halo or AGN activity \citep{giacintucci14minihalos, kokotanekov18}. Due to the bad data quality, we cannot perform the analysis of this system. \citet{birzan20} used \lofar\ data to search for an association between the central radio source and the X-ray cavity in the NW, but found none. A paper presenting the analysis of this system using the \lofar\ international baselines is currently in preparation (Timmerman et al., in preparation).

\paragraph{PSZ2 G040.58+77.12} (A1800) The radio halo is claimed here for the first time. A roundish region of emission without clear optical counterpart is observed in the S.

\paragraph{PSZ2 G045.87+57.70} The presence of diffuse emission in this cluster was originally claimed by \citet{digennaro21fast} using \lotss\ data. We confirm the central emission and following our decision tree we classified it as radio halo.

\paragraph{PSZ2 G046.88+56.48} (A2069) This is a bimodal cluster in which the first detection of megaparsec-scale radio emission was claimed by \citet{farnsworth13}. \citet{drabent15} found extended emission in both components. This is confirmed by our images, in which the radio emission seems to embed the whole system (including the region in-between the two subclusters). The detailed analysis of the \lofar\ observations of this target will be presented by Drabent et al. (in preparation).

\paragraph{PSZ2 G048.10+57.16} (A2061) It hosts a radio halo and a relic in the SW direction, which have already been claimed \citep{rudnick09, vanweeren11nvss, farnsworth13}. In our images, the radio halo remarkably follows the X-ray emission, which is stretched along the SW-NE direction, and is connected with the radio relic with a trail of emission. The detailed analysis of the \lofar\ observations of this target will be presented by Drabent et al. (in preparation).

\paragraph{PSZ2 G049.32+44.37} (A2175) The radio halo is claimed here for the first time.

\paragraph{PSZ2 G053.53+59.52} (A2034) Diffuse radio emission in the center of this cluster was noticed by different authors \citep{kempner01, rudnick09, giovannini09, vanweeren11nvss}. \citet{shimwell16} provided the clearest view of the diffuse radio sources in this system to date using \lofar\ observations. They claimed the presence of a radio halo with filamentary structures and a number of sources in the cluster outskirts with unclear origin. Our new \lotss\ images confirm the structures observed by \citet{shimwell16}. This cluster is displayed in Fig.~\ref{fig:showcases}.

\paragraph{PSZ2 G055.59+31.85} (A2261) Hints of diffuse emission in this system were first noticed with \vla\ 1.4 GHz observations by \citet{venturi08}, but not confirmed with \gmrt\ 235 and 610 MHz data by \citet{kale13}. \citet{sommer17} reanalyzed these radio observations and claimed to detect the presence of megaparsec-scale emission in the cluster, which was classified as a radio halo. The radio halo was confirmed by \lofar\ observations \citep{savini19}. Our images agree with these findings.

\paragraph{PSZ2 G056.77+36.32} (A2244) The radio halo is claimed here for the first time.

\paragraph{PSZ2 G057.61+34.93} (A2249) The presence of a radio relic in the E was recently reported with \lofar\ and \ugmrt\ data by \citet{locatelli20}. The relic is confirmed in our image while we classified as uncertain the emission in the cluster center due to the presence of artifacts from the bright central AGN.

\paragraph{PSZ2 G057.80+88.00} (Coma cluster) It hosts the prototypical radio halo and relic, which have been studied in the past with numerous instruments \citep[\eg,][]{kim89, giovannini91, giovannini93, thierbach03, feretti06coma, bonafede10, brown11coma}. Due to its large angular size, the analysis of \lofar\ data for this cluster requires a special treatment. For this reason, this cluster was excluded from our analysis. Results on this system using \lofar\ observations
have been recently reported by \citet{bonafede21, bonafede22arx}.

\paragraph{PSZ2 G058.29+18.55} It is a double galaxy cluster, comprising RXC J1825.3+3026 (dubbed subcluster E in Table~\ref{tab:cw}) and CIZA J1824.1+3029 (dubbed subcluster W in Table~\ref{tab:cw}). It also known as Lyra complex, in which the presence of a radio halo in RXC J1825.3+3026 was claimed with a targeted \lofar\ observation by \citet{botteon19lyra}. Due to the bad quality of \lotss\ data, we could not perform the analysis of this system.

\paragraph{PSZ2 G063.38+53.44} The candidate radio halo is claimed here for the first time.

\paragraph{PSZ2 G066.34+26.14} The candidate radio halo is claimed here for the first time.

\paragraph{PSZ2 G066.41+27.03} The radio halo is claimed here for the first time.

\paragraph{PSZ2 G067.17+67.46} (A1914) The intricate emission from this system was noticed by early observations with the \vla\ by \citet{bacchi03}. Past \lofar\ observations found the presence of very steep spectrum emission in the cluster due to very bright revived fossil plasma, a radio halo, and a head-tail radio galaxy \citep{mandal19}. Due to the complex emission of this system, we classified it as uncertain. We refer the reader to \citet{mandal19} for more details.

\paragraph{PSZ2 G068.36+81.81} We interpreted the diffuse emission from this cluster as the radio lobes of the central AGN. Therefore, we classified this cluster as NDE.

\paragraph{PSZ2 G069.39+68.05} A faint candidate radio halo is noticed in the lowest-resolution images. An elongated region of emission in the S is classified as candidate radio relic. These two sources are claimed here for the first time.

\paragraph{PSZ2 G071.21+28.86} (MACS J1752.0+4440) It is a textbook example of a double radio relic that has been reported in the past using \gmrt\ and \wsrt\ observations \citep{vanweeren12macsj1752, bonafede12}. A faint radio halo was claimed by \citet{vanweeren12macsj1752} and it is confirmed by our images. This cluster is displayed in Fig.~\ref{fig:showcases}.

\paragraph{PSZ2 G071.39+59.54} Central diffuse emission in form of a radio halo is observed here for the first time. Since the fitting of the radio halo surface brightness profile does not provide reliable results, we classified this emission as RH*.

\paragraph{PSZ2 G072.62+41.46} (A2219) It hosts a radio halo detected in the \nvss\ \citep{giovannini99} and confirmed in deeper \vla\ observation at 1.4 GHz \citep{bacchi03} and 325 MHz \citep{orru07}. Our \lofar\ images of this cluster are affected by bad quality due to the presence of the strong AGN located in the cluster center. Therefore, we could not perform the analysis of this system.

\paragraph{PSZ2 G073.97-27.82} (A2390) Originally, this cluster was classified as a mini-halo by \citet{bacchi03} using \vla\ data at 1.4 GHz. Later, \citet{sommer17} analyzed deeper and wide-band \vla\ data at 1$-$2 GHz and reclassified the emission as radio halo. Past \lofar\ observations revealed that the central radio galaxy may account for most or even all the radio flux that was attributed to the radio halo \citep{savini19}. In this paper, we interpret all the radio emission observed in this system as lobes of the central AGN. Another \lofar\ work highlighted how these radio lobes are filling the X-ray cavities in the ICM \citep{birzan20}.

\paragraph{PSZ2 G076.55+60.29} The X-ray emission of the cluster is offset with respect to the position provided by the PSZ2 catalog. Faint emission in the form of a radio halo is noticed here for the first time. Since the fitting of the radio halo surface brightness profile does not provide reliable results, we classified this emission as RH*.

\paragraph{PSZ2 G077.90-26.63} (A2409) Central diffuse emission in form of a radio halo is observed here for the first time.  Since the fitting of the radio halo surface brightness profile does not provide reliable results, we classified this emission as RH*.

\paragraph{PSZ2 G080.16+57.65} (A2018) A radio relic in the E direction was discovered together with a candidate radio halo by \citet{vanweeren21} in the analysis of the \lotss-DR1 cluster sample. More recently, \citet{paul21} confirmed the presence of the two sources using \ugmrt\ data at 300$-$500 MHz. In this paper, we confirm only the presence of the radio relic.

\paragraph{PSZ2 G080.41-33.24} (A2443) This cluster is known to host complex radio emission with an ultra-steep spectrum in the form of revived fossil plasma \citep{cohen11, clarke13}, which is well recovered in our \lofar\ images.

\paragraph{PSZ2 G080.70+48.31} (A2136) The candidate radio halo is claimed here for the first time.

\paragraph{PSZ2 G081.02+50.57} The radio halo is claimed here for the first time.

\paragraph{PSZ2 G081.72+70.15} (A1838) This system hosts two elongated sources with uncertain origin that are located outside the bulk of the X-ray emission. We did not classified these sources as radio relics as they do not show a sharp surface brightness edge in the \lofar\ images.

\paragraph{PSZ2 G083.29-31.03} The radio halo is claimed here for the first time.

\paragraph{PSZ2 G084.10+58.72} The presence of diffuse emission at the center of this cluster was originally noted by previous \lofar\ studies \citep{digennaro21spectral, digennaro21fast, vanweeren21}, belonging to the \lotss-DR1 area. Here, we confirm the emission and claim the presence of a radio halo.

\paragraph{PSZ2 G084.13-35.41} (A2472) The radio halo is claimed here for the first time. An uncertain source with a wedge-shape is also observed to the S of the radio halo.

\paragraph{PSZ2 G085.23+39.42} The candidate radio halo is claimed here for the first time. The diffuse emission is embedded into two bright structures likely associated with cluster radio galaxies.

\paragraph{PSZ2 G086.58+73.11} The galaxy overdensity seems located toward the SE with respect to the cluster coordinates reported in the PSZ2 catalog. To the S of the optical overdensity, we observed an elongated emission that we classify as candidate radio relic. This relic is claimed here for the first time. The presence of diffuse emission in the central region of the cluster is uncertain due to the artifacts introduced by the source subtraction.

\paragraph{PSZ2 G086.93+53.18} The presence of a radio halo in this cluster has been recently studied with \lofar\ by \citet{digennaro21spectral, digennaro21fast}. This systems belongs to the \lotss-DR1 area and was also reported by \citet{vanweeren21}. Here, we confirm the emission.

\paragraph{PSZ2 G088.53+41.18} (A2208) This cluster hosts a peculiar emission that is likely associated with radio galaxies. This cluster is displayed in Fig.~\ref{fig:showcases}.

\paragraph{PSZ2 G089.39+69.36} The presence of a radio halo in this cluster was recently claimed and studied with \lofar\ and \ugmrt\ data \citep{digennaro21spectral, digennaro21fast}. Here, we confirm the emission.

\paragraph{PSZ2 G089.52+62.34} (A1904) Two radio relics located on the same side of the cluster (NE) have been firstly claimed by \citet{vanweeren21} during the analysis of the \lotss-DR1 cluster sample. We confirm the two detections. Recently, the innermost relic was detected also with the \ugmrt\ at 300$-$500 MHz \citep{paul21}.

\paragraph{PSZ2 G091.79-27.00} The radio relic is claimed here for the first time.

\paragraph{PSZ2 G093.94-38.82} (A2572) This is a system with multiple cluster components in the X-rays, with the main component that we dubbed PSZ2 G093.94-38.82 W and two smaller sub-clumps that we dubbed PSZ2 G093.94-38.82 EN and PSZ2 G093.94-38.82 ES in Table~\ref{tab:cw}. We discover an extended diffuse source in the S of PSZ2 G093.94-38.82 W that has an uncertain origin.

\paragraph{PSZ2 G094.61-41.24} (A2589) Diffuse emission at the center of this cluster is noticed only in the image tapered at 100 kpc resolution. Due to the possible blend of partially subtracted sources at low resolution, we classified the emission as uncertain.

\paragraph{PSZ2 G095.22+67.41} Patches of emission with low significance and uncertain origin are observed in the cluster center and cluster outskirts. The peripheral emission on the E was classified as a candidate radio relic during the analysis of the \lotss-DR1 sample \citep{vanweeren21}.

\paragraph{PSZ2 G095.29+44.13} Central diffuse emission in form of a candidate radio halo is observed here for the first time.  Since the fitting of the radio halo surface brightness profile does not provide reliable results, we classified this emission as cRH*.

\paragraph{PSZ2 G096.43-20.89} The candidate radio relic is claimed here for the first time.

\paragraph{PSZ2 G096.83+52.49} (A1995) Extended emission at the center of this cluster was firstly reported by \citet{giovannini09} with \vla\ 1.4 GHz observations. Our images confirm the presence of a radio halo in this system.

\paragraph{PSZ2 G097.72+38.12} (A2218) Diffuse radio emission in the form of a radio halo in this system was originally reported by \citet{moffet89} and later confirmed by other authors \citep{giovannini00, kempner01}. In our image, the halo is slightly elongated in the E-W direction and it has an extension toward N. This cluster is displayed in Fig.~\ref{fig:showcases}.

\paragraph{PSZ2 G098.30-41.15} The candidate radio halo is claimed here for the first time.

\paragraph{PSZ2 G099.24+42.54} Radio emission with uncertain origin (due to the possible contribution of partially subtracted discrete sources) is noticed in the cluster center. Another emission in the SW with unclear origin is observed.

\paragraph{PSZ2 G099.48+37.72} (A2216) Extended emission is noticed only in the lowest-resolution radio emission. Due to the possible contribution of partially subtracted discrete sources, we classified it as uncertain.

\paragraph{PSZ2 G099.48+55.60} (A1925) This is a double radio relic system and it is claimed here for the first time.

\paragraph{PSZ2 G099.55+34.23} Central diffuse emission in form of a candidate radio halo is observed here for the first time. Since the fitting of the radio halo surface brightness profile does not provide reliable results, we classified this emission as cRH*. Another source elongated in the N-S direction with uncertain origin is located W to the cluster center.

\paragraph{PSZ2 G099.86+58.45} The radio halo in this cluster was claimed by previous \lofar\ observations \citep{cassano19, digennaro21spectral, digennaro21fast, vanweeren21} and is confirmed in our images.

\paragraph{PSZ2 G100.14+41.67} (A2146) This cluster hosts a double radio relic system that has been observed with the \vla\ and \lofar\ \citep{hlavaceklarrondo18, hoang19a2146}. Due to the bad data quality, we could not perform the analysis of this system.

\paragraph{PSZ2 G100.45-38.42} (A2626) Peculiar emission in form of four arcs was observed in this system \citep{gitti04, gitti13a2626, kale17a2626, ignesti17}, which led to the nickname of Kite cluster. Recent \lofar\ observations suggest that the emission originate from the central radio galaxy whose fossil radio plasma has been compressed and revived as a consequence of motions of the ICM \citep{ignesti20a2626}.

\paragraph{PSZ2 G100.96-24.07} The candidate radio halo is claimed here for the first time.

\paragraph{PSZ2 G105.55+77.21} (A1691) Extended emission is noticed only in the lowest-resolution radio image. Due to the possible contribution of partially subtracted discrete sources, we classified it as uncertain.

\paragraph{PSZ2 G106.41+50.82} (A1918) Possible diffuse radio emission is observed at the center of the cluster. However, due to an artifact introduced by the discrete source subtraction, we classified the diffuse emission as uncertain.

\paragraph{PSZ2 G106.61+66.71} Central diffuse emission in form of a radio halo was discovered by \citet{vanweeren21} in the analysis of the \lotss-DR1 cluster sample and was later confirmed by \citet{paul21}. Our images agree with previous results.
 
\paragraph{PSZ2 G107.10+65.32} (A1758) This is a double galaxy cluster (dubbed PSZ2 G107.10+65.32 N and PSZ2 G107.10+65.32 S in Table~\ref{tab:cw}) that has been intensively studied in the past with different instruments \citep{kempner01, giovannini09, venturi13gmrt,  botteon18a1758, botteon20a1758, schellenberger19, vanweeren21}. It hosts a double radio halo system \citep{botteon18a1758}, and the two cluster components are connected with a bridge of radio emission \citep{botteon20a1758}. The northern cluster hosts revived fossil plasma emission and the southern system hosts a radio relic. Our images agree with previous results.

\paragraph{PSZ2 G108.27+48.66} The candidate radio halo is claimed here for the first time.

\paragraph{PSZ2 G109.22-44.01} The candidate radio relic is claimed here for the first time.

\paragraph{PSZ2 G109.97+52.84} The radio halo is claimed here for the first time.

\paragraph{PSZ2 G111.75+70.37} (A1697) The relic and halo in this system were firstly reported by \citet{paul20}. This cluster belongs to the \lotss-DR1 sample presented by \citet{vanweeren21}. Our images confirm the presence of an elongated radio halo and a relic in the NE with a long trail of emission.

\paragraph{PSZ2 G112.48+56.99} (A1767) The radio halo is claimed here for the first time.

\paragraph{PSZ2 G113.29-29.69} Diffuse emission is observed at the center of the cluster. Due to the artifacts introduced by the subtraction of discrete sources, we classified it as uncertain.

\paragraph{PSZ2 G113.91-37.01} The radio halo and relics are claimed here for the first time. This is a double radio relic system. This cluster is displayed in Fig.~\ref{fig:showcases}.

\paragraph{PSZ2 G114.31+64.89} (A1703) Hints of diffuse radio emission in this cluster were noticed with the \vla\ by \citet{owen99rass} and with \lofar\ by \citet{savini18group}. Recently, \citet{vanweeren21} confirmed the presence of a radio halo during the analysis of the \lotss-DR1 sample. Our images agree with previous findings.

\paragraph{PSZ2 G114.99+70.36} (A1682) This systems hosts complex diffuse emission associated with the ICM and radio galaxies that has been studied in the past with \gmrt\ and \lofar\ \citep{venturi08, venturi11, venturi13gmrt, macario13, clarke19, vanweeren21}. Due to the difficulty of disentangling the diffuse emission, we classified it as uncertain. This cluster is displayed in Fig.~\ref{fig:showcases}.

\paragraph{PSZ2 G116.32-36.33} There are two separate X-ray components for this cluster, RX J0027.8+2616 and WHL J002727.0+260707. They are dubbed PSZ2 G116.32-36.33 N and PSZ2 G116.32-36.33 S in Table~\ref{tab:cw} respectively. RX J0027.8+2616 is the target of the \chandra\ observation and it is within the FoV of the \xmm\ observation targeted at WHL J002727.0+260707.

\paragraph{PSZ2 G116.50-44.47} Central diffuse emission in form of a radio halo is observed here for the first time. Since the fitting of the radio halo surface brightness profile does not provide reliable results, we classified this emission as RH*. We also report for the first time a radio relic located in the N.

\paragraph{PSZ2 G118.34+68.79} The presence of a candidate radio halo was claimed during the analysis of the \lotss-DR1 sample \citep{vanweeren21}. Our images confirm the presence of central diffuse emission in form of a candidate radio halo. Since the fitting of the radio halo surface brightness profile does not provide reliable results, we classified this emission as cRH*.

\paragraph{PSZ2 G121.03+57.02} The radio relic in the S of the cluster is located where there is an apparent jump of the X-ray surface brightness and it is claimed here for the first time. The emission in the center has uncertain origin due to the possible residual of discrete source subtraction.

\paragraph{PSZ2 G123.00-35.52} Central diffuse emission in form of a radio halo is observed here for the first time. Since the fitting of the radio halo surface brightness profile does not provide reliable results, we classified this emission as RH*.

\paragraph{PSZ2 G124.20-36.48} (A115) A prominent and bright radio relic in the N periphery of the system was originally claimed with \vla\ 1.4 GHz observations by \citet{govoni01six}. Follow-up studies focused on the connection between relic and shock \citep{botteon16a115}, also exploiting \gmrt\ observations at 610 MHz \citep{hallman18}. Due to the bad quality of the \lotss\ data, we could not study the radio properties of this system. The PSZ2 source comprises two subclusters, which are dubbed PSZ2 G124.20-36.48 N and PSZ2 G124.20-36.48 S in Table~\ref{tab:cw}.

\paragraph{PSZ2G126.28+65.62} A radio halo in this system was claimed by \citet{digennaro21fast} with \lofar\ data at 144 MHz. Follow-up \ugmrt\ observations at 550$-$900 MHz did not show the presence of diffuse radio emission in the cluster possibly due to its steep spectrum \citep{digennaro21spectral}. Owing to the difficulty of separating the contribution of discrete sources from the diffuse emission, we classified this cluster as uncertain.

\paragraph{PSZ2 G133.59+50.68} Central diffuse emission in form of a radio halo is observed here for the first time. Since the fitting of the radio halo surface brightness profile does not provide reliable results, we classified this emission as RH*.

\paragraph{PSZ2 G133.60+69.04} (A1550) The radio halo in this cluster was originally claimed by \citet{govoni12} and recently confirmed during the analysis of the \lotss-DR1 sample \citep{vanweeren21}. We confirm the presence of a radio halo elongated in the E-W direction and the presence of possible revived fossil plasma. This system will be analyzed in detail in a forthcoming publication (Pasini et al., in preparation).

\paragraph{PSZ2 G135.17+65.43} This cluster was recently studied by \citet{vanweeren21} as part of the \lotss-DR1 cluster sample. They found the presence of a halo and relic (source G in \citealt{vanweeren21}). We confirm the previous classification.

\paragraph{PSZ2 G138.32-39.82} The radio halo in this cluster was originally claimed by \citet{kale13} in the context of the Extended \gmrt\ Radio Halo Survey. Our images confirm the detection.

\paragraph{PSZ2 G139.00+50.92} (A1351) The existence of a halo source in this cluster was firstly reported by \citet{owen99rass} and later studied by different authors \citep{giacintucci09a1351, giovannini09}. Our images confirm the presence of radio halo emission that is contaminated by two bright cluster AGN.

\paragraph{PSZ2 G141.05-32.61} The candidate radio halo is claimed here for the first time.

\paragraph{PSZ2 G143.26+65.24} (A1430) The presence of a radio halo in this cluster was claimed by \citet{vanweeren21} during the analysis of the \lotss-DR1 sample, and later confirmed by \citet{paul21}. A dedicated work on this system, composed of two cluster components, using \lofar\ data has been recently published by \citet{hoeft21}.

\paragraph{PSZ2 G143.44+53.66} A number of diffuse radio sources are noted in the cluster center. These sources are claimed here for the first time, and their nature is still uncertain. This cluster is displayed in Fig.~\ref{fig:showcases}.

\paragraph{PSZ2 G144.23-18.19} The candidate radio halo is claimed here for the first time. This is the only target in our sample in which the indication of diffuse emission in the cluster center is noticed in a cluster without redshift/mass.

\paragraph{PSZ2 G144.99-24.64} The candidate radio relic is claimed here for the first time. Another diffuse emission with uncertain origin is observed at the center of our images.

\paragraph{PSZ2 G145.65+59.30} A candidate radio halo was claimed in this system during the analysis of the \lotss-DR1 sample \citep{vanweeren21}. We classified this emission as uncertain as its location is offset with respect to the bulk of the X-ray emission.

\paragraph{PSZ2 G147.88+53.24} The presence of central diffuse emission in the form of a radio halo was recently claimed by \citet{digennaro21fast} and studied with deeper observations in the context of the \lotss\ Deep Fields \citep{osinga21} and with \ugmrt\ data \citep{digennaro21spectral}. Our images confirm the emission (which is classified as candidate radio halo due to the lack of X-ray data).

\paragraph{PSZ2 G148.36+75.23} The radio halo is claimed here for the first time. The diffuse emission surrounds the central cluster AGN. This cluster is displayed in Fig.~\ref{fig:showcases}.

\paragraph{PSZ2 G149.22+54.18} (A1132) The radio halo in this system was claimed with \lofar\ observations by \citet{wilber18a1132} and later studied with deeper observations in the context of the \lotss\ Deep Fields \citep{osinga21}. Our images agree with previous results.

\paragraph{PSZ2 G149.75+34.68} (A665) The presence of diffuse emission coincident with the cluster center was detected in early observations \citep{moffet89, giovannini00, kempner01}, and it was further studied by other authors \citep{feretti04spectral, vacca10, george21gmrt}. Our images show that the emission entirely fills the central volume of the cluster and has a sharp surface brightness edges toward N. This cluster is displayed in Fig.~\ref{fig:example_reimaging}.

\paragraph{PSZ2 G150.56+46.67} The candidate radio halo is claimed here for the first time.

\paragraph{PSZ2 G150.56+58.32} The presence of a radio halo elongated in the NW-SE direction was recently claimed by \citet{vanweeren21} during the analysis of the \lotss-DR1 sample. Our images agree with previous results.

\paragraph{PSZ2 G151.19+48.27} (A959) Diffuse radio emission in this system was reported in \citet{owen99rass}. More recently, this system was studied in detail with \lofar\ and \gmrt\ observations by \citet{birzan19} who confirmed the radio halo and identified a radio relic in the SE. Our images agree with these results.

\paragraph{PSZ2 G154.13+40.19} The candidate radio halo is claimed here for the first time. The diffuse emission surrounds the central cluster AGN. 

\paragraph{PSZ2 G156.26+59.64} The presence of a candidate radio halo in this cluster was reported by \citet{vanweeren21} during the analysis of the \lotss-DR1 sample. Here we confirm the previous classification.

\paragraph{PSZ2 G160.83+81.66} The presence of a radio halo in this system was recently claimed by \citet{digennaro21fast} and followed-up with the \ugmrt\ \citep{digennaro21spectral}. Our images confirm the emission. This is the highest-$z$ radio halo cluster observed to date.

\paragraph{PSZ2 G163.61+34.30} Central diffuse emission in form of a candidate radio halo is observed here for the first time. Since the fitting of the radio halo surface brightness profile does not provide reliable results, we classified this emission as cRH*.

\paragraph{PSZ2 G164.65+46.37} The radio halo is claimed here for the first time. The emission is elongated along the NW-SE direction.

\paragraph{PSZ2 G165.06+54.13} (A990) The radio halo in this cluster was recently claimed by \citet{hoang21a990} using \lofar\ observations. Our images agree with previous results.

\paragraph{PSZ2 G165.46+66.15} (A1240) This cluster hosts a double radio relic system. The two relics are located to the N and S, and were firstly hinted by \citet{kempner01}, and later firmly claimed by \citet{bonafede09double}. This cluster has been the subject of a dedicated study by \citet{hoang18}, who employed \lofar, \gmrt, and \vla\ observations. Our images agree with previous findings.

\paragraph{PSZ2 G166.09+43.38} (A773) The existence of diffuse emission in the central region of the cluster was suggested by \nvss\ and \wenssE\ (\wenss) observations \citep{giovannini99, kempner01}. The cluster was then targeted with the \vla, and the presence of a radio halo was firmly claimed \citep{govoni01six, cuciti21a}. In our images, the halo appears more extended than previously observed.

\paragraph{PSZ2 G166.62+42.13} (A746) The prominent radio relic in the W of this system was firstly claimed by \citet{vanweeren11nvss}. In addition, our images highlight the presence of other two fainter and smaller relics in the E and NE directions. An elongated structure with high surface brightness surrounded by extended diffuse emission is observed toward the cluster center. We classify the extended emission a radio halo, but did not provide a flux density due to the difficulty of disentangling the emission from the bright elongated structure, which possibly traces a radio relic projected on the center of the cluster. A detailed study of this system will be presented in a forthcoming paper. This cluster is displayed in Fig.~\ref{fig:showcases}.

\paragraph{PSZ2 G172.63+35.15} (A655) The radio halo is claimed here for the first time and will be the focus of a forthcoming work (Groeneveld et al., in preparation).

\paragraph{PSZ2 G176.27+37.54} The candidate radio halo is claimed here for the first time. The diffuse emission surrounds the central cluster AGN. 

\paragraph{PSZ2 G179.09+60.12} (A1068) The radio halo is claimed here for the first time, although hints of diffuse emission in the cluster center were noticed in the past by \citet{govoni09} using \vla\ observations at 1.4 GHz. The detailed analysis of this system will be presented in a forthcoming paper (Biava et al., in preparation).

\paragraph{PSZ2 G181.06+48.47} This is a double radio relic system and it is claimed here for the first time.

\paragraph{PSZ2 G183.30+34.98} The candidate radio halo is claimed here for the first time. A diffuse emission elongated in the N-S direction with uncertain origin is also observed to the E of the cluster center.

\paragraph{PSZ2 G184.68+28.91} (A611) The candidate radio halo is claimed here for the first time. The diffuse emission surrounds the central cluster AGN. 

\paragraph{PSZ2 G186.37+37.26} (A697) Hints of diffuse emission located in the cluster center and were firstly claimed by \citet{kempner01} using \wenss\ observations. The radio halo was later confirmed and studied in detail with the \vla, \gmrt\ and \wsrt\ \citep{venturi08, vanweeren11nvss, macario10, macario13}. Our images agree with previous results.

\paragraph{PSZ2 G186.61+62.94} The candidate radio halo is claimed here for the first time. 

\paragraph{PSZ2 G186.99+38.65} The radio halo and radio relic are claimed here for the first time. The radio halo is slightly elongated in the N-S direction while the relic lays in the NE outskirts of the cluster. Curiously, this relic has a mildly convex morphology; to the best of our knowledge, such an unusual curvature has been reported only for the relics in SPT-CL J2023-5535 \citep{hyeonghan20}, the Ant Cluster \citep{botteon21ant}, and Abell 3266 (Riseley et al., in preparation).

\paragraph{PSZ2 G189.31+59.24} (A1033) This cluster hosts revived fossil plasma \citep{degasperin15a1033} and the prototype of Gently Re-Energized Tail \citep[GReET;][]{degasperin17gentle}. In our images, diffuse radio emission in form of a radio halo is observed to the N of the GReET. This radio halo is claimed here for the first time. A dedicated paper on this target is currently in preparation (Edler et al., in preparation).

\paragraph{PSZ2 G190.61+66.46} The radio halo and radio relic are claimed here for the first time. The radio halo is elongated in the E-W direction while the relic is located to the W of the halo emission.

\paragraph{PSZ2 G192.18+56.12} (A961) The radio halo is claimed here for the first time. 

\paragraph{PSZ2 G192.77+33.14} (A671) The candidate radio halo is claimed here for the first time. If confirmed, this would be the radio halo in the lowest-mass cluster observed to date.

\paragraph{PSZ2 G192.90+29.63} The candidate radio halo is claimed here for the first time.
 
\paragraph{PSZ2 G195.60+44.06} (A781) This is a complex system with multiple galaxy cluster components \citep[e.g.,][]{sehgal08} with two components at redshift $z \sim 0.3$ (dubbed PSZ2 G195.60+44.06 E2 and PSZ2 G195.60+44.06 W1 in Table~\ref{tab:cw}) and the other two (dubbed PSZ2 G195.60+44.06 E1 and PSZ2 G195.60+44.06 W1 in Table~\ref{tab:cw}) at redshift $z \sim 0.43$. The presence of a radio halo in the main component of A781 (PSZ2 G195.60+44.06 E2) was disputed in the literature, being observed with the \vla\ at 1.4 GHz \citep{govoni11} but not with the \gmrt\ at lower frequency \citep{venturi08, venturi11, venturi13gmrt}. Deeper observations performed with \lofar\ did not confirm the presence of a radio halo in this system \citep{botteon19a781}. Our images agree with the latter results. The peripheral diffuse emission in the SE was originally believed to trace a radio relic \citep{venturi08, govoni11}, while recently it has been proposed that it results from the interaction between a weak shock and a radio galaxy \citep{botteon19a781}.

\paragraph{PSZ2 G198.46+46.01} The candidate radio relic is claimed here for the first time. It is located in the S and it is elongated in the NE-SW direction. Another diffuse, elongated, emission with uncertain origin is observed toward the cluster center.

\paragraph{PSZ2 G205.90+73.76} The radio halo and double radio relic system are claimed here for the first time. The radio halo is elongated in the N-S direction. The two relics are located in symmetric directions (N-S) with respect to the radio halo emission.

\end{appendix}

\end{document}